\documentclass[preprint2]{emulateapj}

\usepackage{amsmath,natbib,graphicx}
\usepackage[page]{appendix}
\usepackage{epsf}
\input{epsf}
\usepackage{epsfig}
\DeclareGraphicsExtensions{.eps,.jpg,.pdf,.png}

\newcommand{\Tcmb}{\mbox{$T_{\mbox{\tiny CMB}}$}}

\newcommand{\ukcmb}{\mbox{$\mu \mbox{K}_{\mbox{\tiny CMB}}$}}
\newcommand{\kcmb}{\mbox{$\mbox{K}_{\mbox{\tiny CMB}}$}}

\newcommand{\ltsima}{$\; \buildrel < \over \sim \;$}
\newcommand{\ltsim}{\lower.5ex\hbox{\ltsima}}
\newcommand{\gtsima}{$\; \buildrel > \over \sim \;$}
\newcommand{\gtsim}{\lower.5ex\hbox{\gtsima}}
\newcommand{\vell}{{\boldsymbol{\ell}}}

\defcitealias{nagai07}{N07}
\defcitealias{staniszewski09}{S09}

\begin{document}

\title{Sunyaev--Zel'dovich Cluster Profiles Measured With the South Pole Telescope}

\author{ 
T.~Plagge,\altaffilmark{1} B.~A.~Benson,\altaffilmark{2,3}
P.~A.~R.~Ade,\altaffilmark{4} K.~A.~Aird,\altaffilmark{2}
L.~E.~Bleem,\altaffilmark{3,5} J.~E.~Carlstrom,\altaffilmark{3,5,6,7}
C.~L.~Chang,\altaffilmark{3,6} H.-M.~Cho,\altaffilmark{1}  T.~M.~Crawford,\altaffilmark{3,7}
A.~T.~Crites,\altaffilmark{3,7} T.~de~Haan,\altaffilmark{8} M.~A.~Dobbs,\altaffilmark{8}
E.~M.~George,\altaffilmark{1} N.~R.~Hall,\altaffilmark{9}
N.~W.~Halverson,\altaffilmark{10} G.~P.~Holder,\altaffilmark{8}
W.~L.~Holzapfel,\altaffilmark{1} J.~D.~Hrubes,\altaffilmark{2}
M.~Joy, \altaffilmark{11}
R.~Keisler,\altaffilmark{3,5,6} L.~Knox,\altaffilmark{9}
A.~T.~Lee,\altaffilmark{1,12} E.~M.~Leitch,\altaffilmark{3,7}
M.~Lueker,\altaffilmark{1}
D.~Marrone,\altaffilmark{3,13}
J.~J.~McMahon,\altaffilmark{3,6,14} J.~Mehl,\altaffilmark{7}
S.~S.~Meyer,\altaffilmark{2,3,6,7} J.~J.~Mohr,\altaffilmark{15,16,17}
T.~E.~Montroy,\altaffilmark{18,19} 
S.~Padin,\altaffilmark{3,7}
C.~Pryke,\altaffilmark{3,5,6,7} C.~L.~Reichardt,\altaffilmark{1}
J.~E.~Ruhl,\altaffilmark{18} K.~K.~Schaffer,\altaffilmark{3,6}
L.~Shaw,\altaffilmark{8} E.~Shirokoff,\altaffilmark{1}
H.~G.~Spieler,\altaffilmark{12}
B.~Stalder,\altaffilmark{20} Z.~Staniszewski,\altaffilmark{18,19} 
A.~A.~Stark,\altaffilmark{20}
K.~Vanderlinde,\altaffilmark{8} 
J.~D.~Vieira,\altaffilmark{3,5,6} R.~Williamson,\altaffilmark{2}
and O.~Zahn\altaffilmark{21}
}

\altaffiltext{1}{Department of Physics,
University of California, Berkeley, CA 94720, USA}
\altaffiltext{2}{University of Chicago,
5640 South Ellis Avenue, Chicago, IL 60637, USA}
\altaffiltext{3}{Kavli Institute for Cosmological Physics,
University of Chicago,
5640 South Ellis Avenue, Chicago, IL 60637, USA}
\altaffiltext{4}{Department of Physics and Astronomy,
Cardiff University, CF24 3YB, UK}
\altaffiltext{5}{Department of Physics,
University of Chicago,
5640 South Ellis Avenue, Chicago, IL 60637, USA}
\altaffiltext{6}{Enrico Fermi Institute,
University of Chicago,
5640 South Ellis Avenue, Chicago, IL 60637, USA}
\altaffiltext{7}{Department of Astronomy and Astrophysics,
University of Chicago,
5640 South Ellis Avenue, Chicago, IL 60637, USA}
\altaffiltext{8}{Department of Physics,
McGill University,
3600 Rue University, Montreal, Quebec H3A 2T8, Canada}
\altaffiltext{9}{Department of Physics,
University of California, One Shields Avenue, Davis, CA 95616, USA}
\altaffiltext{10}{Department of Astrophysical and Planetary Sciences and Department of Physics,
University of Colorado,
Boulder, CO 80309, USA}
\altaffiltext{11}{Department of Space Science, VP62,
NASA Marshall Space Flight Center,
Huntsville, AL 35812, USA}
\altaffiltext{12}{Physics Division,
Lawrence Berkeley National Laboratory,
Berkeley, CA 94720, USA}
\altaffiltext{13}{National Radio Astronomy Observatory,
520 Edgemont Road Charlottesville, VA 22903, USA}
\altaffiltext{14}{Department of Physics, University of Michigan, 450 Church street, Ann
Arbor, Mi, 48109, USA}
\altaffiltext{15}{Department of Physics,
Ludwig-Maximilians-Universit\"{a}t,
Scheinerstr.\ 1, 81679 M\"{u}nchen, Germany}
\altaffiltext{16}{Excellence Cluster Universe,
Boltzmannstr.\ 2, 85748 Garching, Germany}
\altaffiltext{17}{Max-Planck-Institut f\"{u}r extraterrestrische Physik,
Giessenbachstr.\ 85748 Garching, Germany}
\altaffiltext{18}{Physics Department,
Case Western Reserve University,
Cleveland, OH 44106, USA}
\altaffiltext{19}{Center for Education and Research in Cosmology
and Astrophysics, Case Western Reserve University,
Cleveland, OH 44106, USA}
\altaffiltext{20}{Harvard-Smithsonian Center for Astrophysics,
60 Garden Street, Cambridge, MA 02138, USA}
\altaffiltext{21}{Berkeley Center for Cosmological Physics,
Department of Physics, University of California, and Lawrence Berkeley
National Labs, Berkeley, CA 94720, USA}

\email{tplagge@bolo.berkeley.edu}

\begin{abstract} 

We present Sunyaev--Zel'dovich (SZ) measurements of 15 massive X-ray
selected galaxy clusters obtained with the South Pole Telescope (SPT).  The
SZ cluster signals are measured at 150~GHz,
and concurrent 220~GHz data are used to reduce astrophysical contamination.
Radial profiles are computed using a technique that takes into
account the effects of the beams and filtering.
In several clusters, significant SZ decrements are detected
out to a substantial fraction of the virial radius.
The profiles are fit to the $\beta$-model
and to a generalized Navarro-Frenk-White (NFW) pressure profile,
and are scaled and stacked to probe their average behavior.
We find model parameters that are consistent with previous
studies: $\beta = 0.86$ and $r_{\mathrm{core}}/r_{500} = 0.20$
for the $\beta$-model, and $(\alpha_{\mathrm{n}},\beta_{\mathrm{n}},
\gamma_{\mathrm{n}},c_{500})$=$(1.0,5.5,0.5,1.0)$ for the
generalized NFW model.  Both models fit the SPT data comparably
well, and both are consistent with the average SZ profile out to
beyond $r_{500}$.  The integrated Compton-$y$ parameter $Y_{SZ}$
is computed for each cluster using both model-dependent and model-independent
techniques, and the results are compared to X-ray estimates of cluster parameters.
We find that $Y_{SZ}$ scales with $Y_X$ and gas mass with low scatter.
Since these observables have been found to scale with total mass,
our results point to a tight mass-observable relation for the
SPT cluster survey.

\end{abstract}

\keywords{cosmic background radiation -- cosmology: observations -- galaxies:
 clusters: general -- galaxies: clusters: intracluster medium}

\bigskip\bigskip

\section{Introduction}
\label{sec:intro}

Galaxy clusters are the largest known gravitationally collapsed objects.  They
are believed to have taken nearly a Hubble time to form, and their
abundance as a function of redshift is crucially dependent on the composition and 
expansion history of the universe.  A sufficiently large and well-understood
sample of galaxy clusters can therefore be used to constrain cosmological models.
Several survey campaigns using a variety of techniques are currently underway.
Precision measurements of individual clusters can be used to study the mass--observable relations of these 
surveys, as well as to study the structure of the intracluster medium (ICM).

Well-established techniques for probing galaxy clusters include optical, 
radio, and infrared observations of member galaxies, 
gravitational lensing of background galaxies,
and X-ray measurements of 
bremsstrahlung from hot intra-cluster electron gas.  The 
Sunyaev-Zel'dovich (SZ) effect, which occurs when cosmic microwave 
background (CMB) photons inverse Compton scatter off the hot electrons
of the ICM \citep{sunyaev72,birkinshaw99},
provides a complementary and powerful probe \citep{carlstrom02}.  

The SZ effect can be approximated as the sum of two components: 
that caused by the random thermal motion of the hot scattering electrons,
and that caused by the cluster peculiar velocity relative 
to the Hubble flow.  The former, known as the thermal SZ effect, 
leads to a distortion of the CMB Planck spectrum of the form
\begin{equation}
  f(x) = \left( x \frac{e^x + 1}{e^x - 1} - 4 \right) (1 + \delta_{SZ}(x,T_e)),
\end{equation}
where $x\equiv h \nu / k_B T_{\mathrm{CMB}}$ is the dimensionless frequency,
$T_{\mathrm{CMB}}$ is the temperature of the CMB, $T_e$ is the 
temperature of the ICM electrons, and $\delta_{SZ}$ is a relativistic 
correction\footnote{We adopt the relativistic correction from \citet{itoh00}.}.
For nonrelativistic
electrons, the distortion results in a decrement in the measured
CMB temperature at frequencies below $\sim 217$ GHz and an increment above.
The amplitude of the effect, expressed as a change in temperature
$\Delta T$ relative to the CMB temperature \Tcmb, is given by
\begin{equation}
  \label{eqn:deltsz}
  \frac{\Delta T}{T_{\mathrm{CMB}}} = f(x) y = f(x) \int \sigma_T n_e 
  \frac{k_B T_e}{m_e c^2} dl, 
\end{equation}
where $y$ is the Compton-$y$ parameter, a measure of integrated pressure; 
$\sigma_T$ is the Thomson cross section; $n_e$ and $T_e$ are the electron 
number density and temperature; and the integral is along the line of sight 
through the cluster.  The integrated
$y$ parameter out to a given radius is a quantity proportional to the total
thermal energy of the cluster, and is expected to provide a low scatter
estimate of the total cluster mass \citep{dasilva04,nagai06,motl05,kravtsov06a}.
For typical massive clusters, the kinetic SZ effect produces a signal
that is much smaller than the thermal SZ signal except at 
frequencies near the $\sim 217$~GHz thermal SZ null.  

Much of the past work on the thermal SZ effect 
\citep[e.g.,][]{grego00a,reese02,benson04,halverson09}
has modeled cluster SZ signals by the $\beta -$model
\citep{cavaliere76,cavaliere78}.  The functional
form of this model can be derived from a parameterization of
density under the assumption of isothermality, but since isothermality
is a poor assumption for many clusters---particularly for
merging and cooling core systems---we instead consider the
$\beta -$model simply as a fitting function without a strong physical 
motivation.  Measurements and simulations show significant
departures from this model within the cluster core and
outside $\sim r_{2500},$ the radius at which the
cluster matter density drops to 2500 times the critical density
of the universe \citep{vikhlinin05,piffaretti05,hallman07}.  \citet{laroque06} 
examined two extensions of the $\beta -$model using a sample 
of 38 clusters with both X-ray and SZ data.  As an alternative
to the $\beta -$model, \citet[][hereafter N07]{nagai07} have proposed, 
and \citet{mroczkowski09} and \citet{arnaud09} have 
investigated, a generalization of the Navarro-Frenk-White (NFW) model as a 
parameterization of pressure in clusters.  This model 
is expected to describe the cluster pressure out to 
a significant fraction of the virial radius.

In this work, we describe a set of deep SZ cluster observations 
at 150 and 220~GHz undertaken with the South Pole Telescope (SPT)
during the austral winter of 2008.  Data from the 150~GHz band are used to
measure the thermal SZ signal, and concurrent data from 
the 220~GHz band---which is near the
thermal SZ null---are used to reduce the effect of astrophysical
contamination.  The 220~GHz data are too noisy to allow a 
detection of the kinetic SZ effect, so in this paper we focus
exclusively on the thermal SZ effect.  

Of the 13 highest-luminosity 
REFLEX clusters \citep{boehringer04} in the range of
elevation angles accessible to the SPT, 11 
make up the primary sample for this work; a supplemental sample
of four additional clusters is also included.  Both samples are
listed in Table~\ref{table:clusterlist}.  Taking advantage of the
SPT's wide field imaging capabilities, we make a model independent
estimate of the radial profiles of each cluster and place
constraints on the integrated pressure out to large radius.  
We also fit the profiles both to the $\beta -$model 
and to the generalized NFW model proposed by \citetalias{nagai07}
(hereafter the GNFW model).  Using a stacked analysis, we estimate the slope
of the pressure profile in the cluster outskirts.
Finally, we estimate the integrated $y$ parameters 
for each cluster and compare to X-ray results.

\section{Instrument, Observations, and Data Reduction}
\label{sec:obs-reduc}

\subsection{Instrument}
\label{sec:instrument}

The SPT is an off-axis Gregorian telescope with a 10 meter diameter
primary dish.  One of a new generation of SZ 
instruments (such as ACT \citep{hincks09} and APEX-SZ \citep{halverson09}), 
the SPT has several advantages for cluster observations:
\begin{itemize}
\item Its resolution
($\sim 1'$ at 150~GHz) is matched to cluster scales.
\item It is able to map clusters efficiently due to its 
high sensitivity over a wide field of view ($\sim 1$ $\mathrm{deg}^2$).
\item It observes concurrently in multiple bands, including one
band centered near the 217~GHz SZ null, allowing primary CMB 
anisotropies to be separated from SZ signals.
\item The atmosphere at its observing site, roughly 1~km
from the geographic South Pole, is the best on Earth for 
millimeter astronomy \citep{bussmann05}.
\end{itemize}
The SPT is optimized to image large areas of the CMB sky to arcminute 
resolution with a primary goal of identifying a sample of massive clusters 
out to high redshift.  The first such clusters were reported 
in \citet[][hereafter S09]{staniszewski09}.  

The SPT receiver is equipped with a 960-element array of 
superconducting transition edge sensor (TES) bolometers, each
read out by a frequency-domain-multiplexed system using
superconducting quantum interference device (SQUID) amplifiers.  
This detector array is divided into six triangular ``wedges,'' each of which
is populated with detectors sensitive to radiation within a
single frequency band.  During the 2008 observing season, 
the band centers were $\sim$95~GHz for one wedge, $\sim$150~GHz
for three wedges, and $\sim$220~GHz for the remaining two.
Due to a defect in detector fabrication, few of the 95~GHz 
detectors were active during the 2008 observing season,
and so only the two higher-frequency bands are included in this 
work.  The bands were measured with a Fourier transform spectrometer,
and the bands were found to be $\Delta \nu=35.6$~GHz at $\nu=152.6$~GHz 
and $\Delta \nu =42.7$~GHz at $\nu=220.0$~GHz.  In a typical observation 
used in this paper, the instantaneous per-detector sensitivity was
$\sim 500 \ukcmb \sqrt{\mathrm{s}}$ for the 150~GHz detectors, while
the 220~GHz detectors had sensitivities ranging from
$\sim 1000-1500 \ukcmb \sqrt{\mathrm{s}}$\footnote{Throughout
this work, \kcmb \ refers to equivalent CMB fluctuation
temperature, i.e., intensity divided by $dB_\nu/dT$ (at $T=2.73 \mathrm{K})$.}.
For further details on the telescope and receiver, 
see \citet{ruhl04,padin08}, and \citet{carlstrom09}.

\subsection{Observations}
\label{sec:obs}

The clusters discussed in this work were observed using
constant-elevation scans, which entail sweeping the telescope
at constant angular velocity in azimuth across the field and back,
stepping in elevation, and repeating.  This observing strategy
is similar to that described in \citetalias{staniszewski09}, but
over a much smaller field---approximately 1 deg$^2$.  The azimuthal
scan speed is set such that
$\sim 0^\circ .25$ on the sky is scanned in 1~s, and each
elevation step is $0^\circ .005$.  A complete set of scans across the entire
region, which we refer to as one observation, takes approximately
55 minutes to complete.  The observations are then combined
to produce a single map in each band for each field.  The inner region
of each 150~GHz map, approximately $0^\circ .5 \times 0^\circ .5$ on the sky,
has even coverage to within $10$\%.  Coverage is less uniform at 220~GHz
due to the spatial distribution of the detectors on the focal plane.

One of the clusters in the primary sample, AS~0520, overlaps with the survey
field described in \citetalias{staniszewski09}.  We reprocessed the 2008
150~GHz and 220~GHz survey data with filtering equivalent to that
described in Section \ref{sec:reduction}, and split the survey data into
subsets, each with a weight per pixel equal to that of the
targeted observations.  This allows the survey data to be treated in
a similar manner to the data from the targeted observations for the
purposes of calculating noise estimates.  Adding
the survey data effectively doubles the number of
targeted observations of this cluster.

Between individual observations, we perform a series of
short calibration measurements described in more detail in
\citet{carlstrom09}.  These include measurements of a chopped 
thermal source, $\sim2^\circ$ elevation nods,
and scans across the galactic HII regions RCW38 and MAT5a.  This
series of regular measurements allows us to identify
detectors with good performance, assess relative detector gains, 
and monitor atmospheric opacity.

\begin{table*}
\begin{tiny}
\begin{center}
\caption{Summary of cluster sample.}
\begin{tabular}{|l|l|l|l|l|l|l|l|}
\hline\hline
\rule[-2mm]{0mm}{6mm}
ID & R.A.$^a$ & Decl.$^a$ & $z$ &
  REFLEX $L_{\mathrm{X}}^{\mathrm{d}}$ & $T_e^{\mathrm{e}}$ & $M_{\mathrm{gas}}$ & Ref.$^b$ \\
  &  &  & & ($10^{44}$ erg s$^{-1}$) & (keV) & ($10^{14} M_{\odot}$) & \\
\hline
 \multicolumn{8}{|c|}{Primary sample} \\
\hline
A 2744 &
  $0^\mathrm{h} 14^\mathrm{m}    18^\mathrm{s} .6$ &
  $-30^\circ 23^\prime    15^{\prime\prime} .4$ &
  0.307 &
  12.92 &
   10.1 $\pm$     0.3 & 1.0 $\pm$ 0.2 &
\citet{zhang06} \\
RXCJ0217.2-5244 &
  $2^\mathrm{h} 17^\mathrm{m}    12^\mathrm{s} .6$ &
  $-52^\circ 44^\prime    49^{\prime\prime} .2$ &
  0.343 &
  12.03 &
   10.9$^{\mathrm{c}}$ & - &
\citet{boehringer04} \\
RXCJ0232.2-4420 &
  $2^\mathrm{h} 32^\mathrm{m}    18^\mathrm{s} .8$ &
  $-44^\circ 20^\prime    51^{\prime\prime} .9$ &
  0.284 &
   9.65 &
    7.0 $\pm$     0.3 & 0.9 $\pm$ 0.2 &
\citet{zhang06} \\
AS 0520 &
  $5^\mathrm{h} 16^\mathrm{m}    35^\mathrm{s} .2$ &
  $-54^\circ 30^\prime    36^{\prime\prime} .8$ &
  0.294 &
  13.87 &
    7.5 $\pm$     0.3 & 0.8 $\pm$ 0.2 &
\citet{zhang06} \\
RXCJ0528.9-3927 &
  $5^\mathrm{h} 28^\mathrm{m}    52^\mathrm{s} .5$ &
  $-39^\circ 28^\prime    16^{\prime\prime} .7$ &
  0.284 &
  13.12 &
    7.2 $\pm$     0.4 & 0.9 $\pm$ 0.1 &
\citet{zhang06} \\
AS 0592 &
  $6^\mathrm{h} 38^\mathrm{m}    46^\mathrm{s} .5$ &
  $-53^\circ 58^\prime    18^{\prime\prime} .0$ &
  0.222 &
  10.62 &
    8.0 $\pm$     0.4 & - &
\citet{hughes09} \\
A 3404 &
  $6^\mathrm{h} 45^\mathrm{m}    30^\mathrm{s} .0$ &
  $-54^\circ 13^\prime    42^{\prime\prime} .1$ &
  0.164 &
   7.36 &
    8.1 $\pm$     0.3 & 0.9 $\pm$ 0.1 &
\citet{zhang08} \\
1ES 0657-56 &
  $6^\mathrm{h} 58^\mathrm{m}    30^\mathrm{s} .2$ &
  $-55^\circ 56^\prime    33^{\prime\prime} .7$ &
  0.297 &
  23.03 &
   10.6 $\pm$ 0.2 & 1.8 $\pm$ 0.3 &
\citet{zhang06} \\
RXCJ2031.8-4037 &
  $20^\mathrm{h} 31^\mathrm{m}    51^\mathrm{s} .5$ &
  $-40^\circ 37^\prime    14^{\prime\prime} .0$ &
  0.342 &
  12.04 &
   10.9$^{\mathrm{c}}$ & - &
\citet{boehringer04} \\
A 3888 &
  $22^\mathrm{h} 34^\mathrm{m}    27^\mathrm{s} .1$ &
  $-37^\circ 44^\prime     7^{\prime\prime} .5$ &
  0.151 &
   7.31 &
    7.8 $\pm$     0.4 & 0.8 $\pm$ 0.1 &
\citet{zhang08} \\
AS 1063 &
  $22^\mathrm{h} 48^\mathrm{m}    44^\mathrm{s} .9$ &
  $-44^\circ 31^\prime    44^{\prime\prime} .4$ &
  0.346 &
  30.79 &
   11.1 $\pm$     1.1 & 1.2 $\pm$ 0.1 &
\citet{maughan08} \\
\hline
 \multicolumn{8}{|c|}{Supplemental sample} \\
\hline
RXCJ0336.3-4037 &
  $3^\mathrm{h} 36^\mathrm{m}     18^\mathrm{s} .7$ &
  $-40^\circ 37^\prime    20^{\prime\prime} .0$ &
  0.173 &
   2.53 &
    7.2$^{\mathrm{c}}$ & - &
\citet{boehringer04} \\
RXCJ0532.9-3701 &
  $5^\mathrm{h} 32^\mathrm{m}    55^\mathrm{s} .9$ &
  $-37^\circ 1^\prime    34^{\prime\prime} .5$ &
  0.275 &
   6.94 &
    9.5 $\pm$     0.4 & 0.6 $\pm$ 0.1 &
\citet{zhang06} \\
MACSJ0553.4-3342 &
  $5^\mathrm{h} 53^\mathrm{m}    27^\mathrm{s} .2$ &
  $-33^\circ 42^\prime    53^{\prime\prime} .0$ &
  0.407 &
  - &
   13.1$^{+3.8}_{-2.5}$ & - &
\citet{cavagnolo08} \\
A 3856 &
  $22^\mathrm{h} 18^\mathrm{m}    39^\mathrm{s} .9$ &
  $-38^\circ 53^\prime    43^{\prime\prime} .6$ &
  0.141 &
   3.78 &
    6.7 $\pm$     0.2 & 0.5 $\pm$ 0.1 &
\citet{zhang08} \\
\hline
\end{tabular}
\label{table:clusterlist}
\end{center}
\end{tiny}
$^{\mathrm{a}}$R.A.\ and decl.\ are the right ascension and declination
    angles of the cluster X-ray centroid.\newline
$^{\mathrm{b}}$R.A., decl., redshift, and electron
temperature $T_e$ are taken from these papers and
references therein.  REFLEX luminosities
are from \citet{boehringer04}.  \newline
$^{\mathrm{c}}$For the clusters without published
X-ray temperatures, $T_e$ was estimated using a
scaling relation from luminosity \citep{markevitch98}.\newline
$^{\mathrm{d}}$Luminosity in rest frame 0.1-2.4~keV band corrected
for any missing flux.\newline
$^{\mathrm{e}}$Temperatures are averages over a range of radii:
0.1-0.5$r_{500}$ in \citet{zhang06}, 0.2-0.5$r_{500}$ in
\citet{zhang08}, $r<r_{500}$ in \citet{maughan08}, and
$r<r_{2500}$ with core excised \citet{cavagnolo08}.  The
radial range is unspecified in \citet{hughes09}.
\end{table*}

\begin{table*}
\begin{tiny}
\begin{center}
\caption{Summary of SPT cluster maps.}
\begin{tabular}{|l|l|l|l|l|l|}
\hline\hline
\rule[-2mm]{0mm}{6mm}
ID & R.A.$^a$ & Decl.$^a$ & Observing Time & 150~GHz Depth$^b$ &
 220~GHz Depth$^b$ \\
 & & & (hr) & ($\mu$K$_{\mathrm{CMB}}$) & ($\mu$K$_{\mathrm{CMB}}$) \\
\hline
 \multicolumn{6}{|c|}{Primary sample} \\
\hline
A 2744 &
  $0^\mathrm{h} 14^\mathrm{m}    16^\mathrm{s} .8$ &
  $-30^\circ 23^\prime    22^{\prime\prime} .2$ &
   6.83 &
   12.3 &
   36.2 \\
RXCJ0217.2-5244 &
  $2^\mathrm{h} 17^\mathrm{m}    13^\mathrm{s} .3$ &
  $-52^\circ 44^\prime    49^{\prime\prime} .5$ &
   8.32 &
   11.5 &
   40.8 \\
RXCJ0232.2-4420 &
  $2^\mathrm{h} 32^\mathrm{m}    16^\mathrm{s} .5$ &
  $-44^\circ 21^\prime    15^{\prime\prime} .5$ &
  10.46 &
    9.2 &
   32.6 \\
AS 0520 &
  $5^\mathrm{h} 16^\mathrm{m}    35^\mathrm{s} .4$ &
  $-54^\circ 30^\prime    19^{\prime\prime} .3$ &
  18.48$^{\mathrm{c}}$ &
   10.3 &
   17.4 \\
RXCJ0528.9-3927 &
  $5^\mathrm{h} 28^\mathrm{m}    52^\mathrm{s} .0$ &
  $-39^\circ 27^\prime    57^{\prime\prime} .6$ &
   5.23 &
   14.1 &
   39.4 \\
AS 0592 &
  $6^\mathrm{h} 38^\mathrm{m}    46^\mathrm{s} .1$ &
  $-53^\circ 58^\prime    31^{\prime\prime} .3$ &
   8.32 &
   18.9 &
   70.7 \\
A 3404 &
  $6^\mathrm{h} 45^\mathrm{m}    29^\mathrm{s} .3$ &
  $-54^\circ 13^\prime    20^{\prime\prime} .0$ &
   5.54 &
   14.5 &
   46.5 \\
1ES 0657-56 &
  $6^\mathrm{h} 58^\mathrm{m}    29^\mathrm{s} .7$ &
  $-55^\circ 56^\prime    38^{\prime\prime} .2$ &
  12.59 &
    7.9 &
   22.5 \\
RXCJ2031.8-4037 &
  $20^\mathrm{h} 31^\mathrm{m}    51^\mathrm{s} .1$ &
  $-40^\circ 37^\prime    23^{\prime\prime} .2$ &
   5.23 &
   16.3 &
   49.7 \\
A 3888 &
  $22^\mathrm{h} 34^\mathrm{m}    26^\mathrm{s} .6$ &
  $-37^\circ 44^\prime    57^{\prime\prime} .9$ &
   5.23 &
   15.4 &
   42.6 \\
AS 1063 &
  $22^\mathrm{h} 48^\mathrm{m}    44^\mathrm{s} .9$ &
  $-44^\circ 31^\prime    43^{\prime\prime} .1$ &
  12.21 &
   10.3 &
   29.7 \\
\hline
 \multicolumn{6}{|c|}{Supplemental sample} \\
\hline
RXCJ0336.3-4037 &
  $3^\mathrm{h} 36^\mathrm{m}    16^\mathrm{s} .5$ &
  $-40^\circ 37^\prime    26^{\prime\prime} .6$ &
   5.23 &
   13.1 &
   48.1 \\
RXCJ0532.9-3701 &
  $5^\mathrm{h} 32^\mathrm{m}    55^\mathrm{s} .1$ &
  $-37^\circ 1^\prime    40^{\prime\prime} .8$ &
  10.46 &
   12.8 &
   38.1 \\
MACSJ0553.4-3342 &
  $5^\mathrm{h} 53^\mathrm{m}    24^\mathrm{s} .5$ &
  $-33^\circ 42^\prime    35^{\prime\prime} .4$ &
   5.82 &
   15.4 &
   51.0 \\
A 3856 &
  $22^\mathrm{h} 18^\mathrm{m}    37^\mathrm{s} .6$ &
  $-38^\circ 53^\prime    54^{\prime\prime} .8$ &
   8.72 &
   11.0 &
   34.9 \\
\hline
\end{tabular}
\label{table:clustermapparams}
\end{center}
\end{tiny}
$^a$R.A.\ and decl.\ are the right ascension and declination angles
  where the average SZ decrement within a $1^\prime$ radius is the largest.\newline
$^b$Depth is determined by the standard deviation of the
jackknife maps smoothed with a Gaussian with FWHM=$1^\prime$.\newline
$^c$AS 0520 maps include data from the 2008 $5^\mathrm{h} 30^\mathrm{m}$ survey,
which effectively doubles the amount of targeted observing time.  This value
includes the factor of $\sim 2$.
\end{table*}

\subsection{Data reduction}
\label{sec:reduction}

The data reduction process used for this work is
similar to the preliminary survey pipeline described
in \citetalias{staniszewski09}, with some minor differences
noted below. 

\subsubsection{Data selection and filtering}
\label{sec:cuts}

The first step in the processing pipeline is to identify the data
that will be included in each single-observation map.  For every
observation, a set of well-performing detectors is identified by
applying a set of noise and sensitivity criteria 
\citepalias{staniszewski09}.  On average,
302 detectors at 150~GHz and 165 at 220~GHz pass 
this initial cut.  The data are then divided into individual
azimuth scans. $0.7\%$ of scans have large instantaneous 
pointing offsets or data acquisition problems, and are omitted from 
further processing.  A further set of selection criteria is then
applied to detectors on a scan-by-scan basis (excluding, for example,
data contaminated by cosmic rays).  Data from detectors passing these
cuts, an average of 291 detectors at 150~GHz and 162 at 220~GHz for
each scan, are included in the final maps.

The receiver exhibits some sensitivity to the pulse-tube cooler used to
cool it to $\sim 4$~K, resulting in occasional lines in the 
detector noise power spectra, so a small amount of bandwidth ($< 0.4$\%) is
notch filtered out of the detector timestreams.  The timestreams
are further processed by deconvolving the effect of the detector 
time constants.  We also low-pass filter at 25~Hz, well above the
signal bandwidth.

For every scan, each detector's timestream is then fit
simultaneously to a number of template functions, and
the best fit to each template is subtracted out.
This time-domain filter reduces the effect of low-frequency
noise in the detector timestreams due to readout noise or to
atmospheric fluctuations.   For the analysis presented here, we use as our 
templates a second-order Legendre polynomial and a common mode constructed from
the wedge-averaged timestreams,
resulting in a characteristic filter scale of roughly one half degree.  
This light filtering is sufficient to remove much of the low-frequency
noise, due in part to the outstanding atmospheric conditions at the South Pole.

The common mode template is constructed from the mean across all of the
well-performing detectors in a given wedge using a nominal relative 
calibration.  In \citetalias{staniszewski09}, by contrast, the common 
mode template was constructed from the mean across the entire 
array for each band.  Since the 2008 SPT receiver
contains three 150~GHz and two 220~GHz wedges, the detectors
in the two bands have significantly different distributions on 
the focal plane; thus the array common mode 
removes signals on different spatial scales for each band.  
Since the common mode templates for individual wedges 
are relatively similar to one another, the wedge common mode preserves 
signals on similar spatial scales for each band, simplifying comparisons 
between the 150~GHz and 220~GHz maps.

\subsubsection{Calibration and beams}
\label{sec:calibration}

We take advantage of the superb WMAP5 absolute calibration and
calibrate the SPT data by comparing the 150~GHz 
SPT maps with WMAP5 $V$- and $W$-band maps \citep{hinshaw09}.  
The SPT performed dedicated calibration scans of 
five large fields totaling 1250 deg$^2$ of sky during 2008. 
The WMAP5 maps were then resampled with the SPT pointing information, and 
the resulting timestreams were passed through a simulated SPT analysis 
pipeline to capture the filtering applied during the SPT map-making 
algorithm.  The ratio of the cross-spectra of the SPT maps and
the filtered WMAP maps after correcting for the instrumental beams 
provides an estimate of the relative calibration factor between the 
two experiments.  This procedure, similar to that used for the analysis
of ACBAR data and detailed in \cite{reichardt09a},
results in an absolute calibration accurate to 3.6\% at 150~GHz.

The 150~GHz calibration is transferred to 220~GHz by making use of
the overlapping coverage on SPT's high signal to noise maps of CMB 
fluctuations in large survey fields. 
We calculate the ratio of the cross-spectra of the map at each 
frequency band to the auto-spectra of the CMB-dominated 150~GHz 
map after correcting for the beam and filtering differences. 
The calibration uncertainty of the 220~GHz maps is found to be 7.2\%. 
Because the thermal SZ signal in the 220~GHz map is small, the higher 
calibration uncertainty only marginally increases the errors
on the cluster profiles discussed in this paper.

The 2008 SPT beams were measured by observing bright sources and 
averaging the response for all detectors in a given band.  The
sidelobes of the beam were estimated from observations of planets, 
and---since the detectors show signs of nonlinearity when
pointed directly at planets---the main lobes were estimated
from observations of quasars.
The main lobes are reasonably approximated by two-dimensional elliptical 
Gaussians with average full widths at half maximum (FWHM) of 
1.2 and 1.1 arcmin for 150 and 220~GHz.
These are larger than would be naively calculated from the aperture
diameter and observing wavelength due to the combination of an
underilluminated primary (to reduce spillover) and telescope
pointing uncertainty; see \citet{padin08} for details.  
The sidelobes fall off from $\sim -30 \mathrm{dB}$ at a radius of 
$5^\prime$ to $< -45 \mathrm{dB}$ at $30^\prime$, and account for 
about $15\%$ of the degree-scale beam response.   Errors in this beam estimate 
contribute to the overall calibration uncertainty, and also add small 
($\sim 1$\%) additional uncertainties to the best-fit model parameters.  

\begin{figure*}[ht]
\begin{center}
\includegraphics[width=0.8\textwidth]{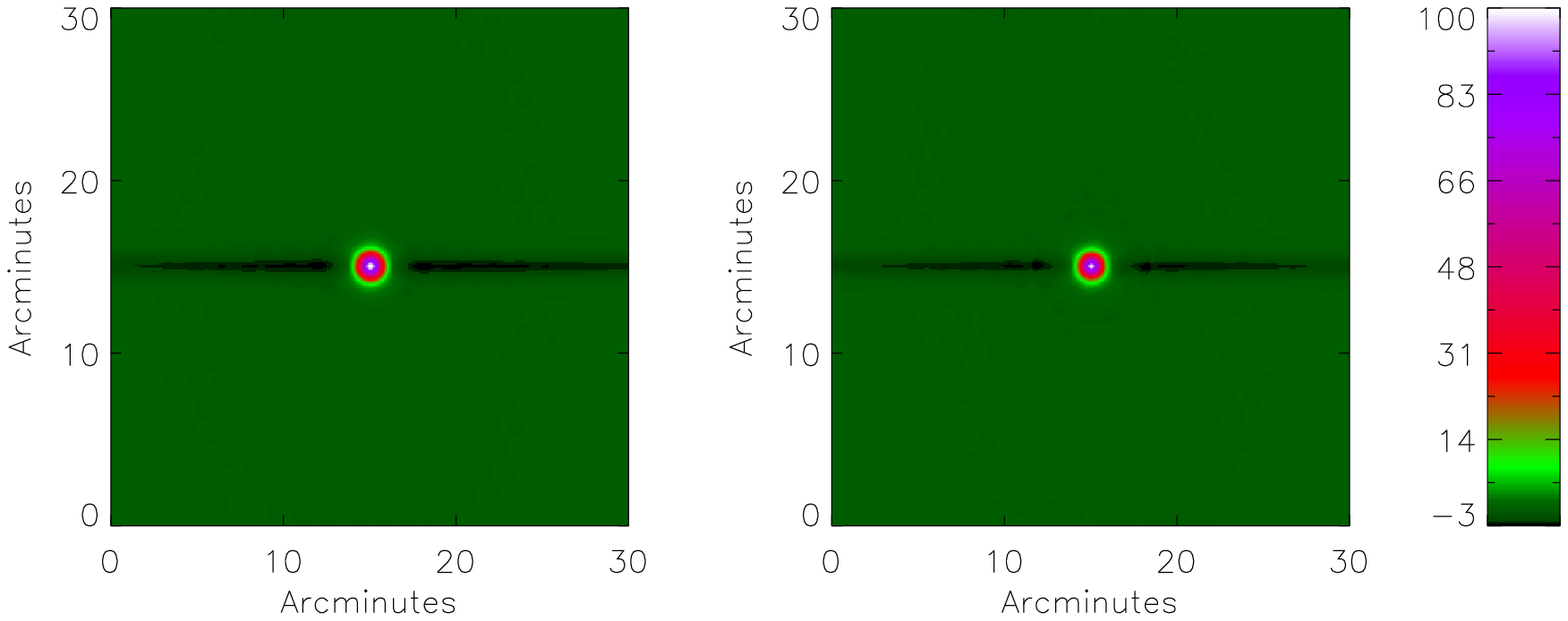}
\caption{Point source transfer functions at 150~GHz (left) and
220~GHz (right).  Units are percent of maximum.  The stripe 
through the center of the maps is the result of the polynomial
template removal.}
\label{fig:xfer}
\end{center}
\end{figure*}

\begin{figure*}
\begin{center}
\includegraphics[width=0.8\textwidth]{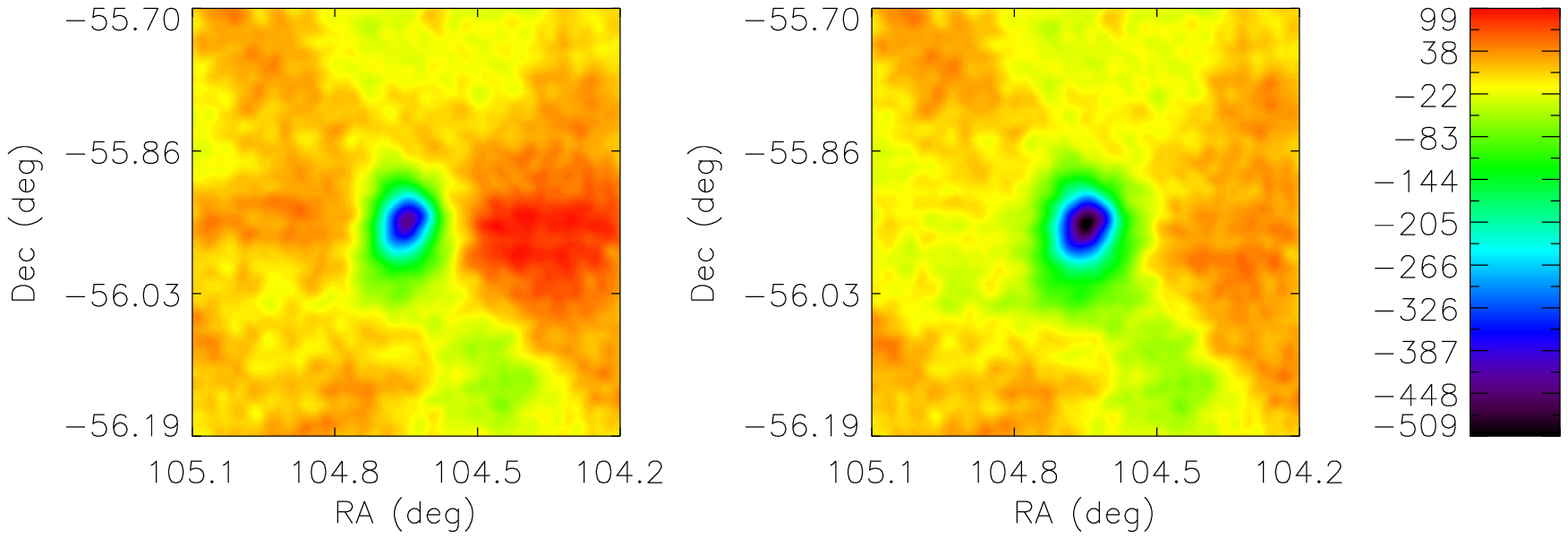}
\caption{Maps of 1ES~0657-56 at 150 GHz, with the cluster
region unmasked (left) and masked (right) for time-domain
filtering.  Units are $\mu K_{\mathrm{CMB}}$.  In the 
masked map, data from a circular region $8^\prime$ in radius
centered on the cluster are not included in the polynomial and wedge common
mode construction. This map gives a clear picture of
the shape and amplitude of the SZ emission, but has slightly higher
noise.  The artifacts of the time-domain filtering 
in the unmasked map are taken into account in the cluster profiles
by the algorithm described in Section~\ref{sec:profiles}.}
\label{fig:masked}
\end{center}
\end{figure*}

We model the response of the instrument to an astronomical source 
by constructing a simulated map of a point source, convolving it with the 
beam, constructing simulated 
timestreams using telescope pointing information, and then running
the timestreams through our processing pipeline.  The output
map normalized to the input signal is referred to as the 
transfer function, and is used to subtract bright point sources
from the cluster maps.  Representative transfer
functions are shown in Figure~\ref{fig:xfer}. 
The negative stripe at constant elevation is due to the polynomial 
template subtraction, and results in apparent temperature 
increments to the east and west of the cluster proportional in
amplitude to the SZ temperature decrement.  This
effect can be ameliorated by masking the cluster 
before applying the time-domain filtering 
(see Figure~\ref{fig:masked}), but doing so 
reduces the effectiveness of the atmospheric 
noise removal.  Since we account for the effects of the 
time-domain filtering on the profiles as described 
in Section \ref{sec:profiles}, we avoid degrading 
the signal to noise of our results and analyze the 
unmasked data.

\begin{figure}
\begin{center}
\includegraphics[width=0.4\textwidth]{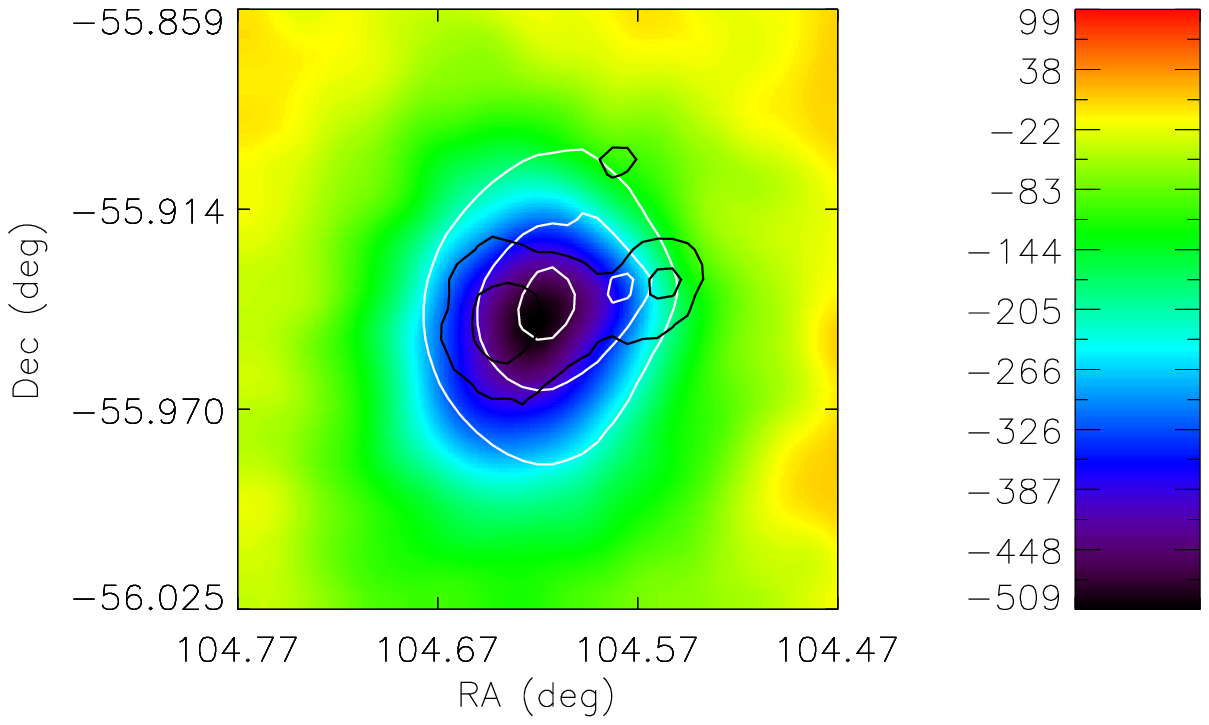}
\caption{150~GHz 1ES~0657-56 masked map with X-ray plasma density
(white) and weak lensing surface mass density (black) 
contours overlaid.  Units are $\mu K_{\mathrm{CMB}}$.  
This cluster is an ongoing merger, and the electron
gas is not in equilibrium within the gravitational potential well. 
The SZ signal tracks the X-ray plasma density more 
closely than the surface mass density.}
\label{fig:bullet_masked}
\end{center}
\end{figure}

\subsubsection{Mapmaking and astrophysical background cleaning}
\label{sec:mapmaking}

For every observation, a map is made for each
observing frequency using the processed data for all detectors
in that band.  Pointing information (R.A.\ and decl.) is
calculated for each detector using focal-plane offsets measured
in observations of the galactic HII regions, and boresight pointing
calculated using data from the telescope pointing readout system, with
a set of corrections described in \citetalias{staniszewski09} that result
in a pointing rms of $7^{\prime\prime}$.
These coordinates are then converted to pixel number using a 
Sanson--Flamsteed projection \citep{calabretta02} with 0.25 arcmin
pixels.  All measurements of a given pixel's brightness
are averaged using inverse-variance weighting based on the
mean of each detector's processed and relative-gain-scaled
power spectrum between 1 and 3~Hz, which for our scan speed
is a frequency range well matched to the scale of SZ cluster signals. 

All of the single-observation maps for a given cluster in a given
band are then co-added to produce a set of single-band maps, which we denote
as $M^{150}$ and $M^{220}$ for the 150~GHz and 220~GHz bands.  Each
single-band map consists of signals from several astrophysical sources: 
the SZ effect, which is at arcminute scales and is much stronger at 150~GHz;
the primary CMB anisotropy, which dominates at large spatial scales
and is the same in both bands; and a background of typically faint 
dusty point sources, 
which are at small angular scales and are stronger at 220~GHz. 
We wish to produce a ``band subtracted map'' $M^{\mathrm{sub}}$, in which
the 220~GHz map is used to remove a fraction of the CMB and 
point source background while leaving the SZ signal intact.  Since
the 2008 SPT receiver was less sensitive to CMB fluctuations at 220~GHz 
than at 150~GHz, and since the atmospheric
noise is worse in the higher-frequency band, this background removal must be
handled carefully to avoid introducing excess noise into the band 
subtracted map.

Using our knowledge of the spatial scales of the CMB and point
source signals, we apply a matched spatial filter $\psi$ to
the 220~GHz map, which we construct by requiring that the variance
in the band subtracted map be minimized 
\citep{haehnelt96,herranz02a,herranz02b,melin06}.
Since the signal and noise are more easily characterized in the spatial 
frequency domain, we adopt the flat sky approximation and construct the
filter as a function of multipole moment $\vell$.  
Denoting the Fourier transform of $X$ as $\tilde{X}$, we can
express the value of the band subtracted map at a given multipole moment 
as
\begin{equation}
  \tilde{M}^{sub}_{\vell} = \tilde{M}^{150}_{\vell} - \psi_{\vell} \tilde{M}^{220}_{\vell}.
\end{equation}
The variance at a given $\vell$, $V_{\vell}$, is then given by
 \begin{align*}
  V_{\vell} = &\left( \begin{array}{cc} 1 & -\psi_{\vell} \end{array} \right) \\
           &\left( \begin{array}{cc} \tilde{S}^{\mathrm{CMB}}_{\vell} + \tilde{S}^{\mathrm{PS}}_{\vell} + \tilde{N}^{150}_{\vell} & 
                                    \tilde{S}^{\mathrm{CMB}}_{\vell} + \alpha \tilde{S}^{\mathrm{PS}}_{\vell} \\
                                    \tilde{S}^{\mathrm{CMB}}_{\vell} + \alpha \tilde{S}^{\mathrm{PS}}_{\vell} & 
                                    \tilde{S}^{\mathrm{CMB}}_{\vell} + \alpha^2 \tilde{S}^{\mathrm{PS}}_{\vell} + \tilde{N}^{220}_{\vell}
                  \end{array} \right) \\
&\left( \begin{array}{c} 1 \\ -\psi_\vell \end{array} \right),
\end{align*}
where $\tilde{S}^{\mathrm{CMB}}_{\vell}$ is the CMB signal, $\tilde{S}^{\mathrm{PS}}_{\vell}$ is the 
point source signal, $\tilde{N}^{X}_{\vell}$ is the noise in band $X$, and
$\alpha$ is a factor corresponding to the spectral index
of the point sources.  The filter that minimizes this variance is given by
\begin{equation}
 \psi_{\vell} = \frac{\tilde{S}^{\mathrm{CMB}}_{\vell} + \alpha \tilde{S}^{\mathrm{PS}}_{\vell}}
                  {\tilde{S}^{\mathrm{CMB}}_{\vell} + \alpha^2 \tilde{S}^{\mathrm{PS}}_{\vell} + \tilde{N}^{220}_{\vell}}.
\end{equation}
Note that both the signal and the noise terms are anisotropic with respect
to azimuth and elevation due to the scan strategy employed in these 
observations, necessitating a two-dimensional filter function.  Errors in either the
signal or noise terms will result in a slightly sub-optimal filter,
and will thus increase the errors in the results, but will not lead to
systematic misestimations of the profiles or cluster parameters.

The signal covariances of the primary CMB anisotropies and undetected point sources
are estimated using simulations based on the best-fit WMAP5 CMB power spectrum
\citep{nolta09} and the \citet{borys03} model for dusty point sources.
We assume that the point sources are Poisson distributed and have a spectral
index of 2.7\footnote{SPT data indicate a steeper spectral index between 
150 and 220~GHz \citep{hall09}, but we adopt this value---extrapolated from
higher-frequency measurements---so that our estimate of
point source power is an upper limit.  Our results are insensitive to
this choice.}.  The power spectrum of the source count 
distribution is computed using the formalism in \citet{white04}, 
and is inflated by a factor of 40\% to account for lensing by our 
massive cluster targets \citep{lima09}.  Even with this enhancement,
the dusty sources are predicted to contribute an rms of $\sim$4~$\mu$K 
to the band subtracted maps, well below the map noise level.  
We disregard contamination due to radio sources, as explicit simulations 
demonstrate \citep{sehgal10} that they are not likely to significantly 
fill in the 150~GHz SZ decrements of massive $z \ltsim 0.4$ clusters,
and as in most cases we are able to remove bright sources from existing catalogs
(see Section~\ref{sec:sources}).  For both the CMB and the dusty point sources,
we generate 300 simulated maps and apply the same time-domain filtering
that was applied to the SPT maps.  We find $\tilde{S}^{\mathrm{CMB}}_{\vell}$
and $\tilde{S}^{\mathrm{PS}}_{\vell}$ by taking the mean of the two-dimensional spatial 
power spectra over the simulated maps. 

The instrumental and atmospheric noise properties of the maps are
estimated using the two-dimensional power spectra of noise
maps \citep{sayers09,halverson09}.  Under the assumption of
stationarity in the map basis, the noise covariance matrix in band $X$,
$\tilde{N}_{\mathrm{noise}}^{\mathrm{X}}$, is diagonal in the 
spatial frequency domain and equal to the noise power spectrum.
For each observation, we produce one map using only data from 
the left-going scans, and one map using only data from the 
right-going scans.  We then multiply one half of the $2n$ maps 
by $-1$ and co-add to produce a jackknife noise map.
We repeat this process for $m >> n$ combinations of observations,
computing the two-dimensional spatial power spectrum for each individual jackknife 
map.  The average of these power spectra is our estimate of
$\tilde{N}_{\mathrm{noise}}^{\mathrm{X}}$.  

\citet{sayers09} and \citet{halverson09} use this noise estimate to
construct a covariance matrix and to fit models directly to 
their cluster maps.  However, since we have relatively few independent 
observations of each cluster, our noise estimate is less 
well-constrained.  We cannot combine data from different 
clusters to improve the noise estimate due to the fact that the noise
is non-stationary between the observations.  The amplitude of the
atmospheric noise varies with time and as a function of elevation
angle, and the clusters in this sample are at a wide range of
elevation angles.  We circumvent this limitation by computing
the projected radial profiles of the clusters, for which 
the covariance matrix has fewer degrees of freedom and is thus better
constrained by our data.  

\section{Sample selection and cluster maps}
\label{sec:maps}

During the austral winter of 2008, the SPT observed 11 of the
13 highest-luminosity clusters in the REFLEX survey within
the range of declination angles observable by the SPT
($-30^\circ \gtsim$decl.$\gtsim -70^\circ$).  These clusters
form the primary sample used in this work.  One of the two REFLEX
clusters that were omitted, RXCJ1253.6-3931, is associated with a
radio bright PMN source that would have been difficult to remove from the SZ maps.
The other, RXCJ1234.2-3856, was omitted due to time constraints. 
Of the next three brightest REFLEX clusters, RXCJ0532.9-3927
is a part of the supplemental sample; RXCJ2011.3-4037
was found by \citet{maughan08} to have an electron temperature of
just 3.8~keV, and was not detected at greater than $3\sigma$ by the
SPT; and RXCJ1317.1-3821 is associated with a bright radio source.
From the remaining 2008 SPT targeted cluster observations, four
additional clusters were selected to form our supplemental 
sample.  These clusters were chosen because of the 
availability of archived X-ray satellite data, which
will be relevant for future work.  They also contribute
to our understanding of pressure profiles, and demonstrate
the ability of the SPT to make high signal to noise maps of
clusters with lower X-ray luminosities.

The positions and depths of the final co-added maps are listed in 
Table~\ref{table:clustermapparams}, and the images can be found in the
Appendix.  Four maps are shown for each cluster:
the 150 and 220~GHz single-band maps, the band subtracted map, and
a jackknife band subtracted map computed by multiplying half of the observations
by $-1$ before co-adding.  Each map is smoothed by a Gaussian with an 
FWHM of $1^\prime$.  All clusters are detected with high 
significance.  Since the maps each have distinct noise
properties, we do not expect the detection significance to 
provide quantitative information about cluster parameters.

As discussed in Section \ref{sec:calibration}, the time-domain
processing creates large-scale distortions in the maps visible
as temperature increments to the east and west of the clusters.
No mask has been applied to the data used to construct the maps 
in the Appendix.
Figure~\ref{fig:bullet_masked} shows the 1ES~0657-56 SZ map constructed
from data with the cluster masked before time-domain filtering.  Shown
as contours in Figure~\ref{fig:bullet_masked} are publicly available 
X-ray plasma surface density and weak lensing surface density 
data\footnote{http://flamingos.astro.ufl.edu/1e0657/public.html
\citep{clowe06}}.  As expected, once the effects of the beam and 
time-domain filtering are mitigated, the SZ signal closely tracks 
the electron plasma density. 

\subsection{Source contamination}
\label{sec:sources}
  
We attempt to remove bright point sources from the SPT maps.  For
sources more than $2^{\prime}$ away from the cluster center, we
identify and remove all sources detected at $5\sigma$ or greater
in either frequency band.
A point source within $10^{\prime\prime}$ of quasar J001341.2-300926 
was detected at $> 5\sigma$ significance in the 
map for A~2744, $16^\prime .5$ from the cluster center, and has been removed 
from the maps by subtracting the point source transfer function scaled
to the best-fit amplitude.  The SPT measured fluxes 
for this source are 17.7 mJy at 150~GHz and 27.8 mJy at 220~GHz.
In the AS 0592 map, a point source corresponding to
the radio source SUMSS~J063845-535824 was removed.  This
source is $\sim 7^{\prime}$ from the cluster center, and the 
fluxes at 150~GHz and 220~GHz are 21.8~mJy and 13.4~mJy, 
respectively.  

Sources within $2^{\prime}$ of the cluster center may fill in the 
cluster SZ decrement without being directly detected.
We searched existing radio source catalogs within $2^{\prime}$ of our
cluster locations using the NED 
database\footnote{http://nedwww.ipac.caltech.edu/}. 
Two SUMSS sources with $843$~MHz fluxes greater than 10~mJy were identified:
SUMSS~J021714-524529, a 24.5~mJy source near RXCJ0217.2-5244; 
and SUMSS~J065837-555718, a 79.4~mJy source near 1ES~0657-56.  
\citet{colafrancesco07} note that the latter source has been
found to have a power law SED with $\alpha=-0.9$, and would therefore
have fluxes at 150~GHz and 220~GHz that are well below our noise levels.
Insufficient data exist to extrapolate the SUMSS~J021714-524529 
flux to 150 GHz.  We assume that its flux contributes 
negligibly to the SZ flux at 150 GHz, as would be the case 
for typical radio source SEDs.  We note that RXCJ0217.2-5244 is 
not included in our scaling relation analysis.

Dusty galaxies may also appear as bright point sources in the SPT maps.
\citet{wilson08} report a dusty point source approximately $1^\prime$
to the east of 1ES~0657-56, with flux density of $13.5 \pm 0.5$~mJy 
at 270~GHz.  Assuming a spectral index of 2.7, we would expect to 
see 7.8 and 2.8~mJy in the 220 and 150~GHz SPT maps.  The SPT 
does not make a significant detection of 
a source at these coordinates in either band.  We measure a 
temperature increment of $34\pm 33$~$\mu$K$_{\mathrm{CMB}}$ in our
220~GHz map (smoothed by a $1^\prime$ FWHM Gaussian).  By finding
a best-fit amplitude for our point source transfer function, we infer a
220~GHz flux density of $4.2 \pm 4.5$ mJy.  At 150~GHz, the 
strong SZ decrement near the source position prevents us from
determining the source flux.  We find from simulations that a source with the 
predicted fluxes at these coordinates would have a $<5$\% effect
on the corresponding bins in the band subtracted cluster profile.

\section{Projected radial profiles}

\subsection{Profile computation}
\label{sec:profiles}

To proceed with measurements of global cluster properties
from the maps, it is useful to approximate the clusters as azimuthally
symmetric and to calculate their projected radial profiles, which are
expected to take a universal form \citep{arnaud09}.  We do so by
computing transfer functions for a set of concentric annuli
with inner and outer radii set to the boundaries of the radial 
bins for which we wish to estimate an average SZ decrement.  
These transfer functions are then used to compute a profile transfer function 
matrix, which is inverted and applied to the maps in order to 
account for the effects of the beam and the time-domain
filtering.  This technique also allows
us to characterize the correlated errors between bins.

The cluster center coordinates must be determined in order to generate 
radial profiles.  We define the center position by fitting a
$\beta$-model with $\beta=0.86$ to the map with the center R.A.\ and decl.,
central temperature decrement, and core radius free to vary.  We then
marginalize over the central decrement and core radius to determine
the cluster center.  The fit uncertainties on
the center position vary from $1^{\prime\prime}$ to $4^{\prime\prime}$, 
indicating very high signal to noise detections, and the results
are consistent when the generalized NFW model
is used in place of the $\beta -$model.  The difference between 
the X-ray centroid (which depends primarily on density) and the 
SZ center position (which depends on integrated pressure) is small for 
most clusters, but reaches $40-50^{\prime\prime}$ for RXCJ0232.2-4420,
MACSJ0553.4-3342, and A~3888.  An X-ray-derived pressure map for 
RXCJ0232.2-4420 is shown in \citet{finoguenov05}, and the
SPT centroid falls within the elongated central high-pressure region.

\begin{figure*}[ht]
\begin{center}
\includegraphics[width=0.8\textwidth]{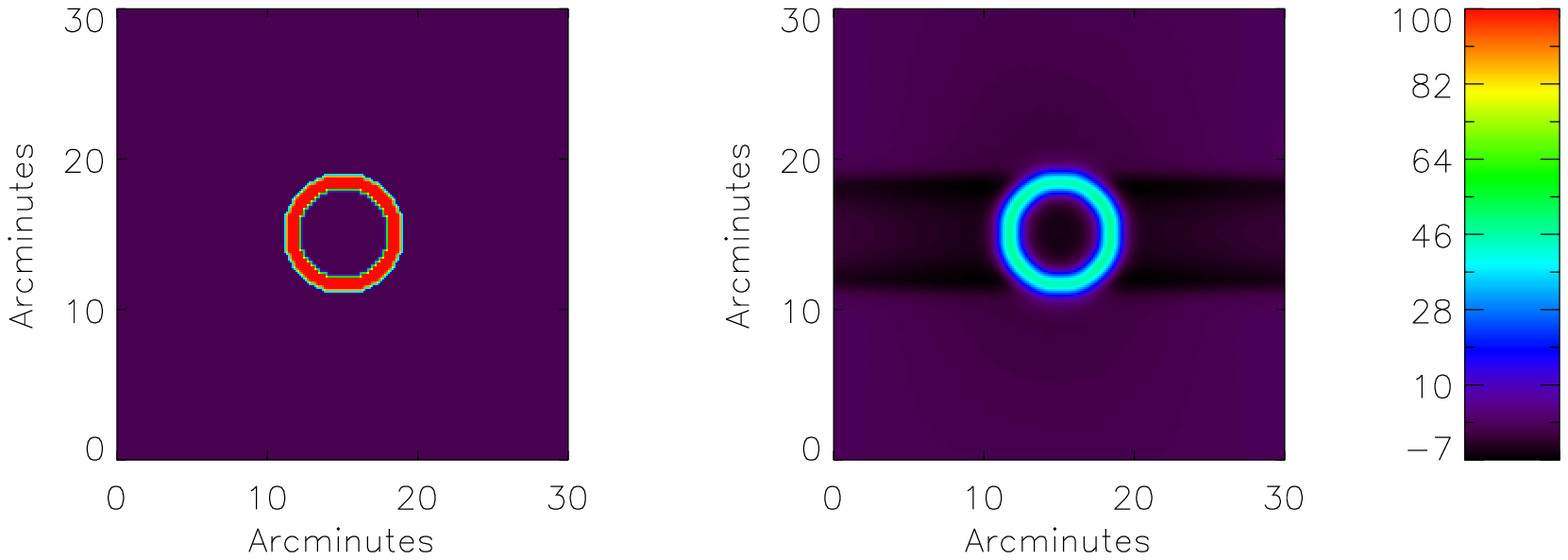}
\caption{Bin $b_3$ (left) and its 150~GHz annular transfer function map (right).
Units are percent of maximum.  The annular transfer function map represents
the response of the SPT to a corresponding annulus of constant temperature,
taking into account the beam and the time-domain filtering.}
\label{fig:ring3}
\end{center}
\end{figure*}

For each band subtracted cluster map $M^{\mathrm{sub}}$, we divide the 
$0^\circ .5 \times 0^\circ .5$ region surrounding the center position
into radial bins $b_i$.  The bin spacing is defined so as to measure
the profile of the cluster center with beam-scale resolution, and to
keep the number of bins small enough so that the noise properties
can be well constrained.  Each bin is then assigned a radius $r_i$ by 
finding the average radius over all pixels within the bin.  For bin $b_i$
and band $X$, we define $T_i$ to be the mean SZ temperature decrement
on the sky within the bin.  We construct a set of simulated maps 
corresponding to each bin in each band, in which the ``temperature'' for
map $i$ is set to 1 inside bin $b_i$ and 0 elsewhere.  These maps are 
then convolved with the appropriate beam.  To account for the effect of
the time-domain filtering, we construct simulated timestreams from these
maps using the telescope pointing data from each observation, 
and apply to them the same polynomial and common mode filtering that
was applied to the data.  We then produce from these filtered
timestreams a set of $n_{\mathrm{bins}}$ processed maps---the annular transfer 
function maps $F^{\mathrm{sub}}_i$ (see Figure~\ref{fig:ring3} for an 
example)---which show the response of the instrument 
at each pixel location to an annulus of constant temperature corresponding 
to bin $b_i$.  Finally, these maps are Hanning apodized, Fourier transformed, 
and represented as vectors $\tilde{F}^{\mathrm{sub}}_i$ of 
length $n_{\mathrm{pixels}}$.  From these vectors, we
construct an $n_{\mathrm{pixels}} \times n_{\mathrm{bins}}$ matrix 
\begin{equation}
\tilde{B}^{\mathrm{sub}}_{ji} = (\tilde{F}^{\mathrm{sub}}_i)_j ,
\end{equation}
the annular transfer function matrix, where $\tilde{B}^{\mathrm{sub}}_{ji}$ 
gives the response in the $j$th Fourier mode of the band subtracted map 
to an impulse in bin $b_i$.  

\begin{figure}
\begin{center}
\includegraphics[width=0.4\textwidth]{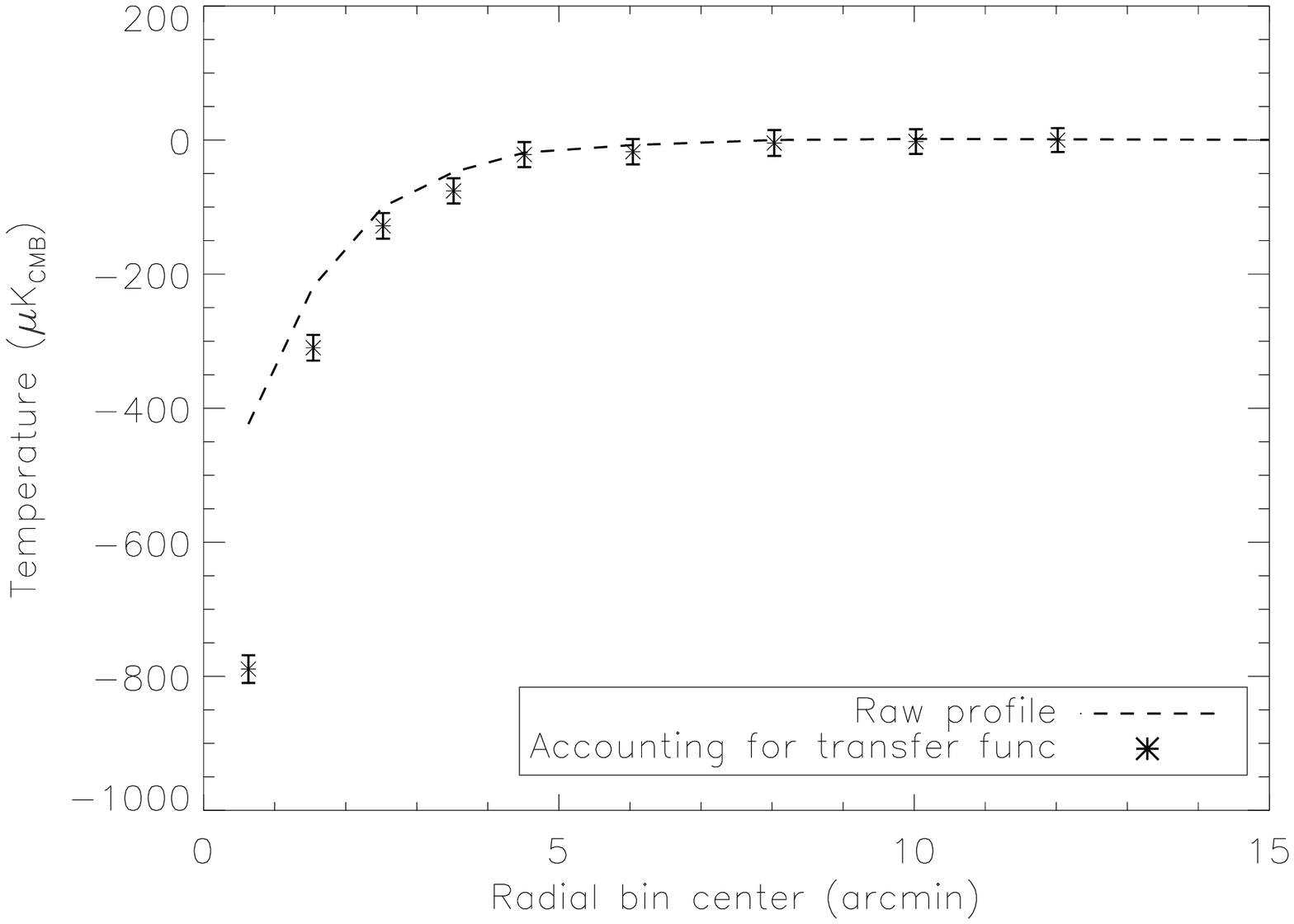} 
\caption{
AS~1063 profile with and without beam and time-domain filtering correction.
The uncorrected profile is computed by averaging the pixels in the 
band subtracted map within the radial bins, and the corrected 
profile is computed as described in Section~\ref{sec:profiles}.
}
\label{fig:decon}
\end{center}
\end{figure}

Using this formalism, we can write a set of equations relating
the Fourier transform of the Hanning apodized measured SPT
map $\tilde{M}^{\mathrm{sub}}_j$ to the sky temperatures $T_i$:
\begin{equation}
  \tilde{M}^{\mathrm{sub}}_j = \tilde{B}^{\mathrm{sub}}_{ji} T_i .
\end{equation}
Since there are many more map pixels than radial bins, this is an
overconstrained system of linear equations, for which the
least-squares solution is given by
\begin{equation}
  T = (\tilde{B}^T \tilde{W} \tilde{B})^{-1} \tilde{B}^T \tilde{W} \tilde{M} ,
\end{equation}
where we have suppressed the superscript denoting band subtracted 
quantities and where $\tilde{W}$ is the weight matrix.  
The weights are given by
$\tilde{W} = 1/(\tilde{N} + \tilde{S}^{\mathrm{CMB}} + \tilde{S}^{\mathrm{PS}})$, 
where $\tilde{N}$ is the detector and atmospheric
noise and $\tilde{S}^{\mathrm{CMB}}$ and $\tilde{S}^{\mathrm{PS}}$ are
the CMB and dusty point source signals.  These terms are all computed
for the band subtracted map using the method described in 
Section~\ref{sec:mapmaking}.  Weighting the Fourier modes by their
noise reduces the profile errors by 20\%-30\% compared with uniform
weights.  The covariance matrix for the profile is given 
by $(\tilde{B}^T W \tilde{B})^{-1}$, plus small additional
terms to account for calibration and beam uncertainties.  Figure~\ref{fig:decon}
shows the importance of this procedure: the amplitudes and shapes of 
the profiles are badly misestimated if the beams and time-domain filtering
are not taken into account.

Since the SPT measures temperature differences, the cluster maps 
have arbitrary constant offsets.  These manifest 
themselves as offsets in the profiles, and could bias the model
fits if they were not taken into account.
We include a constant offset as a parameter in our
model fitting, and marginalize over this parameter in order
to determine the quoted central decrements and angular scale radii.  
This marginalization ensures that our uncertainty in the 
zero point of the map is correctly accounted for in the model
fits.  The profiles shown in the Appendix are 
plotted with the offset fixed such that the temperature
decrement in the outermost radial bin is zero. 

\subsection{Model fitting}
\label{sec:fits}

We make use of the cluster profiles and their noise 
covariance matrices to perform maximum likelihood
fits to proposed models of the ICM.  Our likelihood
function is given by
\begin{equation}
 \mathcal{L}(T
 | \mathcal{M}) = \frac{1}{\sqrt{(2\pi)^n \mathrm{det}(N_{\mathrm{noise}})}}
  \mathrm{exp}\left(-\frac{\chi^2}{2}\right) ,
\end{equation}
where $n=10$ is the number of radial bins, $T$ is the 
measured profile,
$\mathcal{M}$ is the proposed model, and $\chi^2$ is
given by
\begin{equation}
 \chi^2 =  (\mathcal{M} - T)^T
  (N_{\mathrm{noise}})^{-1}
  (\mathcal{M} - T).
\end{equation}
We will consider two models: the 
$\beta$-model and the GNFW model.

The $\beta -$model for thermal SZ cluster signals is 
motivated by the widely used $\beta -$model parameterization 
of the three-dimensional electron number density: 
\begin{equation}
  n_e (\boldsymbol{r}) = n_{\mathrm{e}_0} \left( 
       1 + \frac{r^2}{r^2_{\mathrm{core}}} \right)^{-3\beta / 2},
\end{equation}
where $n_{\mathrm{e}}$ is the electron number density, 
$n_{\mathrm{e}_0}$ is the 
number density at the cluster center, $r$ is the radius from
the cluster center, $r_{\mathrm{core}}$ is the core radius
of the gas distribution, and $\beta$ is the power law index
at large radii.  Using this density function and assuming isothermality, 
Equation \ref{eqn:deltsz} can be integrated along the line
of sight through the cluster to give a simple
analytic formula for the SZ decrement:
\begin{equation}
 \label{eqn:betamodel}
  \Delta T_{\mathrm{SZ}} = \Delta T_0 
    \left( 1 + \frac{\theta^2}{\theta^2_{\mathrm{core}}}
    \right)^{ (1 - 3 \beta)/2},
\end{equation}
where $\theta$ is the angular distance from the cluster 
center, given by $\theta = r / D_A$ where $D_A$ is the 
angular diameter distance (computed assuming that 
$(\Omega_M , \Omega_\Lambda , h) = (0.3, 0.7, 0.7)$), 
$\Delta T_0$ is the central temperature decrement, 
and $\theta_{\mathrm{core}}$ is the 
angle corresponding to $r_{\mathrm{core}}$. 
Although we do not necessarily expect the clusters in our sample
to be isothermal, we nevertheless adopt Equation~\ref{eqn:betamodel} as
a convenient parameterization of the SZ signal.

Two of the $\beta -$model parameters, $\beta$ and 
$r_{\mathrm{core}}$, are highly degenerate and difficult to
constrain with SZ data alone \citep{grego01}.
We fix $\beta$ at $0.86$, the overall best-fit value 
for our sample (see Section~\ref{sec:stacked}).
Typical X-ray cluster analyses yield values of $\beta$ between
about 0.6 and 0.8 \citep[][and citations therein]{laroque06}, but
while the X-ray surface brightness is relatively insensitive to ICM temperature
\citep{mohr99}, the SZ signal is proportional to the integrated
pressure of the ICM.  Therefore, any radial trend in temperature
would lead to systematic differences between X-ray and SZ profiles.
Our finding of $\beta \sim 0.86$ suggests that the ICM temperature
tends to decrease with increasing radius, in qualitative
agreement with direct measurements of cluster temperature profiles
\citep[e.g.,][]{pratt07} and with hydrodynamical cluster simulations.
\citet{hallman07} find that $\beta$ values for SZ profiles are 
systematically higher than X-ray $\beta$ values by a factor 
of 1.21, consistent with the SPT results.

We also fit to the GNFW model, which parameterizes the three-dimensional
electron pressure as
\begin{equation}
  P_{\mathrm{e}}(\boldsymbol{r}) = \frac{P_{\mathrm{e}_0}}{
    (r/r_{\mathrm{s}})^{\gamma_{\mathrm{n}}} 
    \left( 1 + (r/r_{\mathrm{s}})^{\alpha_{\mathrm{n}}}
    \right)^{(\beta_{\mathrm{n}}-\gamma_{\mathrm{n}})/\alpha_{\mathrm{n}}}},
\end{equation}
where $P_{e_0}$ is the pressure at 
the cluster center; $r$ is the radius from the cluster center; 
$r_s = r_{500}/c_{500}$ is a scaling radius set by $r_{500}$, the 
radius within which the average cluster density falls to 500 times the critical
density of the universe at the cluster redshift, and $c_{500}$, a
parameter characterizing the gas concentration; and 
$\alpha_{\mathrm{n}}$, $\beta_{\mathrm{n}}$, and $\gamma_{\mathrm{n}}$ 
set the slope at intermediate ($r \sim r_{\mathrm{s}}$),
large ($r > r_{\mathrm{s}}$), and small ($r < r_{\mathrm{s}}$) radii, 
respectively.  We fix $(\alpha_{\mathrm{n}}, \beta_{\mathrm{n}}, 
\gamma_{\mathrm{n}},c_{500})$ to $(1.0, 5.5, 0.5, 1.0)$, the 
parameters found to be the best fit to our stacked cluster profile
(see Section~\ref{sec:stacked}).  These values are very similar to
those found by \citetalias{nagai07} when fitting to {\it Chandra} data,
$(\alpha_{\mathrm{n}}, \beta_{\mathrm{n}}, 
\gamma_{\mathrm{n}},c_{500}) = (0.9, 5.0, 0.4, 1.3)$\footnote{These 
values are taken from \citet{mroczkowski09}, and are updated from 
the results in \citetalias{nagai07}.}.  As the 
SZ effect is directly proportional to integrated pressure, we 
calculate $\Delta T_{\mathrm{SZ}}$ given the pressure profile by numerically
integrating the above function along the line of sight.  As before, we 
define $\theta_{s} = r_{s} / D_A$.  Note that fitting for $r_{s}$
with $c_{500}$ fixed is equivalent to fitting for $r_{500}$.  However,
because the quality of the GNFW fit varies across our sample, 
we will later make use of a scaling relation to estimate $r_{500}$.

Both models thus contain two free parameters: a central temperature 
decrement, and a scaling radius ($\theta_{\mathrm{core}}$ for the 
$\beta -$model and $\theta_{s}$ for the GNFW model). 
As discussed in Section~\ref{sec:profiles}, an offset parameter is 
also included.

\subsection{The effect of CMB and atmospheric noise}
\label{sec:cmbatm}

The model fitting technique described in Section \ref{sec:fits} 
makes use of the full profile noise covariance matrices, including 
off-diagonal terms that describe correlations.  This is important
because the noise in the SPT maps can introduce spurious structure 
in the profiles that is highly correlated between bins.
Most of these correlations arise from primary CMB anisotropies
and atmospheric noise, the combination of which
dominates the uncertainties in the profiles.

The large angular scale confusion due to primary CMB anisotropies
represents a fundamental limitation for single-band measurements 
of SZ radial profiles.  The SPT 220~GHz data
allow us to reduce the level of confusion; however, the 220~GHz
maps have significantly more atmospheric noise than do the
150~GHz maps.  Timestream filtering can help to mitigate 
atmospheric noise, but filtering too heavily
will remove CMB signal.  We chose our level of
time-domian filtering by balancing these
two effects, in order to maximize the CMB signal to noise
at 220~GHz.  

The matched spatial filter described in Section \ref{sec:mapmaking}  
is constructed to downweight noise-contaminated modes in the 220~GHz
maps, and thus to increase the effectiveness of the CMB subtraction.  
Other spatial filters can be applied to the 220~GHz maps in order 
to change the balance between CMB and atmospheric confusion in the
band subtracted map.  We constructed several alternative 
filters which improved the CMB subtraction at the expense of
introducing atmospheric noise.  In all cases, the resulting profiles
and fits were found to be consistent but the errors were increased: by 
construction, the matched filter yields the best signal to noise in the
band subtracted maps.  Optimal band subtraction reduces
the large angular scale confusion by a factor of 2 or more, 
depending on the depth of the 220~GHz map.

\begin{figure}
\begin{center}
\includegraphics[width=0.40\textwidth]{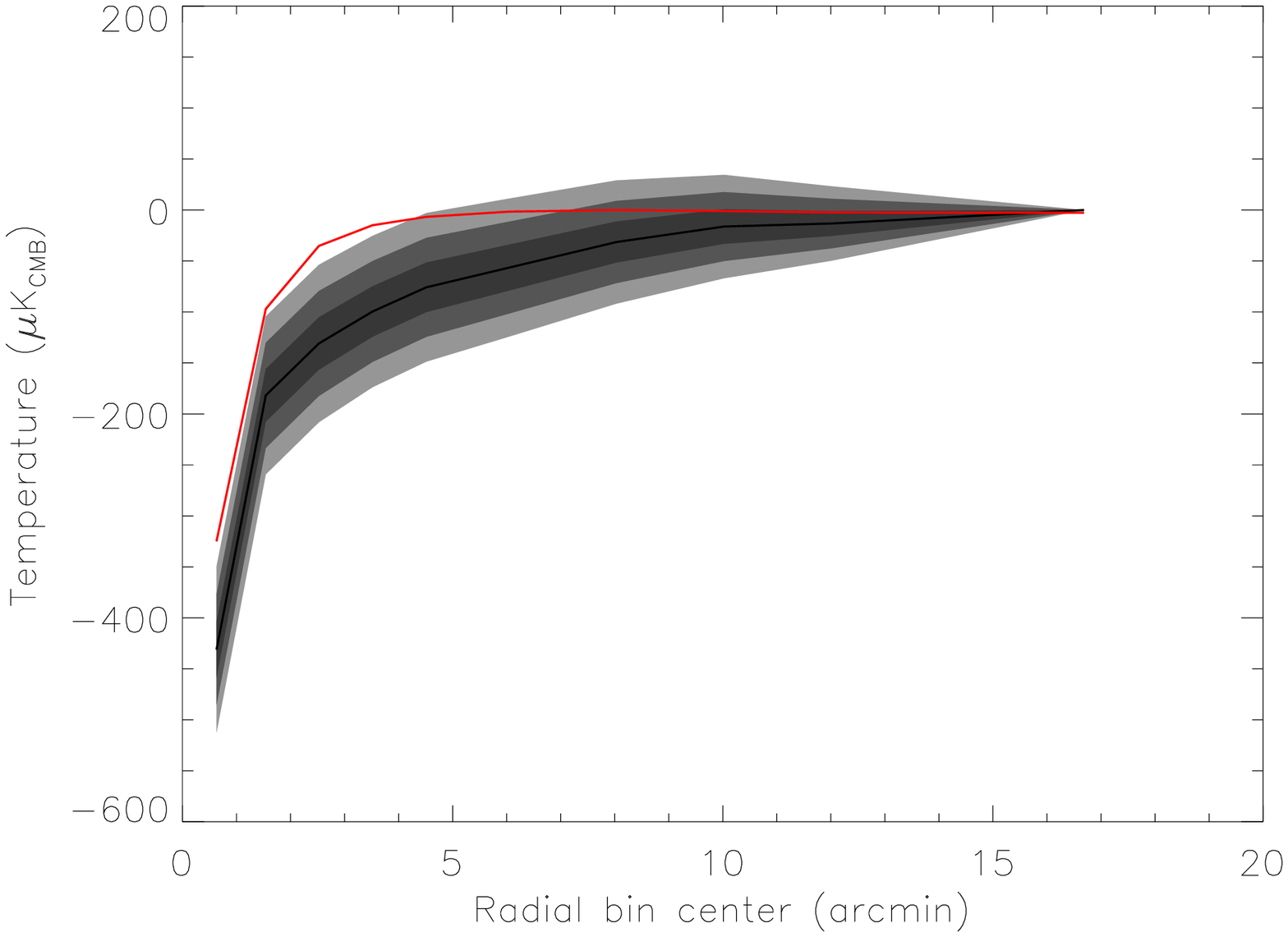}
\caption{Measured profile (black) and best-fit 
$\beta$-model (red) for RXCJ2031.8-4037.  The noise level in the
220~GHz map for this cluster is among the highest in the sample, 
so the noise correlations between bins are large.  Shown in gray are the 
results of varying the first component of the measured profile by 
1, 2, and 3$\sigma$ in the eigenbasis of the noise covariance matrix.  }
\label{fig:rxcj2031_profscatter}
\end{center}
\end{figure}

Nevertheless, large-scale correlated noise due to atmospheric
fluctuations and unsubtracted CMB remains present in the
band subtracted maps.  Figure~\ref{fig:rxcj2031_profscatter} 
shows the effect of this confusion on one band subtracted
profile with particularly low 220~GHz signal to noise.  
While this cluster is an extreme example, the same issue
persists across the entire sample.  As in earlier SZ works, 
we find that model fits are in some respects more robust 
indicators of global cluster properties than are the measured 
profiles, since the choice of a realistic model imposes 
sensible constraints on less well constrained modes.
We therefore make extensive use of model fitting when 
interpreting the SPT profiles and integrated Componizations.

\section{Results}

\subsection{Profiles and model fits}

The projected radial profiles of the band subtracted maps can be found in 
the Appendix.  Dashed vertical lines in the profile
plots denote estimates of $r_{500}$ determined from
the temperature--radius scaling relation in \citet{vikhlinin06}.  The 
error bars on the profiles are determined using Monte Carlo 
methods, and as discussed in Section~\ref{sec:cmbatm}, 
are strongly correlated between bins.

The best-fit central decrement and scaling radius for each cluster
are listed in Table~\ref{table:fits}, along with the minimum $\chi^2$
and the logarithmic ratio of the 
marginal likelihoods of the GNFW and $\beta -$models:
\begin{equation}
  \label{eqn:marglikeratio}
  R \equiv \log_{10}\left(\int \mathcal{L}_{\mathrm{GNFW}} d\theta_{s} d\Delta T_0
  / \int \mathcal{L}_{\mathrm{\beta}} d\theta_{\mathrm{core}} d\Delta T_0 \right)
\end{equation}
where $\mathcal{L}_{\mathrm{GNFW}}$ is the likelihood function for the GNFW
model and $\mathcal{L}_\beta$ is the likelihood function for the
$\beta -$model.  This quantity can be interpreted as a  
likelihood ratio, with a value of 10 corresponding to a 10:1 
preference for the GNFW model.  For the recent mergers 1ES~0657-56
and A~2744, neither model is a good fit, though the $\beta$-model 
is strongly favored.  
The $\beta -$model yields a larger marginal likelihood for over half 
of the sample, but the SPT data do not definitively favor one model
over the other. 

\begin{table*}
\begin{center}
\caption{Best-fit Model Parameters}
\begin{tiny}
\begin{tabular}{|l|l|l|l|l|l|l|l|}
\hline\hline
\rule[-2mm]{0mm}{6mm}
 & \multicolumn{2}{|c|}{Isothermal $\beta -$model} & \multicolumn{2}{|c|}{GNFW Model} & \multicolumn{3}{|c|}{ } \\
\hline
ID & $\Delta T_0$ ($\mu$K$_{\mathrm{CMB}}$) & $\theta_{\mathrm{core}}$ ($\prime$) 
   & $\Delta T_0$ ($\mu$K$_{\mathrm{CMB}}$) & $\theta_{s}$ ($\prime$) & $R$ 
   & $\chi^2$ & PTE \\
\hline
 \multicolumn{8}{|c|}{Primary sample} \\
\hline
A 2744 &
$        -549 \pm           27$ & 
$   1.51 \pm    0.06$ & 
$        -670 \pm           33$ & 
$   7.87 \pm    0.51$ & 
$ -50.79$ &
$  19.50$ &
$   0.01$ \\ 
RXCJ0217.2-5244 &
$        -214 \pm           29$ & 
$   0.64 \pm    0.13$ & 
$        -262 \pm           34$ & 
$   3.12 \pm    0.63$ & 
$  -0.84$ &
$   5.24$ &
$   0.63$ \\ 
RXCJ0232.2-4420 &
$        -510 \pm           26$ & 
$   0.99 \pm    0.05$ & 
$        -626 \pm           32$ & 
$   5.01 \pm    0.31$ & 
$  17.72$ &
$   9.68$ &
$   0.21$ \\ 
AS 0520 &
$        -217 \pm           13$ & 
$   1.22 \pm    0.12$ & 
$        -268 \pm           16$ & 
$   6.80 \pm    0.86$ & 
$  16.33$ &
$   9.16$ &
$   0.24$ \\ 
RXCJ0528.9-3927 &
$        -335 \pm           20$ & 
$   1.78 \pm    0.14$ & 
$        -412 \pm           25$ & 
$  10.03 \pm    1.24$ & 
$   5.53$ &
$   7.46$ &
$   0.38$ \\ 
AS 0592 &
$        -529 \pm           31$ & 
$   0.79 \pm    0.06$ & 
$        -658 \pm           40$ & 
$   3.77 \pm    0.29$ & 
$  -3.51$ &
$   3.00$ &
$   0.89$ \\ 
A 3404 &
$        -472 \pm           29$ & 
$   0.95 \pm    0.08$ & 
$        -588 \pm           38$ & 
$   4.52 \pm    0.43$ & 
$   1.20$ &
$   4.09$ &
$   0.77$ \\ 
1ES 0657-56 &
$        -932 \pm           43$ & 
$   1.47 \pm    0.03$ & 
$       -1133 \pm           52$ & 
$   7.56 \pm    0.22$ & 
$-270.59$ &
$  65.61$ &
$   0.00$ \\ 
RXCJ2031.8-4037 &
$        -511 \pm           42$ & 
$   0.65 \pm    0.07$ & 
$        -616 \pm           48$ & 
$   3.24 \pm    0.35$ & 
$  -3.76$ &
$  11.66$ &
$   0.11$ \\ 
A 3888 &
$        -516 \pm           26$ & 
$   1.35 \pm    0.07$ & 
$        -634 \pm           32$ & 
$   6.79 \pm    0.49$ & 
$ -28.25$ &
$   8.64$ &
$   0.28$ \\ 
AS 1063 &
$       -1062 \pm           50$ & 
$   0.86 \pm    0.02$ & 
$       -1321 \pm           62$ & 
$   4.14 \pm    0.12$ & 
$  -2.54$ &
$  28.61$ &
$   0.00$ \\ 
\hline
 \multicolumn{8}{|c|}{Supplemental sample} \\
\hline
RXCJ0336.3-4037 &
$        -306 \pm           30$ & 
$   0.70 \pm    0.10$ & 
$        -388 \pm           40$ & 
$   3.20 \pm    0.46$ & 
$  -5.11$ &
$  28.33$ &
$   0.00$ \\ 
RXCJ0532.9-3701 &
$        -449 \pm           24$ & 
$   1.00 \pm    0.07$ & 
$        -543 \pm           29$ & 
$   5.66 \pm    0.52$ & 
$  21.25$ &
$  35.43$ &
$   0.00$ \\ 
MACSJ0553.4-3342 &
$        -680 \pm           34$ & 
$   0.96 \pm    0.05$ & 
$        -849 \pm           44$ & 
$   4.54 \pm    0.26$ & 
$ -18.63$ &
$  28.33$ &
$   0.00$ \\ 
A 3856 &
$        -256 \pm           26$ & 
$   0.71 \pm    0.11$ & 
$        -313 \pm           31$ & 
$   3.49 \pm    0.55$ & 
$  -0.01$ &
$   2.17$ &
$   0.95$ \\ 
\hline
\end{tabular}
\end{tiny}
\label{table:fits}
\tablecomments{The $\beta$-model fit is performed with $\beta$ fixed
to 0.86, and the GNFW fit is performed with the parameters
$(\alpha_{\mathrm{n}}, \beta_{\mathrm{n}},
\gamma_{\mathrm{n}},c_{500})$ fixed to $(1.0, 5.5, 0.5, 1.0)$.
The sixth column quantifies the
relative goodness of fit in terms of the marginal likelihood 
ratio (see the text).  Positive values indicate that the GNFW 
model is preferred, and negative values 
indicate that the $\beta -$model is 
preferred.  The seventh column is the $\chi^2$ 
for the preferred model, and the eighth is the probability 
to exceed this value of $\chi^2$ for seven degrees of freedom.  } 
\end{center}
\end{table*}

\subsection{Stacked profiles}
\label{sec:stacked}

\begin{figure}
\begin{center}
\includegraphics[width=0.4\textwidth]{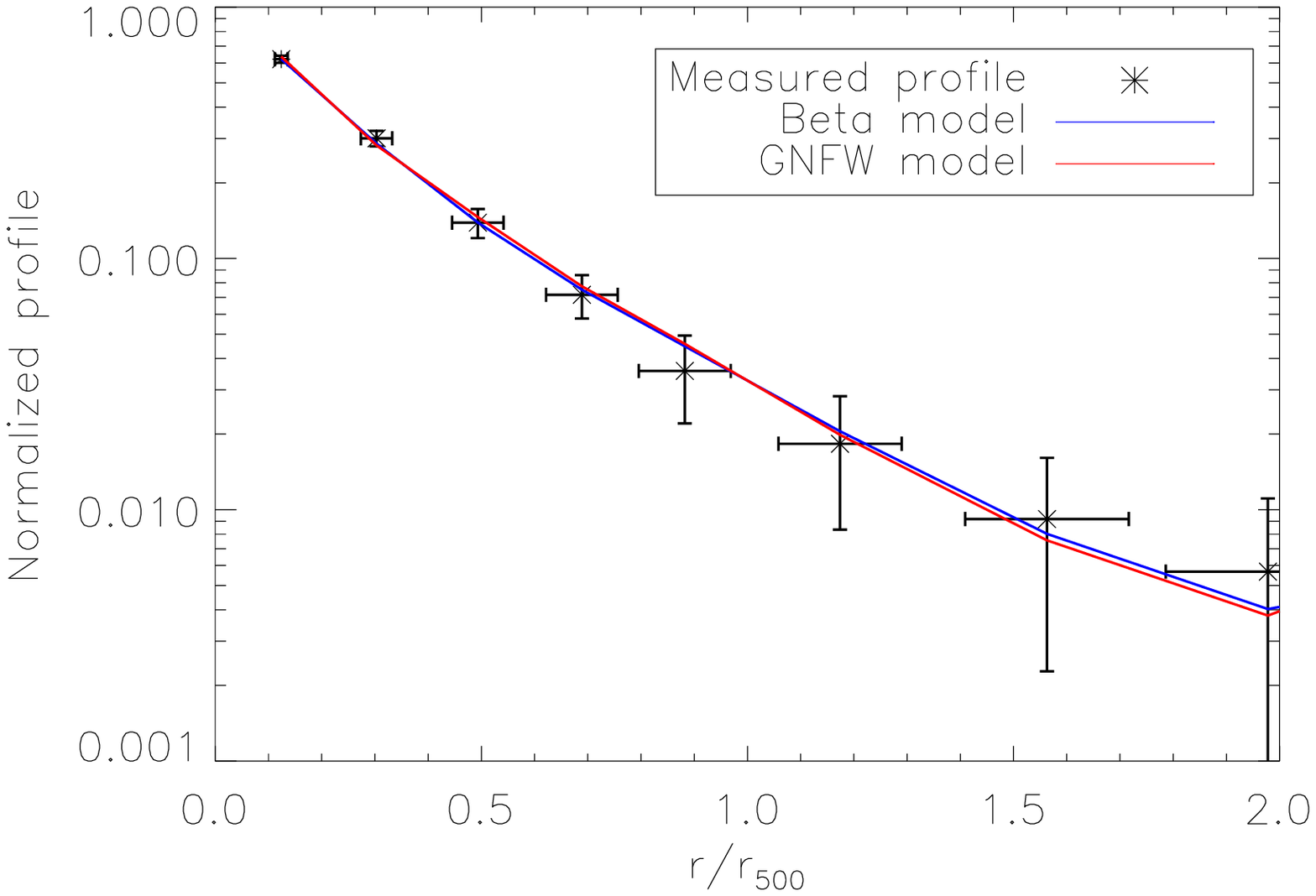}
\caption{Stacked radial profile and maximum likelihood
$\beta$-model and GNFW model fits.  The plot extends
to $2 r_{500}$, which is approximately equal to the 
average virial radius for the sample.  The best-fit
model profiles are computed from simulated processed
maps in the same way that the cluster profiles are
computed (see Section~\ref{sec:profiles}), and are
shown as a continuous line only for the purpose of
visualization.}
\label{fig:stacked}
\end{center}
\end{figure}

Each individual cluster profile contains too little information to 
place a tight constraint on the pressure behavior at large radius.
We proceed by scaling the profiles to $r_{500}$, as determined from
the temperature--radius scaling relation in \citet{vikhlinin06}, and 
analyzing the data set as a whole---by constructing a stacked profile, and
by fitting models to the sample.  The stacked profile, shown in
Figure~\ref{fig:stacked}, is determined by averaging the scaled profiles 
weighted by their noise.  Each radial bin in the 
stacked profile is assigned a radius and a
temperature from the average of the scaled profiles, and
an uncertainty is assigned by the standard error on the mean. 

We fit models to the sample by evaluating the likelihood function
for each individual cluster at each point in a grid of parameter space,
and then by multiplying the cluster likelihoods together to produce
a joint likelihood.  This allows us to determine average values for
the structure parameters in our two models.  For the $\beta$-model,
we allow $\beta$ and $r_{\mathrm{core}}/r_{500}$ to vary.  We allow
the central decrement and overall constant offset of each profile to 
vary as well, and we marginalize over these parameters.  The cluster
center position is held fixed, leaving 
six remaining degrees of freedom per cluster.
We find that for the full sample, the marginalized values of 
$\beta$ and $r_{\mathrm{core}}/r_{500}$ are $0.86 \pm 0.09$ 
and $0.20 \pm 0.05$.  Excluding the supplemental sample from 
the fit does not appreciably change the results.  The 
minimum $\chi^2$ for the fit is 338 for 90 degrees of freedom,
indicating a relatively poor fit.  If $\beta$ is fixed
to a typical X-ray value of 2/3, the minimum $\chi^2$ 
is 362.  If 1ES~0657-56 is excluded from the fit, the
minimum reduced $\chi^2$ changes from $338/90=3.76$ to $296/84=3.52$, 
and the marginalized values of $(\beta, r_{\mathrm{core}}/r_{500})$ change to
$(0.78 \pm 0.08, 0.17 \pm 0.04)$.

For the generalized NFW model, we fit for the structure parameters
$\alpha_{\mathrm{n}}$, $\beta_{\mathrm{n}}$, $\gamma_{\mathrm{n}}$,
and $c_{500}$, again fixing the center positions and marginalizing 
over the central decrements of
all of the clusters.  The likelihood-weighted mean values are 
$(\alpha_{\mathrm{n}}, \beta_{\mathrm{n}},\gamma_{\mathrm{n}},
c_{500})=(1.0,5.5,0.5,1.0)$ and the marginalized values are
$(1.2 \pm 0.5, 5.0 \pm 2.0, 0.7 \pm 0.3, 1.3 \pm 0.5)$.
The minimum $\chi^2$ for the fit is 468, compared to 472 for 
the \citetalias{nagai07}
parameter values of $(\alpha_{\mathrm{n}},\beta_{\mathrm{n}},
\gamma_{\mathrm{n}},c_{500})=(0.9,5.0,0.4,1.3)$---the change
in $\chi^2$ is small due to strong degeneracies between the 
parameters.  If 1ES~0657-56 is excluded from the GNFW
fit, the marginalized parameter values do not change appreciably, but
the minimum reduced $\chi^2$ drops from $468/75=6.24$ to $318/70=4.54$.

\begin{figure}
\begin{center}
\includegraphics[width=.4\textwidth]{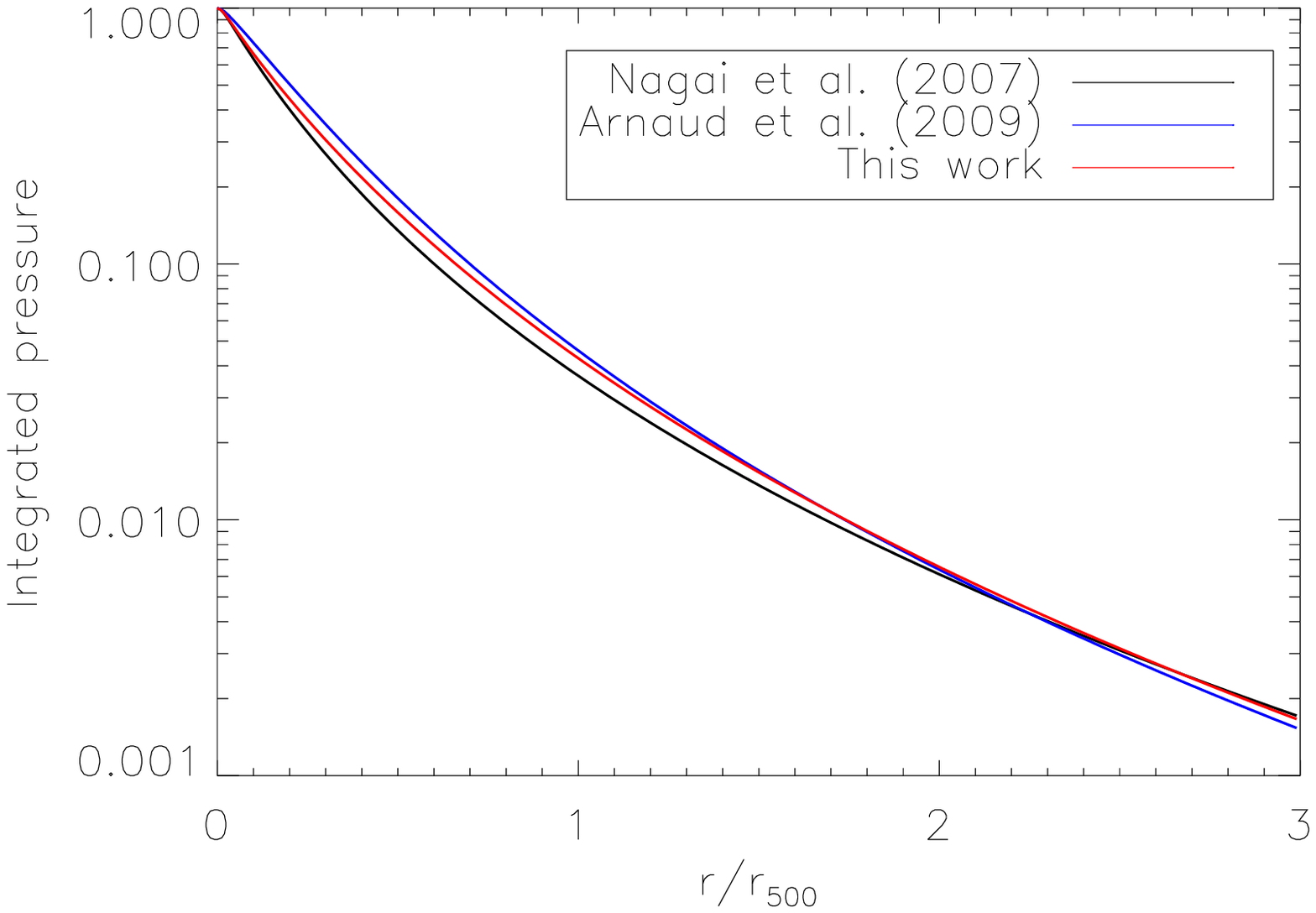} 
\end{center}
\caption{GNFW models with best-fit slope
and concentration parameters from \citetalias{nagai07}, 
\citet{arnaud09}, and this work.  Note that the samples in each
work differ in characteristic redshift, morphology, and temperature.
These results are indicative of the 
consistency between X-ray and SZ pressure measurements. }
\label{fig:models}
\end{figure}

In Figure~\ref{fig:models}, the SPT maximum likelihood GNFW model 
with $(\alpha_{\mathrm{n}},\beta_{\mathrm{n}},
\gamma_{\mathrm{n}},c_{500})$=$(1.0,5.5,0.5,1.0)$ is compared to
the $(0.9,5.0,0.4,1.3)$ model found by \citetalias{nagai07}
and the $(1.05,5.49,0.31,1.18)$ model found
by \citet{arnaud09}.  \citetalias{nagai07}
use a sample of nearby relaxed clusters, while \citet{arnaud09}
use the REXCESS representative sample of $z<0.2$ clusters.  Two systems---A~3888
and A~3856---are in both the SPT and REXCESS samples.  On average,
however, the samples are quite different, with the SPT clusters tending
to be hotter and more distant.  The similarity between the 
resulting profiles is encouraging for future applications of 
self-similar pressure models.

The SPT best-fit $\beta$-model and GNFW model
are shown in Figure~\ref{fig:stacked}.  Since the models are
very similar, the fits are of roughly equal quality.
The logarithmic ratio of marginal likelihoods 
(see Equation~\ref{eqn:marglikeratio})
for the GNFW model compared to the $\beta$-model is $R=-322$,
indicating a preference for the latter.  This preference is
due primarily to 1ES~0657-56, and to a lesser extent to A~2744.
Both of these clusters are major mergers.  Excluding the former brings
the marginal likelihood ratio to $R=-51$, and excluding both brings
the ratio to $R=-0.6$.  Stronger model discrimination will require
even better multi-band measurements to large angular scales, where 
atmospheric noise and astrophysical contamination limit the 
constraining power of the data.

\subsection{Integrated $y$ parameter and scaling relations}
\label{sec:xray_comp}

The integrated Compton-$y$ parameter $Y_{\mathrm{SZ}}$ is expected 
to serve as a good proxy for mass, and to be related to other physical
properties of clusters by scaling relations.
$Y_{\mathrm{SZ}}$ can be computed either using the model fits 
discussed in Section \ref{sec:fits}, or using the model-independent
radial profiles.  Since the large angular scale noise
contributes strongly to the model-independent estimate, we consider
the model-dependent results to be slightly more robust.  
The average ratio across the sample of $Y_{\mathrm{SZ}}$ computed from the 
$\beta$-model fit, to $Y_{\mathrm{SZ}}$ computed from the GNFW
model fit, is $1.06 \pm 0.08$ at $r_{2500}$ and 
$1.19 \pm 0.20$ at $r_{500}$.  This is due to the fact that
a $\beta=0.86$ model falls off more gradually at intermediate radius 
than does the GNFW model with our best-fit structure parameters.
Since the $\beta$-model is slightly preferred over the GNFW model
across our sample, we use the $\beta$-model fits to compute our 
model-dependent results; the relations derived from the GNFW
values are consistent to within $1\sigma$.

Both the model-dependent and the model-independent estimates of
the integrated Compton-$y$ parameter are 
given in Table~\ref{table:ysz}.  Uncertainties are determined
via Monte Carlo methods, and the apertures are again determined from 
the temperature--radius scaling relation.  Since the slope of the
temperature--radius scaling relation is relatively shallow, the aperture
radii vary only by $\sim 10$\% across our sample.  Although our
integration radii are slightly different than those used in the X-ray
analyses, we find that this choice makes a negligible difference in
the integrated $Y$ parameters---even integrating out to a fixed radius 
of 1~Mpc does not significantly increase the scatter in the scaling relations.  

The $\beta$-model and model-independent estimates at
$r_{2500}$ are in relatively good agreement, with a ratio of model-dependent
to model-independent $Y_{SZ}$ of $0.99 \pm 0.22$.  Significant 
scatter is seen at $r_{500}$, where the ratio is $1.30 \pm 0.56$.  
The model-independent SZ signal is less 
well constrained at larger radius due to its lower amplitude and
to the large angular scale confusion discussed in 
Section~\ref{sec:cmbatm}.  The model-dependent estimates are
less affected by confusion, as noisier modes are downweighted
in the model fitting process.  We therefore use the $\beta$-model
estimates to explore scaling relations with X-ray values, which are
also derived in a model-dependent manner.

\citet{halverson09} have recently observed 1ES~0657-56 with the APEX-SZ
experiment, and have fit an ellipsoidal $\beta$-model to their 150~GHz
SZ map.  Since we assume azimuthal symmetry and fit models to the radial 
profiles rather than the maps, we cannot perform a directly 
comparable fit.  However, we can compare our
respective estimates of $Y_{\mathrm{SZ}}$.  The APEX-SZ best-fit
ellipsoidal model yields $(Y_{\mathrm{SZ},2500},Y_{\mathrm{SZ},500})
=(30 \pm 5, 66 \pm 18) \times 10^{-11}$ sr, which is
consistent at the $\sim 1\sigma$ level with both the model-dependent
and model-independent SPT results.

\begin{table*}
\begin{center}
\caption{Integrated Compton-$y$ parameters.} 
\small
\begin{tabular}{|l|l|l|l|l|l|l|l|l|}
\hline\hline
\rule[-2mm]{0mm}{6mm}
 ID & $r_{2500}$ & $r_{500}$ 
 & \multicolumn{2}{|c|}{$\beta -$model}
 & \multicolumn{2}{|c|}{GNFW Model}
 & \multicolumn{2}{|c|}{Model-independent} \\
\hline
   & (Mpc) & (Mpc) &
  $Y_{\mathrm{SZ},2500}$ & $Y_{\mathrm{SZ},500}$ &
  $Y_{\mathrm{SZ},2500}$ & $Y_{\mathrm{SZ},500}$ &
  $Y_{\mathrm{SZ},2500}$ & $Y_{\mathrm{SZ},500}$ \\
\hline
 \multicolumn{9}{|c|}{Primary sample} \\
\hline
A 2744 &
$   0.63$ & $   1.40$ &
$   20.3 \pm     0.7$ &
$   49.2 \pm     2.3$ &
$   19.8 \pm     1.0$ &
$   47.8 \pm     3.8$ &
$   19.0 \pm     1.4$ &
$   38.1 \pm     6.8$ \\
RXCJ0217.2-5244 &
$   0.61$ & $   1.36$ &
$    3.3 \pm     0.4$ &
$    6.3 \pm     1.0$ &
$    3.1 \pm     0.4$ &
$    5.2 \pm     1.0$ &
$    3.6 \pm     0.9$ &
$    4.8 \pm     2.7$ \\
RXCJ0232.2-4420 &
$   0.52$ & $   1.17$ &
$   11.2 \pm     0.4$ &
$   24.8 \pm     1.1$ &
$   10.8 \pm     0.4$ &
$   23.1 \pm     1.5$ &
$   12.9 \pm     0.9$ &
$   26.7 \pm     4.3$ \\
AS 0520 &
$   0.54$ & $   1.21$ &
$    5.7 \pm     0.4$ &
$   13.7 \pm     1.4$ &
$    5.9 \pm     0.6$ &
$   14.4 \pm     2.1$ &
$    4.7 \pm     0.8$ &
$    7.2 \pm     2.7$ \\
RXCJ0528.9-3927 &
$   0.53$ & $   1.19$ &
$   11.1 \pm     0.7$ &
$   31.0 \pm     2.9$ &
$   11.5 \pm     0.9$ &
$   33.0 \pm     4.3$ &
$    9.8 \pm     0.9$ &
$   22.1 \pm     4.4$ \\
AS 0592 &
$   0.58$ & $   1.30$ &
$   12.9 \pm     0.6$ &
$   24.7 \pm     1.6$ &
$   11.6 \pm     0.7$ &
$   18.7 \pm     1.7$ &
$   14.7 \pm     2.0$ &
$   26.0 \pm     9.8$ \\
A 3404 &
$   0.60$ & $   1.35$ &
$   18.2 \pm     1.3$ &
$   34.6 \pm     3.1$ &
$   15.9 \pm     1.5$ &
$   24.5 \pm     3.1$ &
$   13.5 \pm     3.0$ &
$   14.0 \pm     5.4$ \\
1ES 0657-56 &
$   0.65$ & $   1.44$ &
$   36.2 \pm     0.6$ &
$   84.9 \pm     1.8$ &
$   35.1 \pm     0.7$ &
$   81.0 \pm     2.4$ &
$   30.0 \pm     1.0$ &
$   45.9 \pm     4.7$ \\
RXCJ2031.8-4037 &
$   0.64$ & $   1.43$ &
$    8.4 \pm     0.5$ &
$   16.1 \pm     1.4$ &
$    7.7 \pm     0.5$ &
$   12.8 \pm     1.3$ &
$   14.3 \pm     1.6$ &
$   37.0 \pm     8.0$ \\
A 3888 &
$   0.59$ & $   1.33$ &
$   30.1 \pm     1.5$ &
$   60.8 \pm     3.7$ &
$   28.7 \pm     2.0$ &
$   51.9 \pm     5.2$ &
$   24.6 \pm     3.8$ &
$   36.1 \pm    14.7$ \\
AS 1063 &
$   0.65$ & $   1.44$ &
$   23.3 \pm     0.4$ &
$   48.3 \pm     1.1$ &
$   21.5 \pm     0.5$ &
$   39.7 \pm     1.5$ &
$   25.5 \pm     1.1$ &
$   44.5 \pm     5.3$ \\
\hline
 \multicolumn{9}{|c|}{Supplemental sample} \\
\hline
RXCJ0336.3-4037 &
$   0.56$ & $   1.26$ &
$    7.5 \pm     0.7$ &
$   13.7 \pm     1.7$ &
$    6.2 \pm     0.8$ &
$    8.9 \pm     1.7$ &
$   11.0 \pm     2.5$ &
$   27.8 \pm    11.8$ \\
RXCJ0532.9-3701 &
$   0.62$ & $   1.38$ &
$   12.8 \pm     0.7$ &
$   26.6 \pm     1.9$ &
$   13.6 \pm     1.0$ &
$   27.6 \pm     3.2$ &
$   18.1 \pm     1.7$ &
$   42.4 \pm     8.0$ \\
MACSJ0553.4-3342 &
$   0.69$ & $   1.52$ &
$   15.8 \pm     0.5$ &
$   34.3 \pm     1.5$ &
$   14.5 \pm     0.6$ &
$   28.8 \pm     2.0$ &
$   14.7 \pm     1.4$ &
$   20.5 \pm     5.1$ \\
A 3856 &
$   0.55$ & $   1.23$ &
$    7.3 \pm     0.8$ &
$   13.1 \pm     1.7$ &
$    6.2 \pm     0.9$ &
$    8.4 \pm     1.7$ &
$    7.7 \pm     2.9$ &
$    9.8 \pm     6.4$ \\
\hline
\end{tabular}
\label{table:ysz}
\normalsize
\tablecomments{Units for $Y_{SZ}$ are
$10^{-11}$~sr.  Both model-dependent and
model-independent estimates are provided (see the text).  $r_{2500}$
and $r_{500}$ are determined from the temperature--radius
scaling relation.} 
\end{center}
\end{table*}

\begin{figure*}[ht]
\begin{center}
\begin{tabular}{lr}
\includegraphics[width=0.45\textwidth]{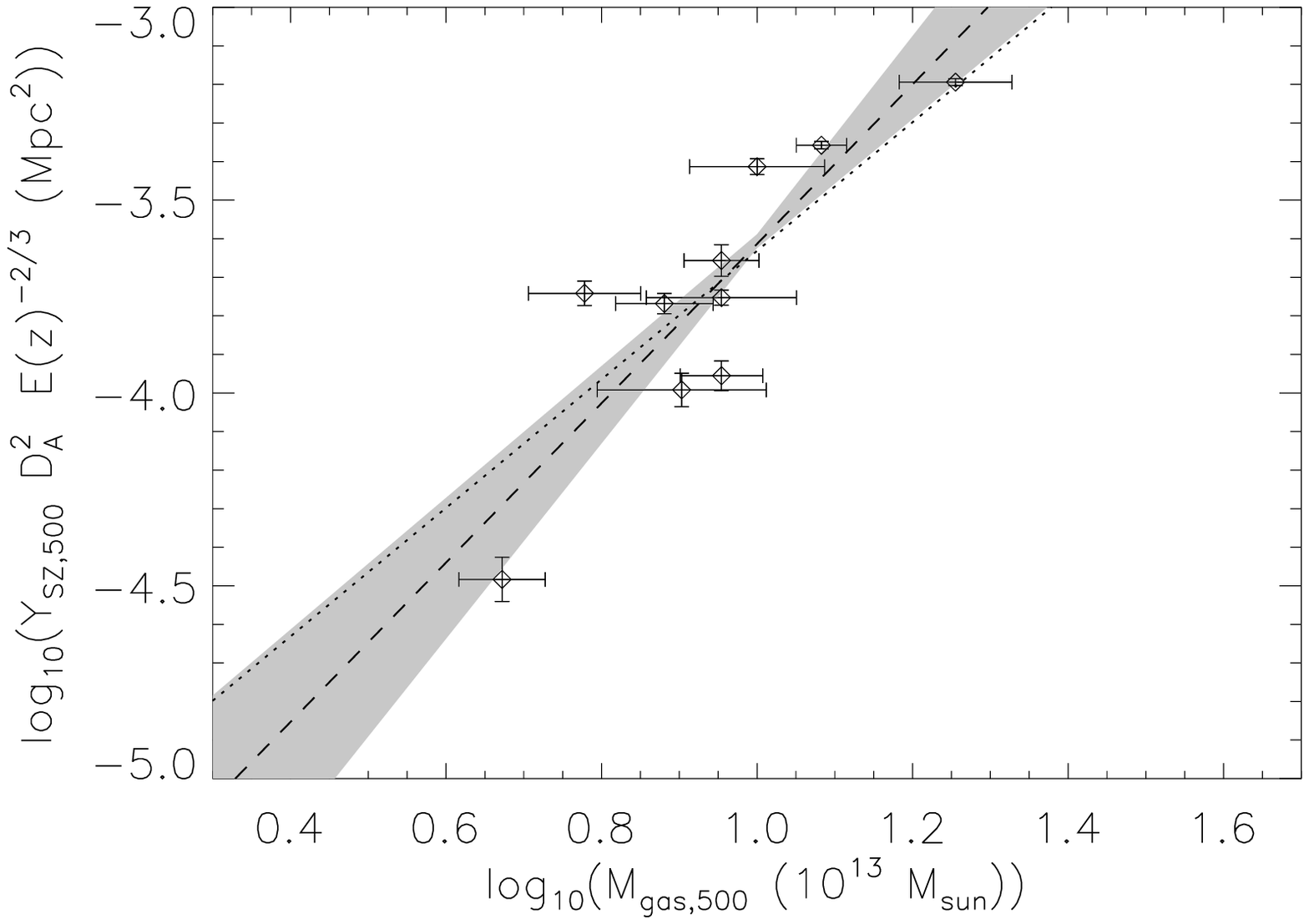} &
\includegraphics[width=0.45\textwidth]{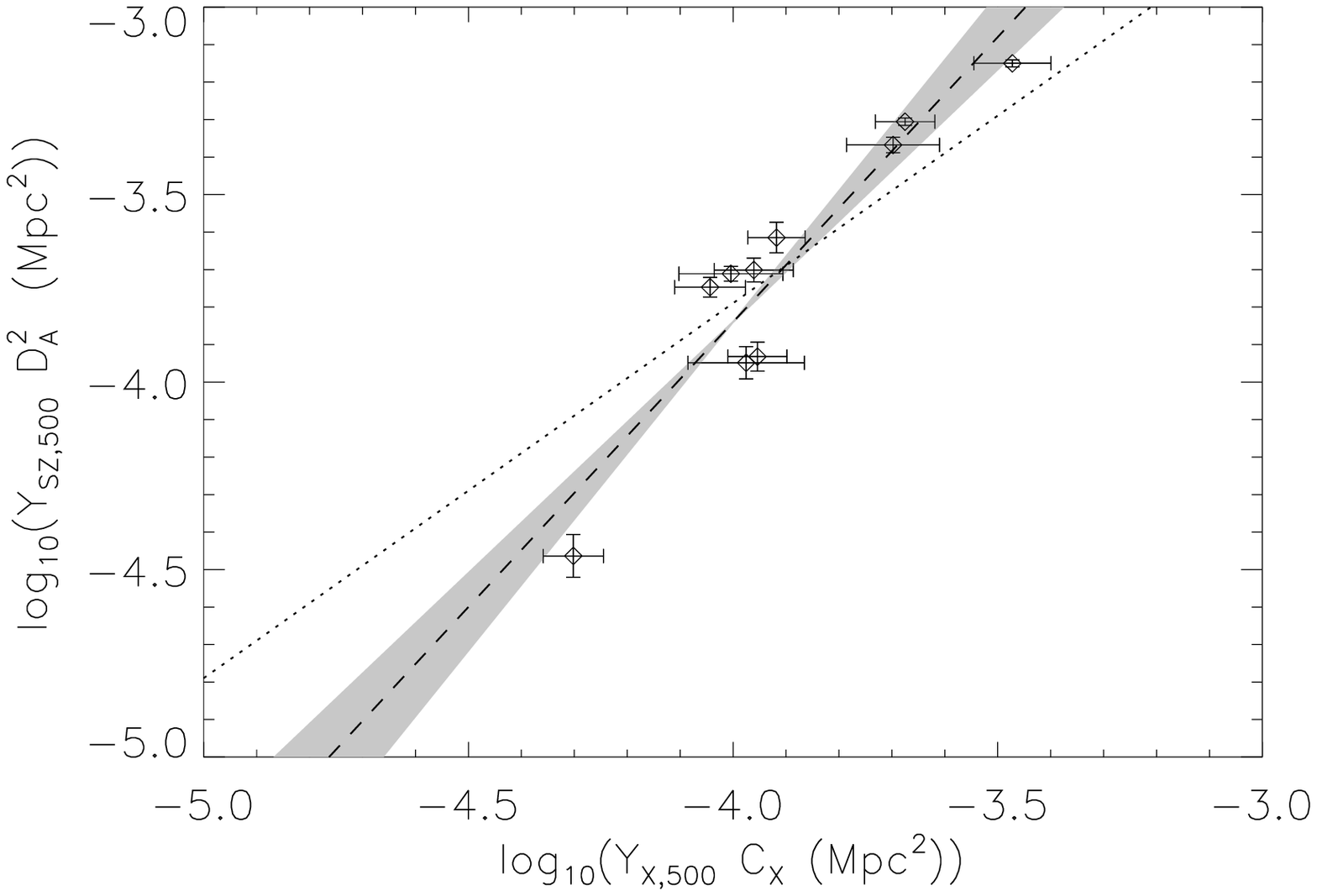} \\
\end{tabular}
\caption{Comparisons between SZ and X-ray measurements within $r_{500}$:
$Y_{SZ}$--$M_{\mathrm{gas}}$ (left) and $Y_{SZ}$--$Y_{X}$ (right).  X-ray values
are taken from the references in Table~\ref{table:clusterlist}, and
clusters without published values in these sources are omitted.  SZ
values are taken from the $\beta$-model estimates in 
Table~\ref{table:ysz}.  In the right panel, $C_X=C (\sigma_T/m_ec^2)/\mu_em_p$
where $C$ is a constant accounting for spherical versus cylindrical integration
\citep{bonamente08}.  The dashed lines represent the best-fit power laws, and
the shaded area represents the 1$\sigma$ allowed region.
The dotted lines represent the best bit powers law with the slopes
fixed to the expected self-similar values of 5/3 (left) and 1 (right). }
\label{fig:xray}
\end{center}
\end{figure*}

The $\beta$-model estimates of $Y_{\mathrm{SZ}, 500}$ are 
plotted versus X-ray estimates of gas mass and $Y_{X}$
in Figure~\ref{fig:xray}. 
$Y_{\mathrm{X}} = M_{\mathrm{gas}} kT_e$ is an X-ray observable 
analogous to $Y_{\mathrm{SZ}}$ \citep{kravtsov06a}.  
Only clusters with published values of
gas mass and electron temperature from the references
in Table~\ref{table:clusterlist} are included in the
plots.  Several clusters have multiple published X-ray
estimates of electron temperature and gas mass.  For our
plots in Figure~\ref{fig:xray}, we chose to adopt the
\citet{zhang06,zhang08} results where available, since these
samples had the largest overlap with our own.
For different choices of published X-ray values,
we find scaling relation parameters that are consistent at the $2\sigma$ level.

In order to compare $Y_{\mathrm{X}}$ and $Y_{\mathrm{SZ}}$, we convert 
the former from a spherically defined quantity into a cylindrical 
projection analogous to $Y_{\mathrm{SZ}}$ as described in 
\citet{bonamente08}.  This method requires an estimate of the ratio
of the cluster pressure within a cylindrical and spherical volume, which
we calculate assuming the best-fit SZ profile for each cluster.  Since
this ratio has large uncertainties in the model-independent case,
we report scaling relations only for our model-dependent $Y_{\mathrm{SZ}}$
estimates.  Temperature structure in clusters typically causes $Y_{\mathrm{X}}$ 
to be biased high by $\sim 10$\% \citep{arnaud09,vikhlinin06}; however, we
consider this bias to be subdominant to the other uncertainties in this comparison. 

$Y_{\mathrm{SZ}}$ is also expected to scale with electron temperature and 
total mass.  However, the reported errors in the X-ray estimates of 
electron temperature vary widely across our sample, and different analyses
of the same clusters often produce inconsistent results.  The X-ray 
total mass estimates are determined under the assumption of hydrostatic 
equilibrium, and the errors are quite large.  For this reason,
we do not include the total mass and electron temperature scaling relations
in this work.  In a future paper, we will pursue a
joint SZ and X-ray analysis which can more appropriately
incorporate these uncertainties.  

The $Y_{\mathrm{SZ}}$--gas mass plot is in a form readily comparable to the 
scaling relation given in \citet{bonamente08}:
\begin{equation}
  Y_{\mathrm{SZ}} D_A^2 \propto f_{\mathrm{gas}}^{-2/3} M_{\mathrm{gas}}^{5/3} E(z)^{2/3},
\end{equation}
where $D_A$ is the angular diameter distance, $f_{\mathrm{gas}}$ is the 
gas mass fraction (which we assume to be constant), $E(z)$ is given by 
$\sqrt{\Omega_M (1+z)^3 + \Omega_\Lambda}$, and $M_{\mathrm{gas}}$ 
is the cluster gas mass.  We fit for the gas mass scaling relation
in log--log space using the method described in \citet{marrone09},
assuming a scaling in the form $Y = \alpha + \beta X$ and allowing
for intrinsic scatter.  We find $\alpha=-5.73 \pm 0.43$ and
$\beta=2.12 \pm 0.45$.  The best-fit intrinsic scatter in this scaling 
relation, in terms of a percentage change in $M_{\mathrm{gas}}$ at 
fixed $Y_{SZ}$, is $14 \pm 10$\%.
If we substitute $Y_{SZ,2500}$ for $Y_{SZ,500}$, we find a similar
scaling relation with $M_{\mathrm{gas},500}$:
$\alpha = -5.92 \pm 0.41$, $\beta = 1.97 \pm 0.44$, and a 
scatter of $15$\%~$\pm 11$\%.

The gas masses of two clusters in our sample, 
1ES~0657-56 and RXCJ0232.2-4420, are estimated both from {\it Chandra} data
in \citet{maughan08} and from XMM data in \citet{zhang06}.  In
both cases, the \citet{maughan08} estimates 
($2.31^{+0.01}_{-0.01} \times 10^{14} M_{\odot}$ 
for 1ES~0657-56 and $1.20^{+0.01}_{-0.02} \times 10^{14} M_{\odot}$ 
for RXCJ0232.2-4420) are
larger than the \citet{zhang06} results we have adopted.  If we instead 
adopt the larger values, we find $\beta=1.85 \pm 0.43$, consistent
with the self-similar expectation.

The $Y_{SZ}$--$Y_{X}$ relation is expected to have a small scatter
and to have a slope of 1.0.  Our best-fit power law parameters
are $\alpha=2.42 \pm 0.87$ and $\beta=1.56 \pm 0.21$, with an
intrinsic scatter of $12$\%~$\pm 12$\% in $Y_X$.  Substituting
$Y_{SZ,2500}$ for $Y_{SZ,500}$ gives a scaling relation
of $\alpha=1.69 \pm 0.78$, $\beta=1.46 \pm 0.20$, and
a scatter of $9\%$~$\pm 9$\%.  While these slopes are 
inconsistent with the expected value of 1.0, 
the discrepancy is largely due to the three lowest signal to noise 
data points: A~3856, AS~0520, and A~3404.  In the gas mass scaling
relation, we found that adopting independent published values of
gas mass introduced significant variations in the slope and offset.
Since $Y_X$ is computed from the same $M_{\mathrm{gas}}$ estimates, 
comparable systematic uncertainties and offsets are expected.
We therefore hesitate to draw conclusions before performing a 
more uniform X-ray analysis.

We consider these $Y_{SZ}$--gas mass and $Y_{SZ}$--$Y_X$
scaling relations to be reasonably consistent with the self-similar
values, and we find no measurable scatter within the limits
of our statistical and systematic errors. 
Previous studies \citep[e.g.,][]{kravtsov06a,arnaud07,nagai07,vikhlinin09}
have found that $M_{\mathrm{gas}}$ and $Y_X$ scale with the total cluster mass
with low scatter, and these results indicate that $Y_{SZ}$ as measured
with the SPT should behave similarly.
Future joint analyses of X-ray and SZ data will further 
constrain these scaling relations, and will provide better
limits on any departure from self-similarity.  

\section{Conclusions}
\label{sec:conclusions}

In this paper, we have summarized SZ observations of 15
bright galaxy clusters taken with the SPT.  The 
SZ signals were measured at 150~GHz, and concurrent 220~GHz measurements
were used to remove astrophysical contamination.  Radial profiles were
computed for each cluster using a technique that accounts for the
effects of the beam and the time-domain filtering, and simultaneously
characterizes the correlated errors due to detector and atmospheric
noise.  

The profiles were fit both to a $\beta -$model and to
a generalized NFW model \citepalias{nagai07}.
By scaling and stacking the SPT profiles, and allowing the
structure parameters of the models to vary, we obtained self-similar
pressure profiles that agree closely with previous X-ray and SZ results.
For the $\beta -$model, we found $\beta = 0.86 \pm 0.09$ and 
$r_{\mathrm{core}}/r_{500}=0.20 \pm 0.05$.  For the GNFW model, we found 
$(\alpha_{\mathrm{n}},\beta_{\mathrm{n}},
\gamma_{\mathrm{n}},c_{500})$=$(1.0,5.5,0.5,1.0)$, similar to
the parameters found by \citetalias{nagai07} and \citet{arnaud09}.
The SPT data do not strongly prefer one model over the other.
Our measurements show no significant difference between the
stacked SZ profile and the models out to beyond $r_{500}$.  This is
the first SZ-measured constraint on cluster pressure profiles 
at such large radii.

We also estimated the integrated Compton-$y$ parameter $Y_{SZ}$ for each 
cluster using both model-dependent and model-independent techniques.
The scaling relations between $Y_{SZ}$ and the X-ray-determined
$M_{\mathrm{gas}}$ and $Y_X$ were found to be reasonably consistent with
the self-similar values.  We find an intrinsic scatter 
in the $Y_X$--$Y_{SZ}$ scaling relation of $9$\%~$\pm 9$\% at 
$r_{2500}$ and $12$\%~$\pm 12$\% at $r_{500}$.  Previous studies have indicated that
gas mass and $Y_X$ scale with low scatter to the total cluster mass.
These results indicate that $Y_{SZ}$ as measured with the SPT
should behave similarly, which is encouraging for the ongoing SPT cluster
SZ survey.

The data presented in this work demonstrate the utility of SZ 
measurements for characterizing the ICM out to large radii.  
In future works, we will expand the SPT sample of X-ray luminous 
clusters, and will include 95~GHz data in addition to 150 and 220~GHz.
The addition of a third frequency band will allow us to better
remove astrophysical backgrounds such as the CMB, thus improving 
our estimates of thermal SZ signals at large radii.  Future analyses will 
also combine SZ measurements with X-ray cluster observations in 
order to separately estimate the temperature and density of the 
ICM, and to determine the cluster gas mass fractions.

\acknowledgments

The South Pole Telescope is supported by the National Science Foundation through grants ANT-0638937 and ANT-0130612.  Partial support is also provided by the
NSF Physics Frontier Center grant PHY-0114422 to the Kavli Institute of Cosmological Physics at the University of Chicago, the Kavli Foundation and the Gordon and Betty Moore Foundation.
The McGill group acknowledges funding from the National Sciences and
Engineering Research Council of Canada, the Quebec Fonds de recherche
sur la nature et les technologies, and the Canadian Institute for
Advanced Research. The following individuals acknowledge additional support:
B.\ Benson and K.\ Schaffer from a KICP Fellowship; J.\ McMahon from a Fermi Fellowship;  Z.\ Staniszewski from a GAAN Fellowship; A.~T.\ Lee from the Miller Institute for Basic Research in Science,
University of California, Berkeley; and N.~W.\ Halverson from an Alfred P. Sloan Research
Fellowship.

\bibliography{spt.bib}

\newpage
\appendix
\section{Cluster Maps and Profiles}
\label{sec:appendix_maps}

Shown below are the SPT maps and profiles for each of the clusters
in this sample.  Four maps are shown for each cluster: the 150~GHz map
(upper left), the 220~GHz map (upper right), the
band subtracted map (lower left),
and a jackknife band subtracted map (lower right).  The pixel size is
$0^\prime .25$, and each map is smoothed with a
Gaussian with an FWHM of
$1^\prime$.  The 220~GHz maps contain
more noise than the 150~GHz maps due to lower 
sensitivity and higher-amplitude atmospheric 
fluctuations in the 220~GHz band.  The band subtracted
map is calculated by applying a matched filter to the
220~GHz map and subtracting the result from the
150~GHz map.  The jackknife band subtracted map is calculated by
multiplying half of the single-observation maps by
$-1$ before co-adding.  The locations of bright point 
sources that were removed are marked by crosses.  
The visible structure to the 
east and west of the brighter clusters is due to the 
timestream polynomial subtraction (see the text), and
is accounted for in the radial profiles.

The profiles are computed from the band subtracted
maps using our knowledge of the SPT beams and 
time-domain filtering.  Each profile point represents
the average SZ temperature decrement within a given
radial bin.  The dashed vertical line is an estimate 
of $r_{500}$.  The algorithm used to derive the profiles
is described in Section~\ref{sec:profiles}.

\begin{figure*}[ht!]
\begin{tabular}{cc}
\includegraphics[width=.4\textwidth]{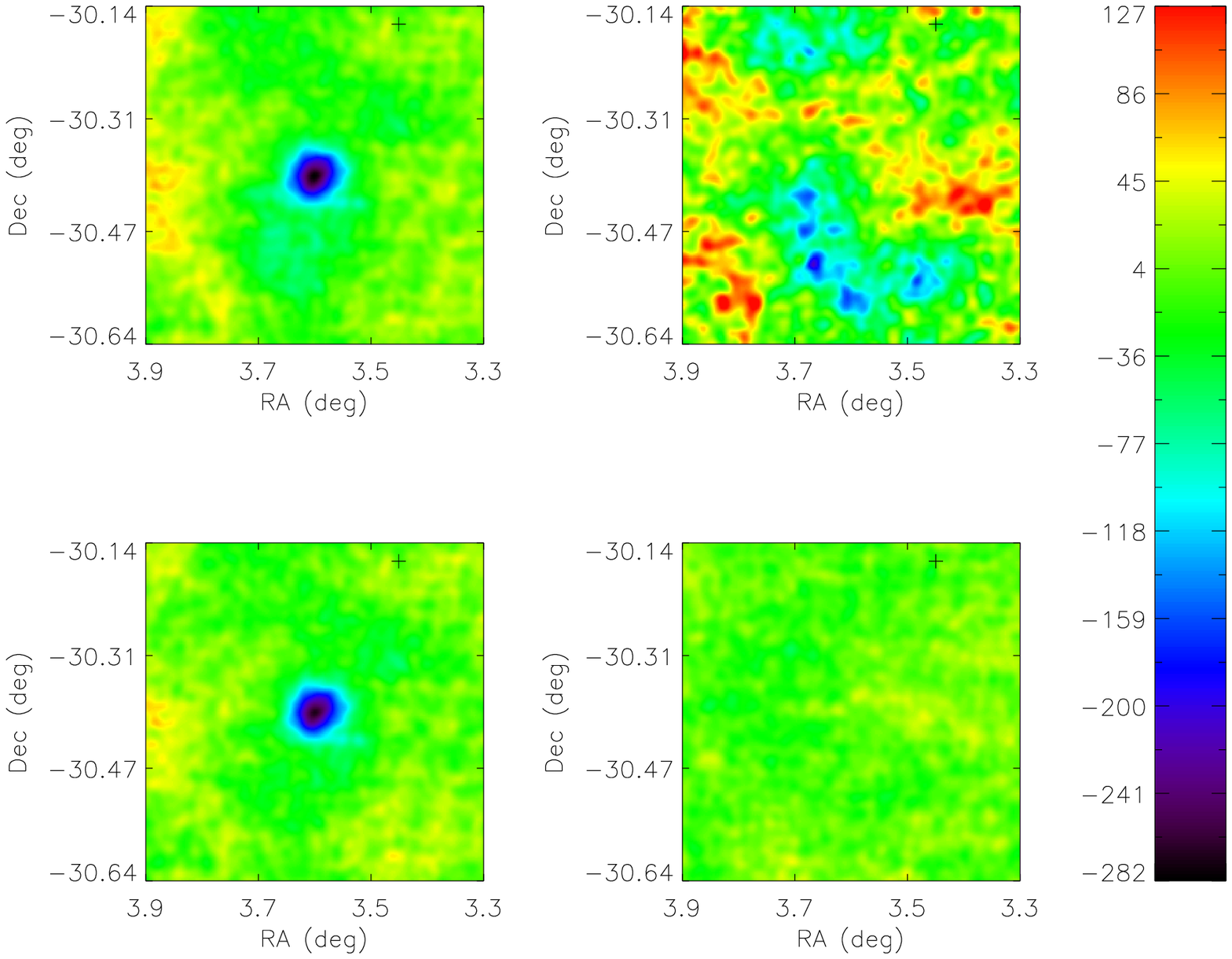} &
\includegraphics[width=.4\textwidth]{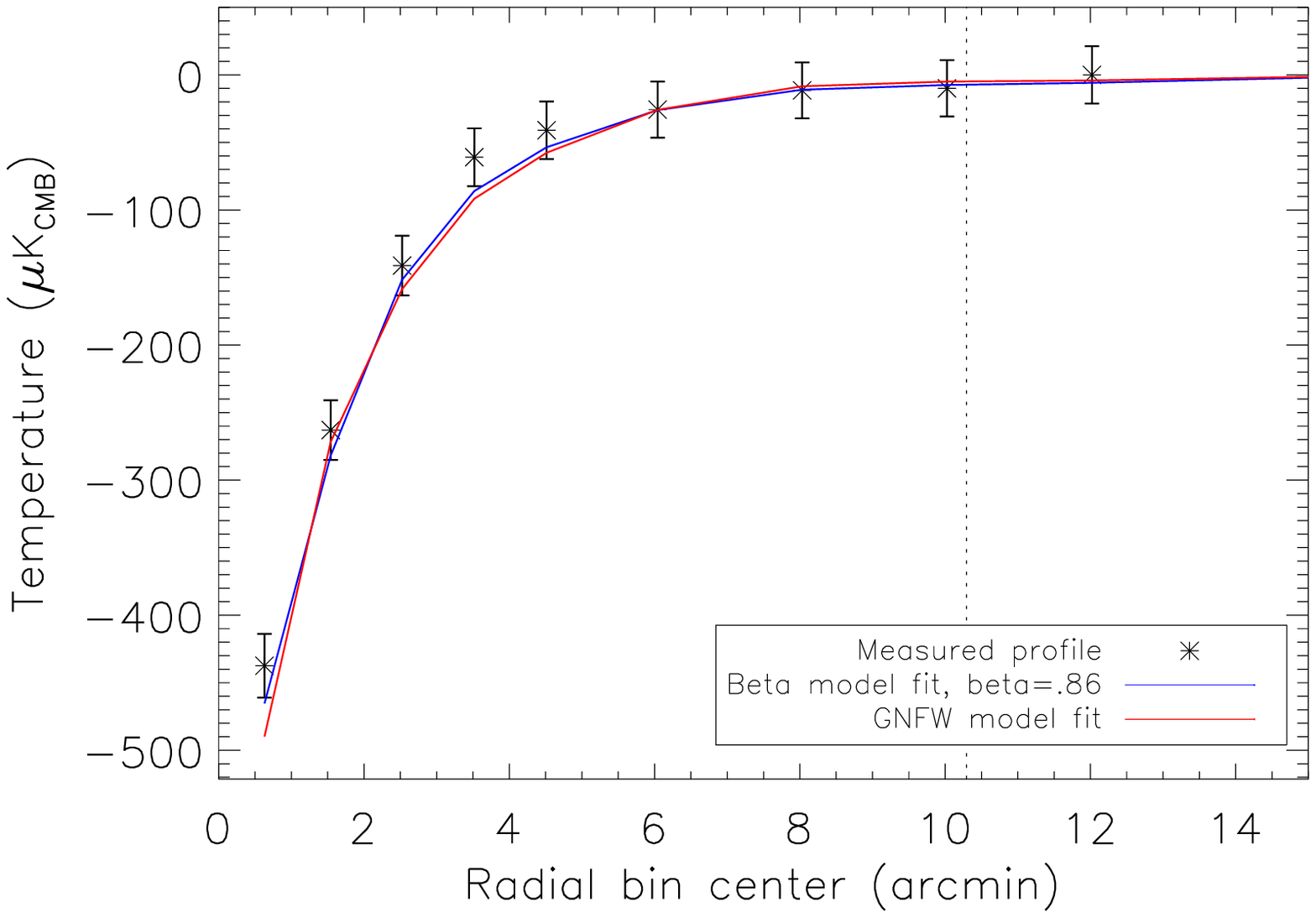} \\
\end{tabular}
\caption{A~2744 maps (left) and profile (right).  The
four maps are 150~GHz (upper left), 220~GHz (upper right),
band subtracted (lower left), and jackknife (lower right).  
Units are $\mu K_{\mathrm{CMB}}$.}
\end{figure*}

\begin{figure*}[ht!]
\begin{tabular}{cc}
\includegraphics[width=.4\textwidth]{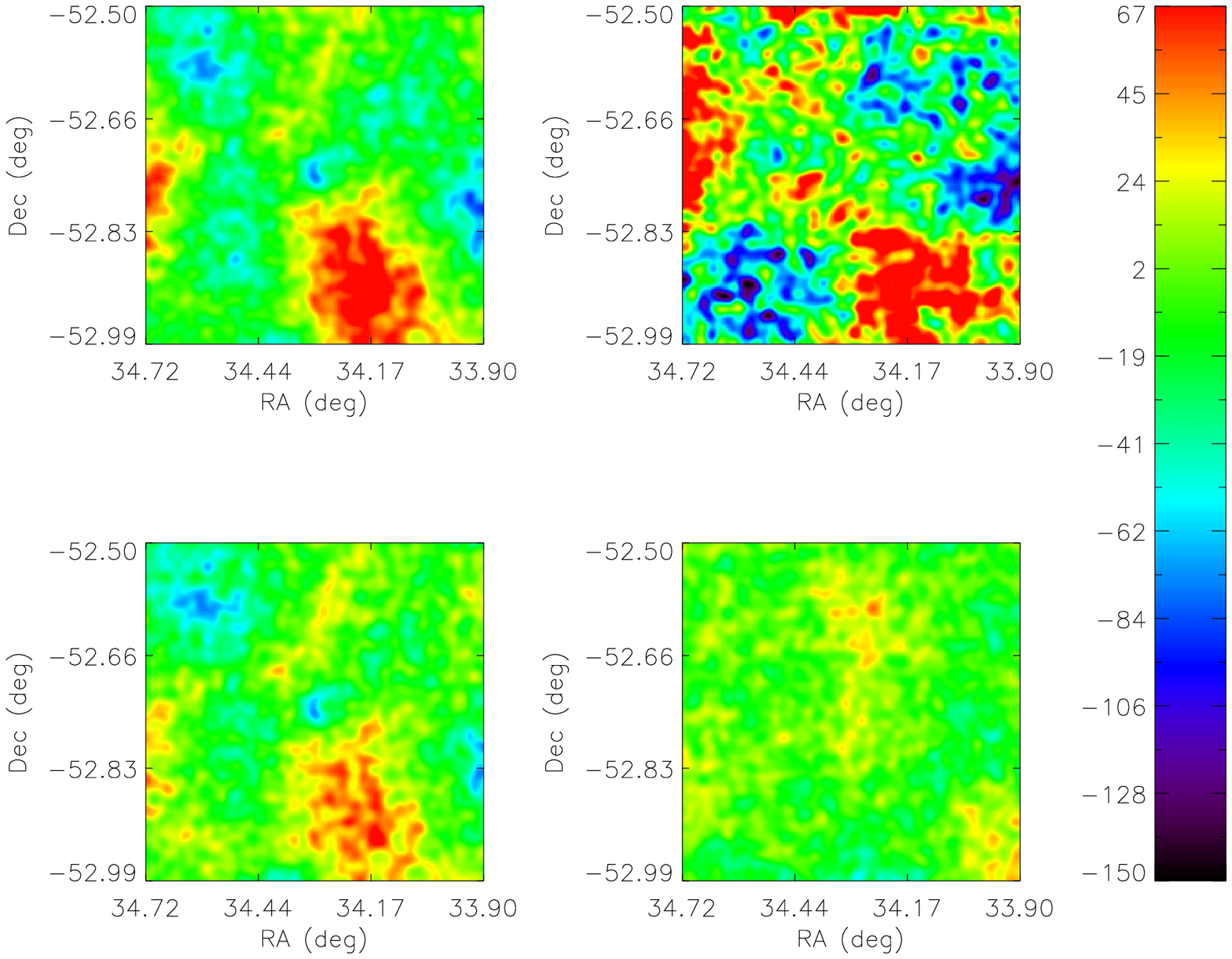} &
\includegraphics[width=.4\textwidth]{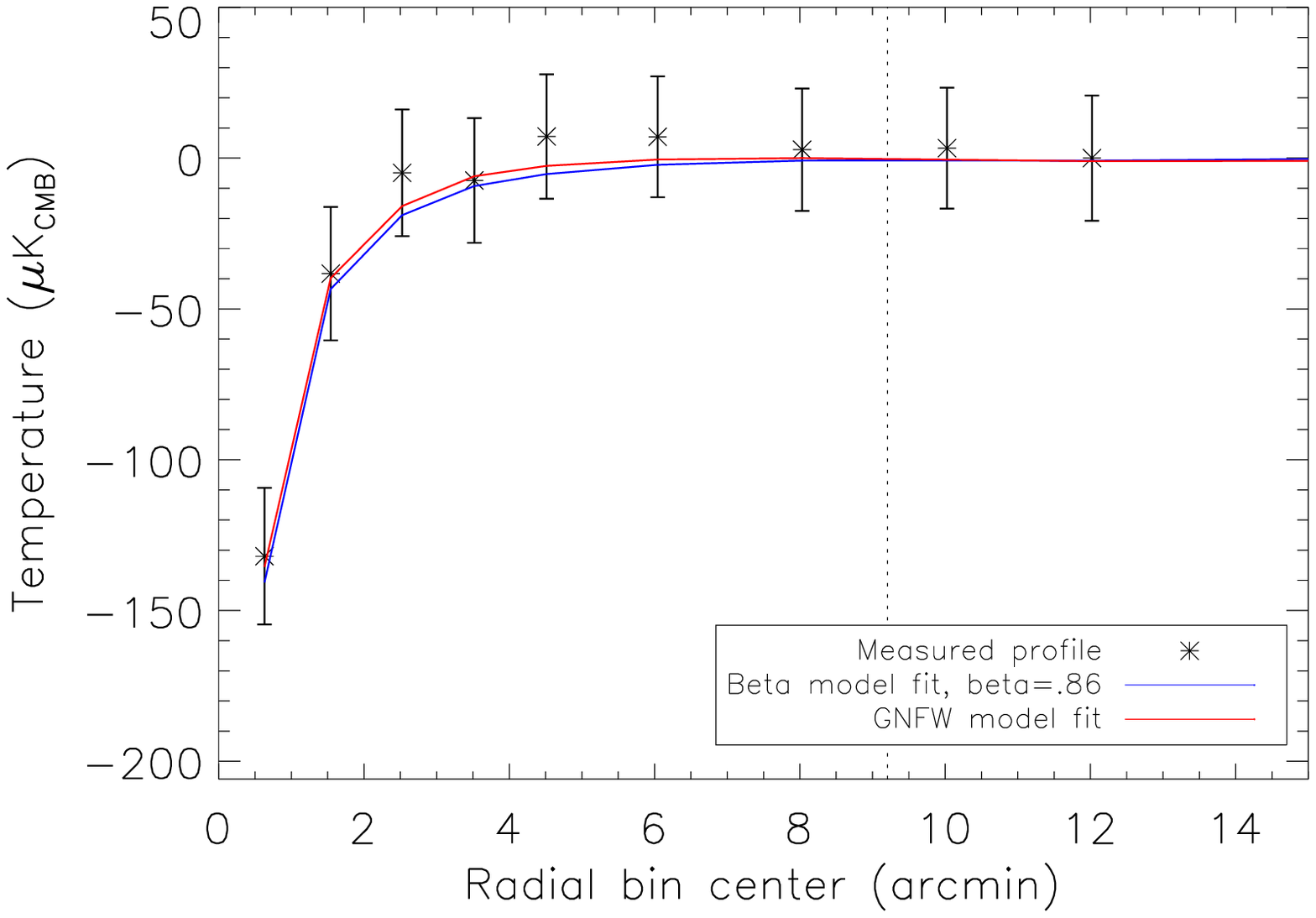} \\
\end{tabular}
\caption{RXCJ0217.2-5244 maps (left) and profile (right).  Units
are $\mu K_{\mathrm{CMB}}$.}
\end{figure*}

\begin{figure*}[ht!]
\begin{tabular}{cc}
\includegraphics[width=.4\textwidth]{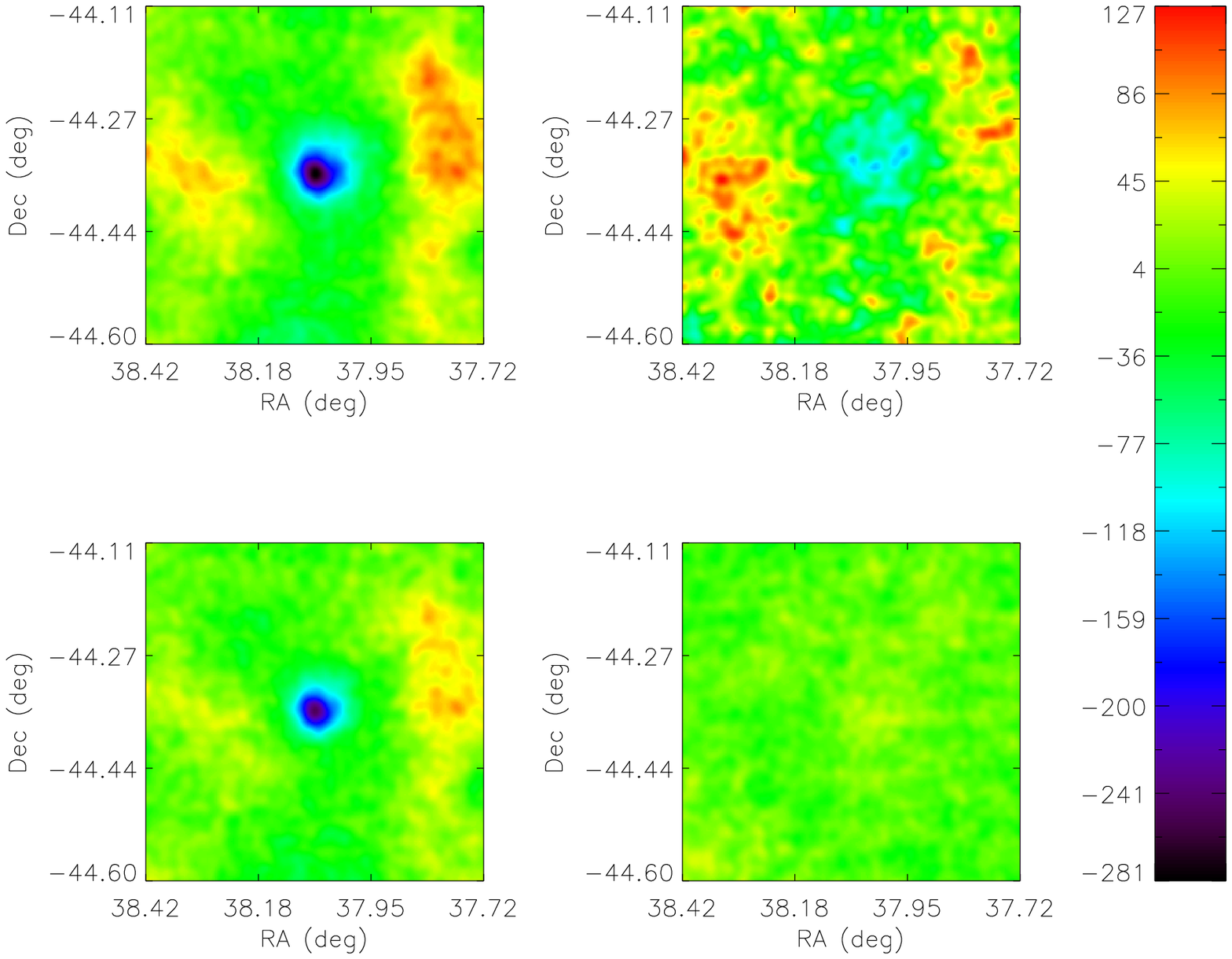} &
\includegraphics[width=.4\textwidth]{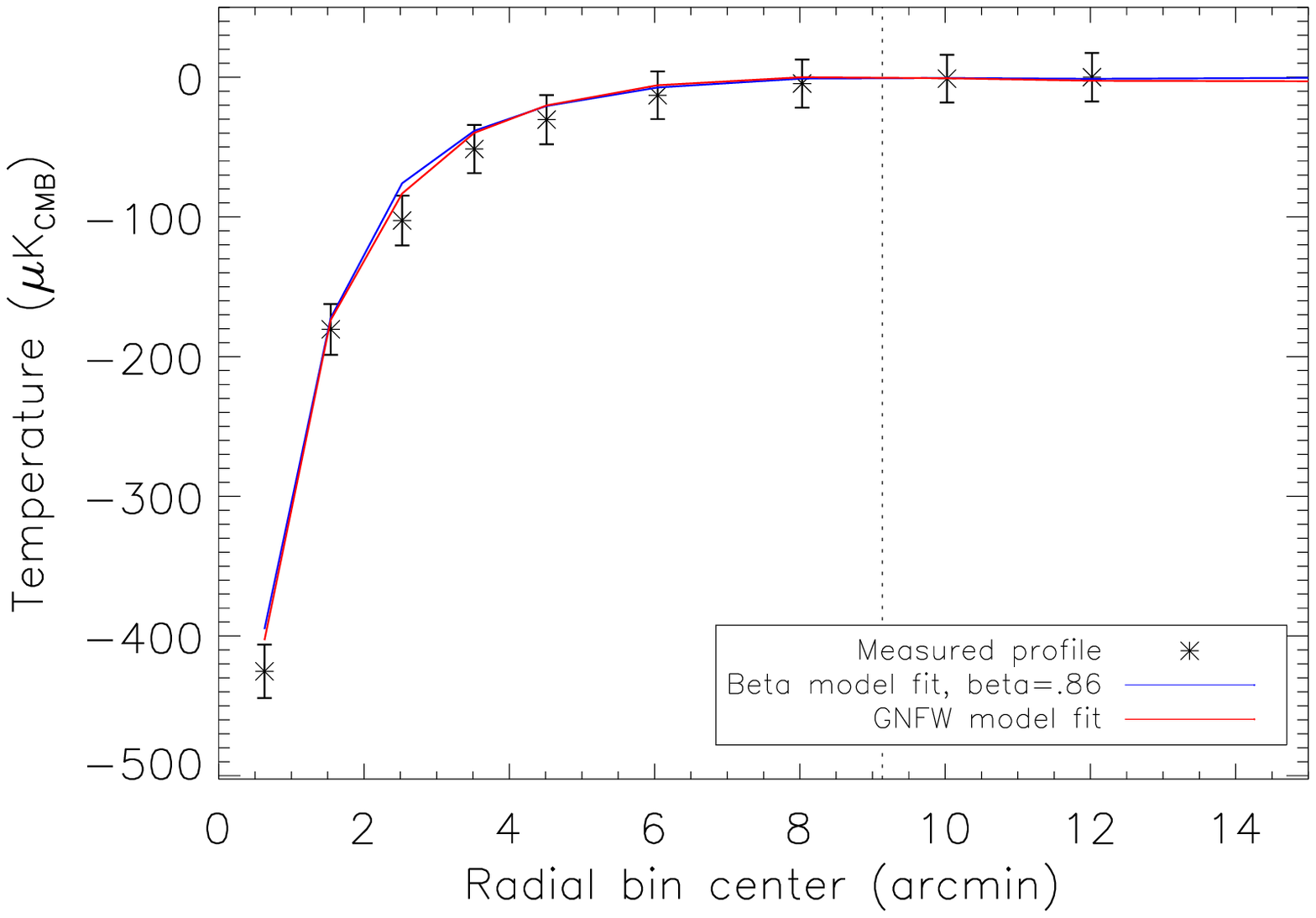} \\
\end{tabular}
\caption{RXCJ0232.2-4420 maps (left) and profile (right).  Units
are $\mu K_{\mathrm{CMB}}$.}
\end{figure*}

\begin{figure*}[ht!]
\begin{tabular}{cc}
\includegraphics[width=.4\textwidth]{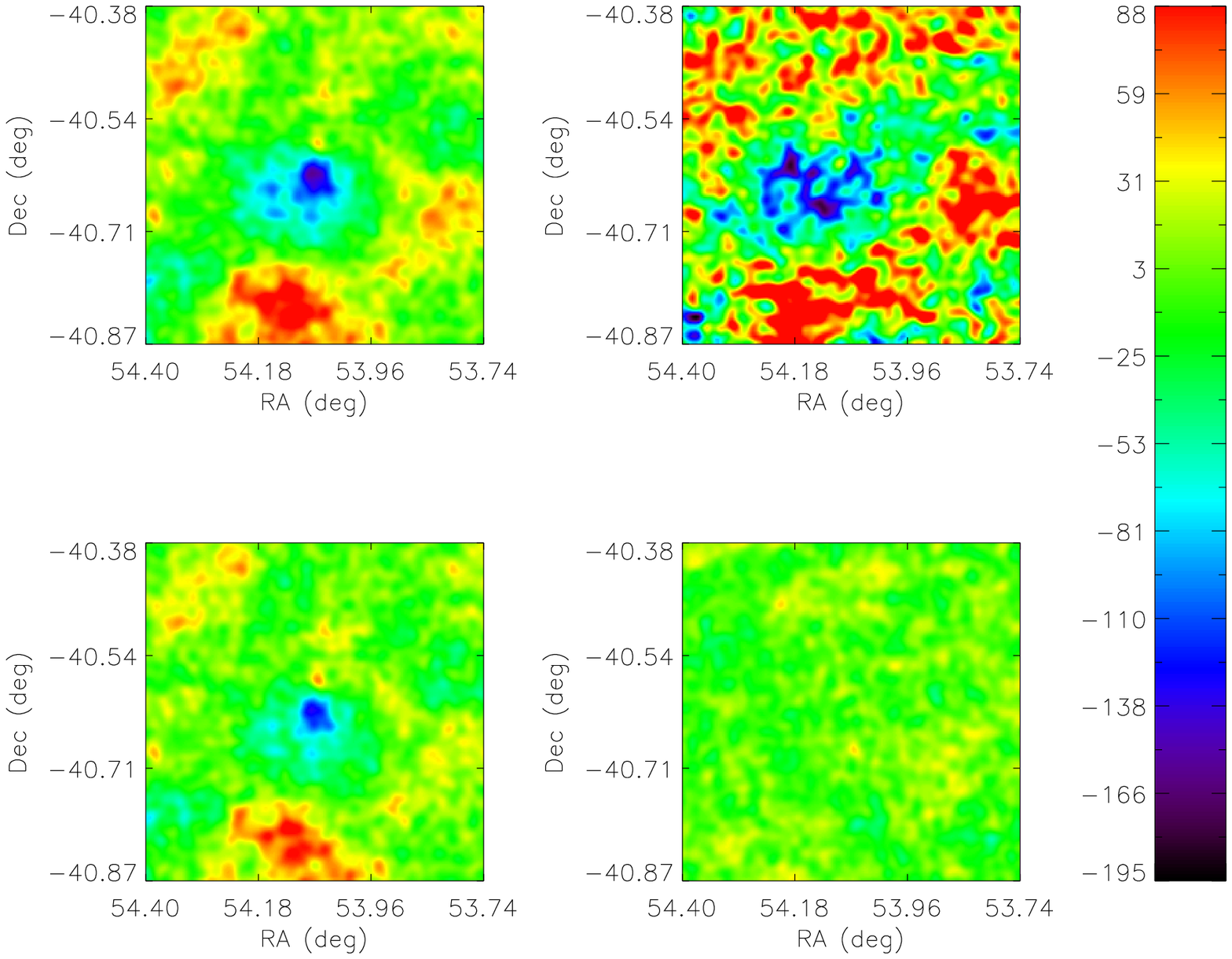} &
\includegraphics[width=.4\textwidth]{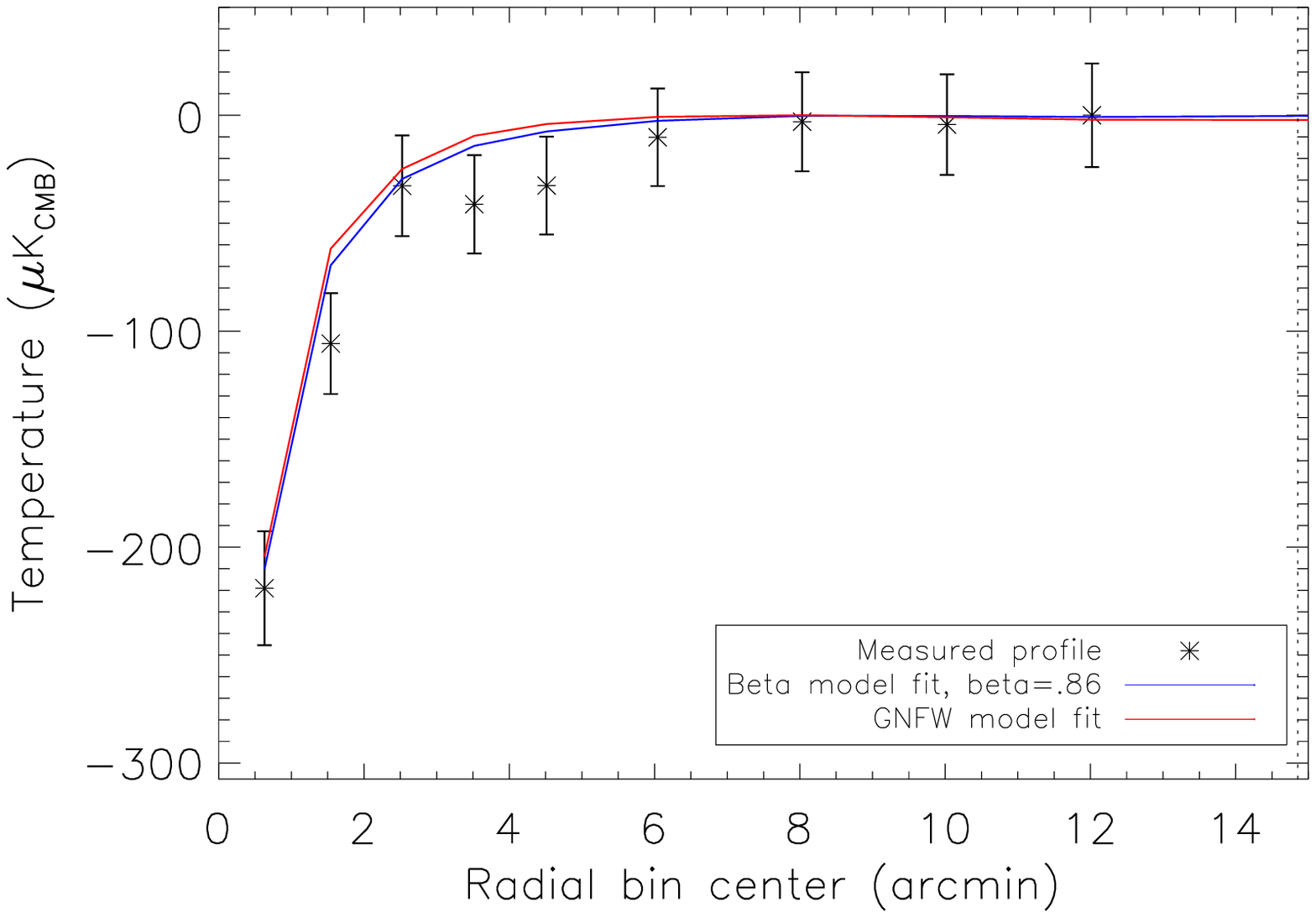} \\
\end{tabular}
\caption{RXCJ0336.3-4037 maps (left) and profile (right).  Units
are $\mu K_{\mathrm{CMB}}$.}
\end{figure*}

\begin{figure*}[ht!]
\begin{tabular}{cc}
\includegraphics[width=.4\textwidth]{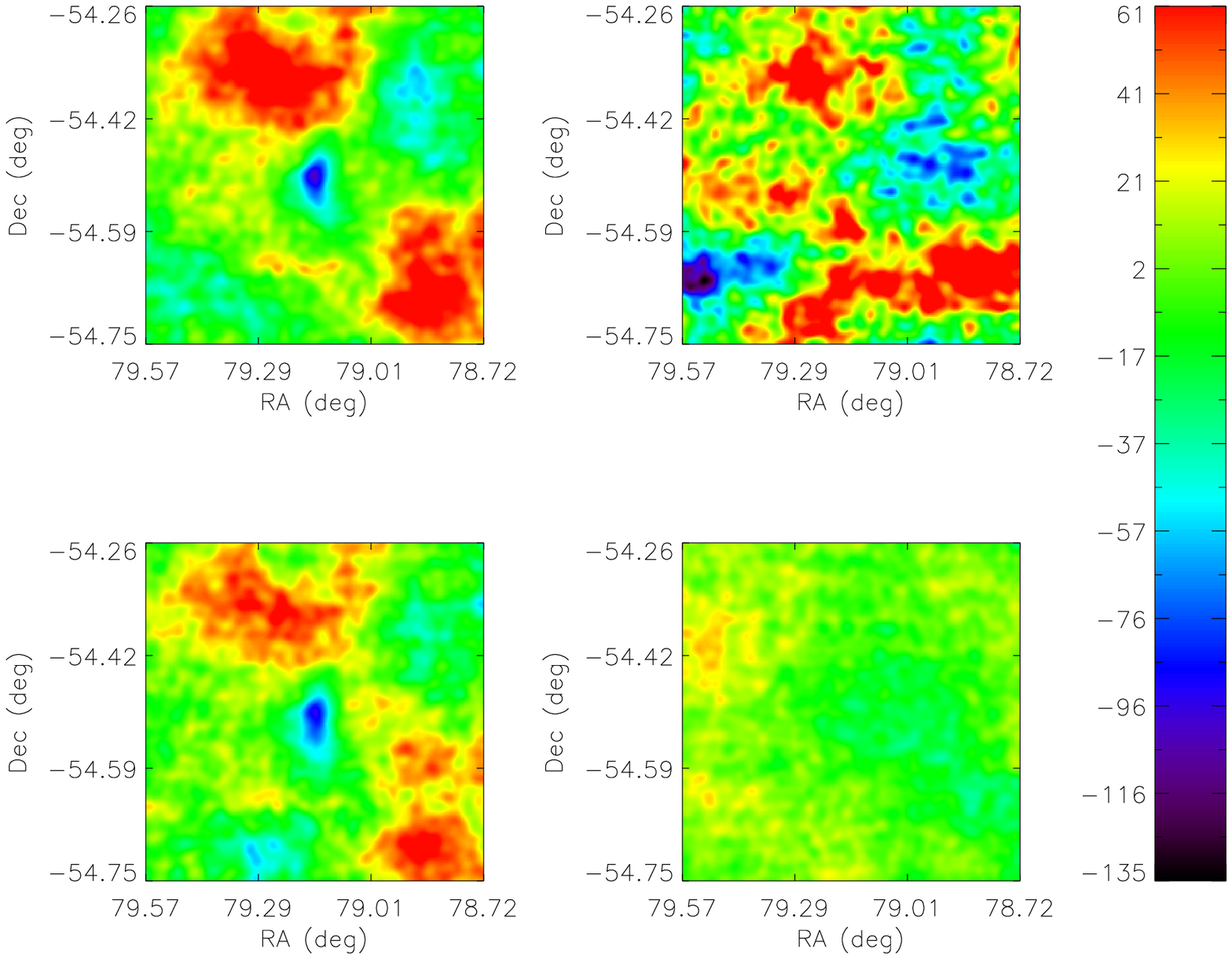} &
\includegraphics[width=.4\textwidth]{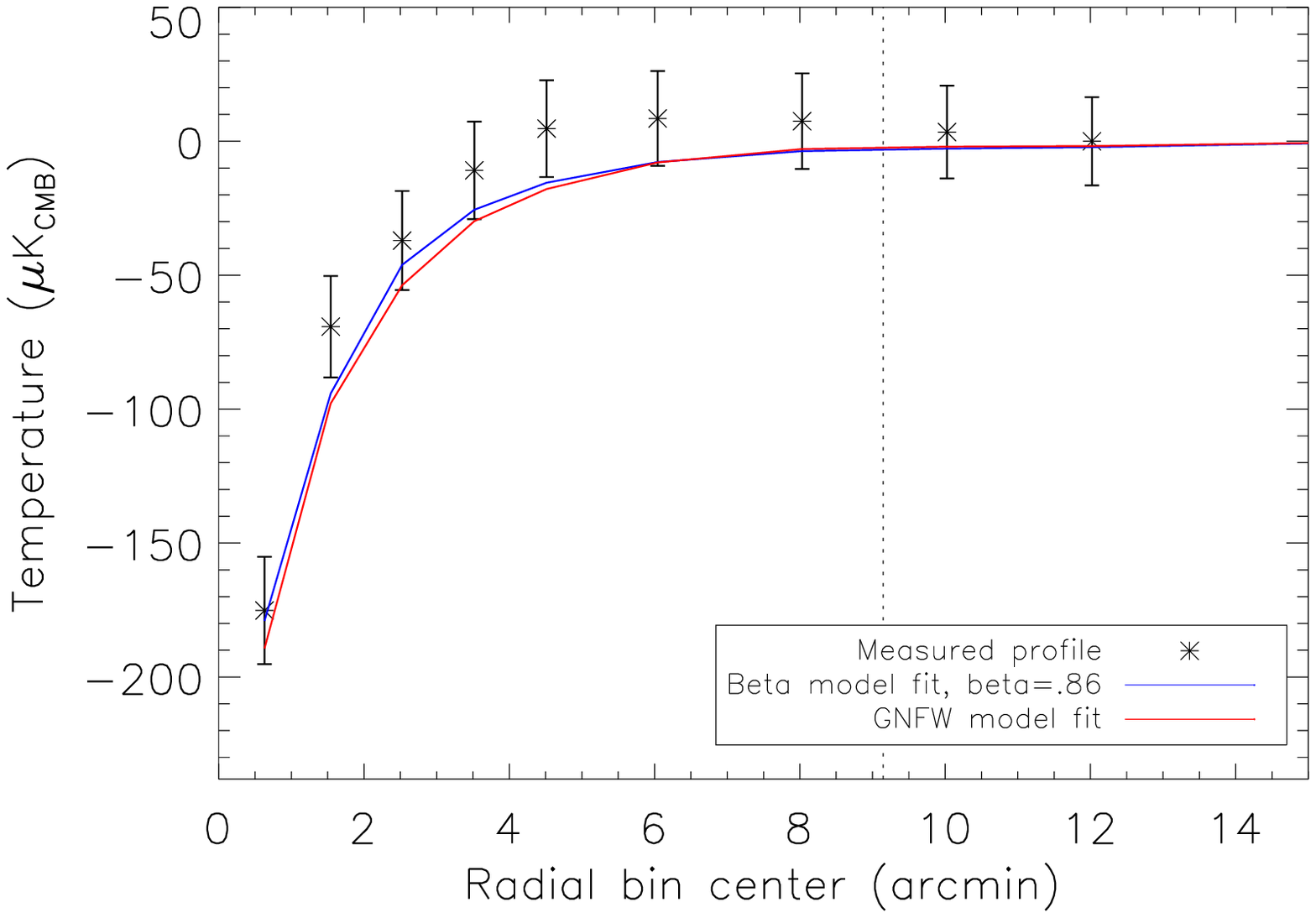} \\
\end{tabular}
\caption{AS~0520 maps (left) and profile (right).  Units
are $\mu K_{\mathrm{CMB}}$.}
\end{figure*}

\begin{figure*}[ht!]
\begin{tabular}{cc}
\includegraphics[width=.4\textwidth]{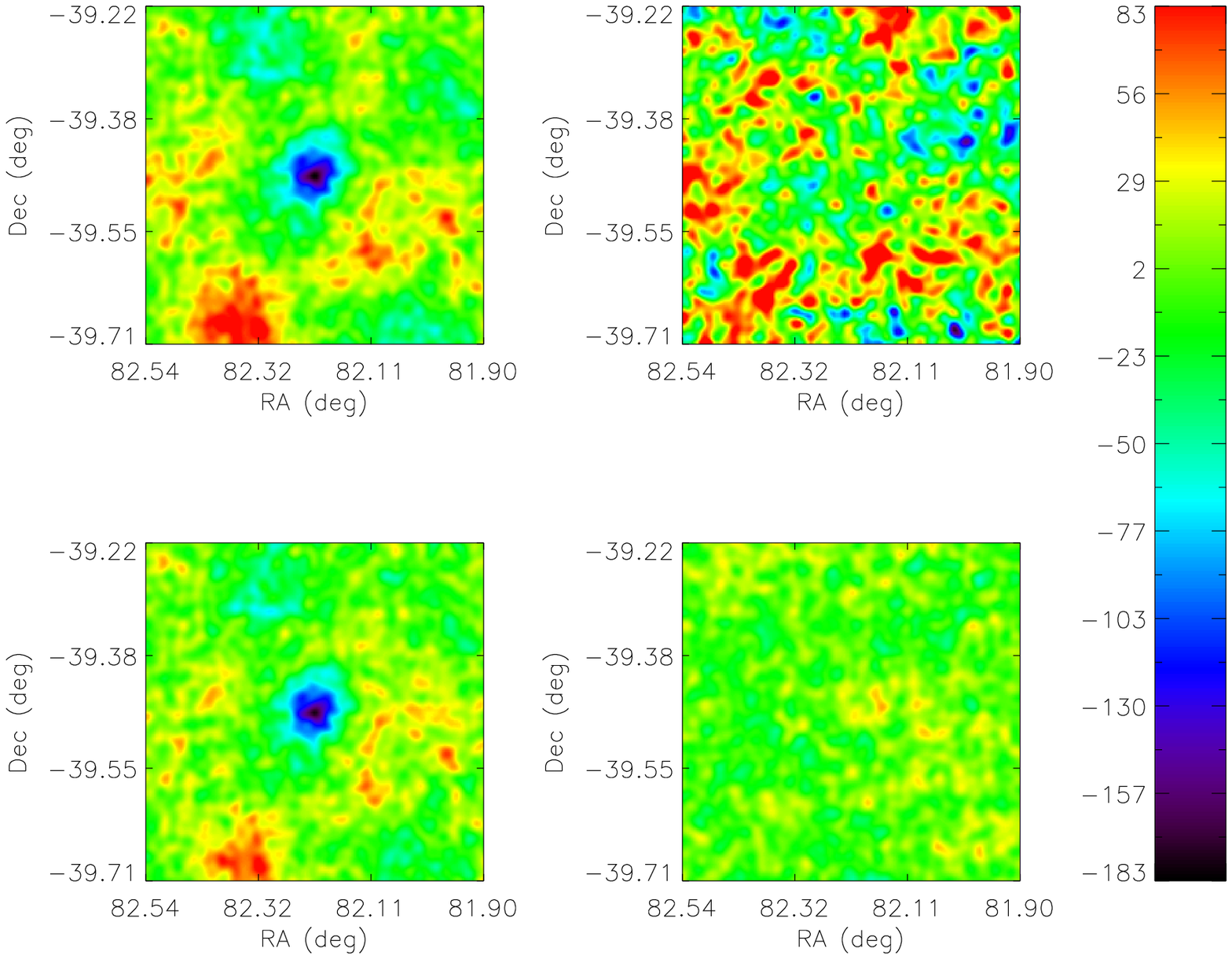} &
\includegraphics[width=.4\textwidth]{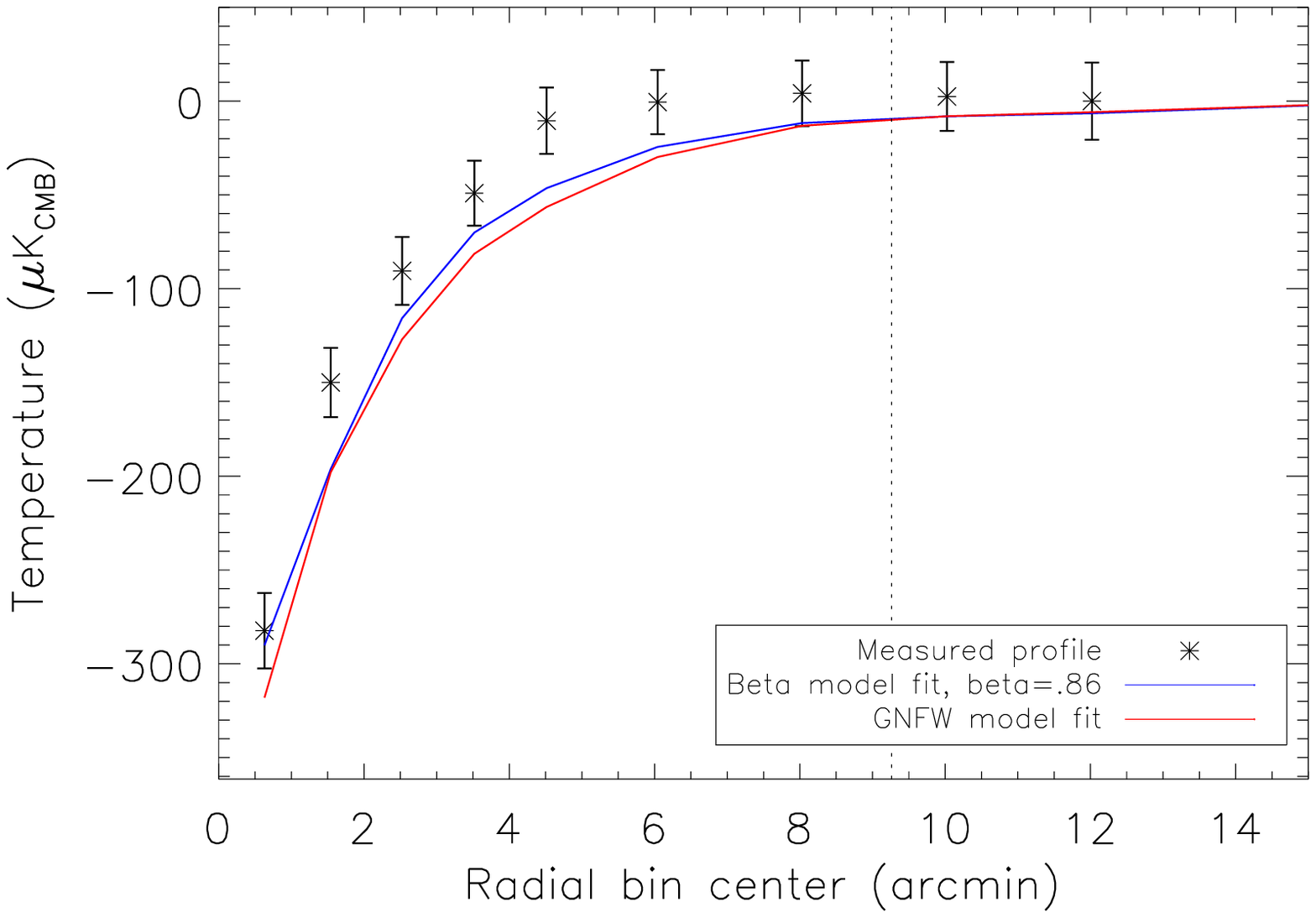} \\
\end{tabular}
\caption{RXCJ0528.9-3927 maps (left) and profile (right).  Units
are $\mu K_{\mathrm{CMB}}$.}
\end{figure*}

\begin{figure*}[ht!]
\begin{tabular}{cc}
\includegraphics[width=.4\textwidth]{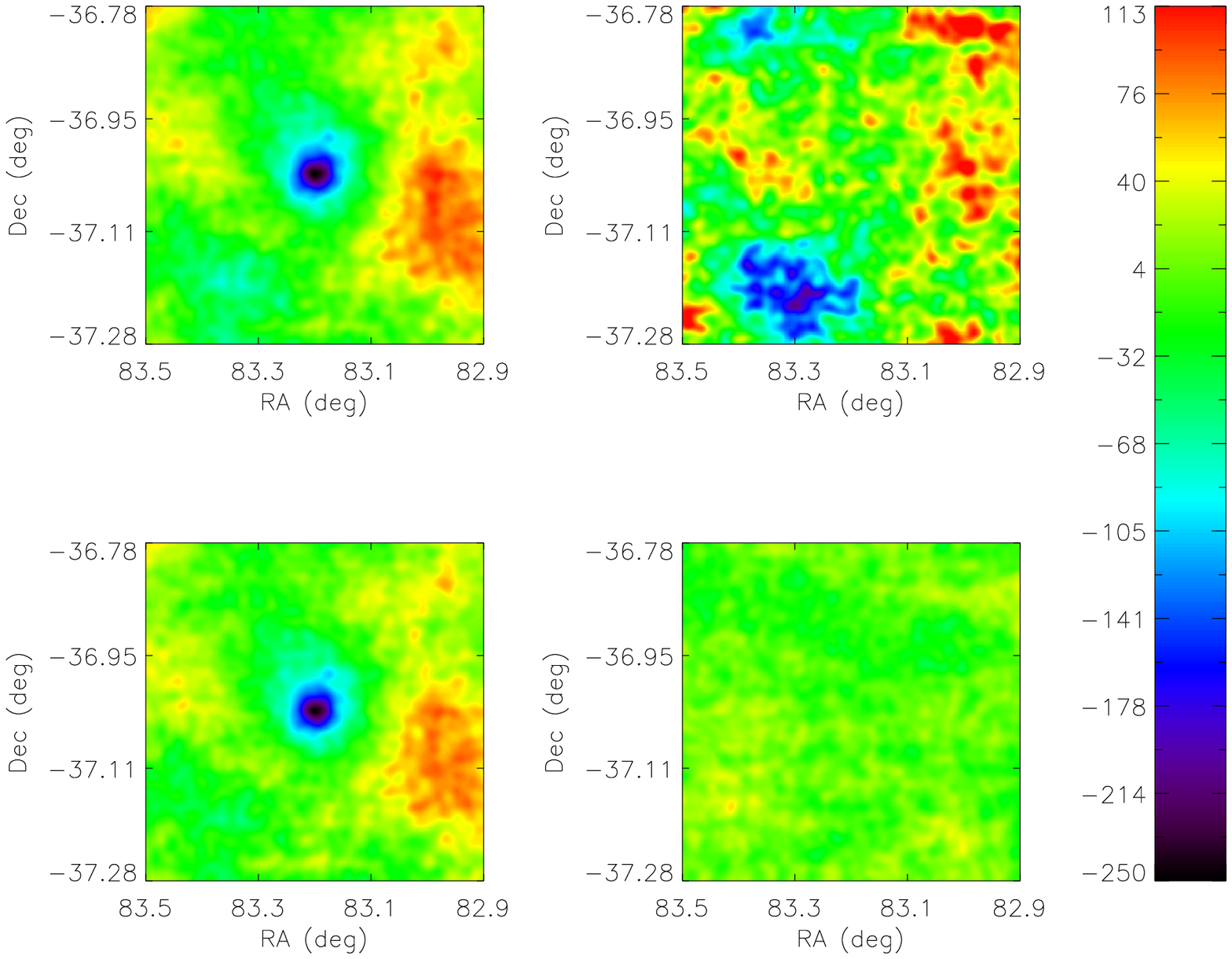} &
\includegraphics[width=.4\textwidth]{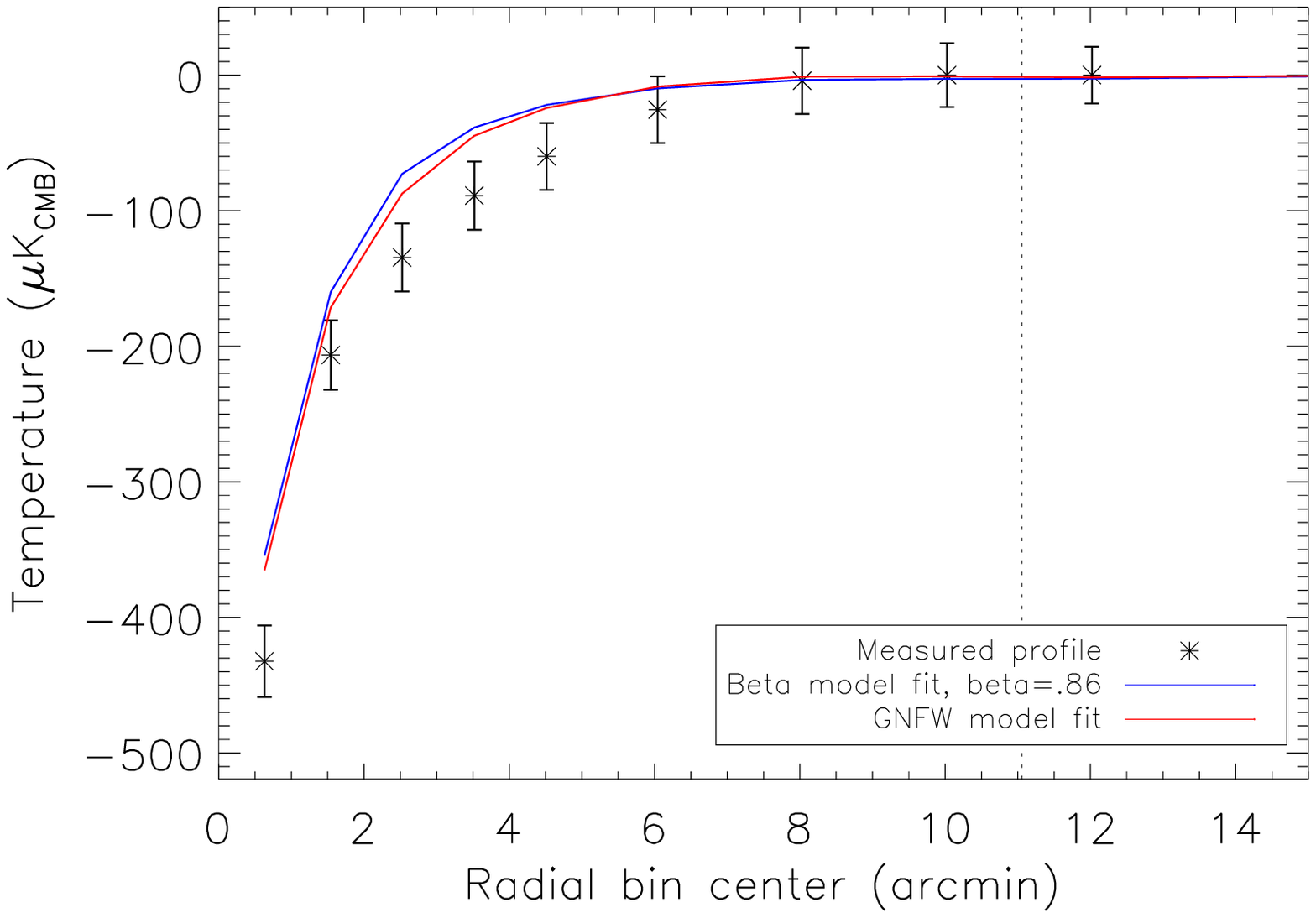} \\
\end{tabular}
\caption{RXCJ0532.9-3701 maps (left) and profile (right).  Units
are $\mu K_{\mathrm{CMB}}$.}
\end{figure*}

\begin{figure*}[ht!]
\begin{tabular}{cc}
\includegraphics[width=.4\textwidth]{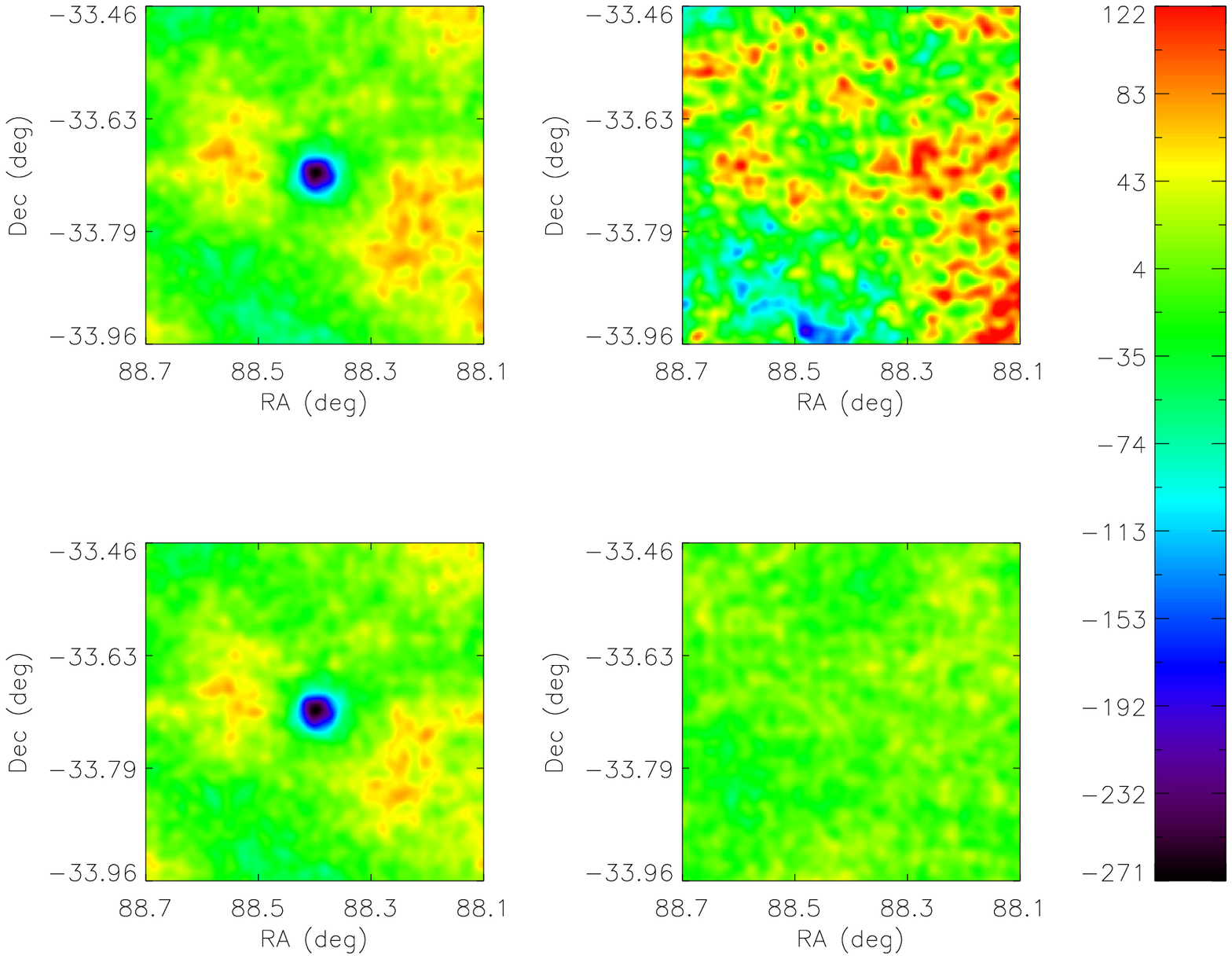} &
\includegraphics[width=.4\textwidth]{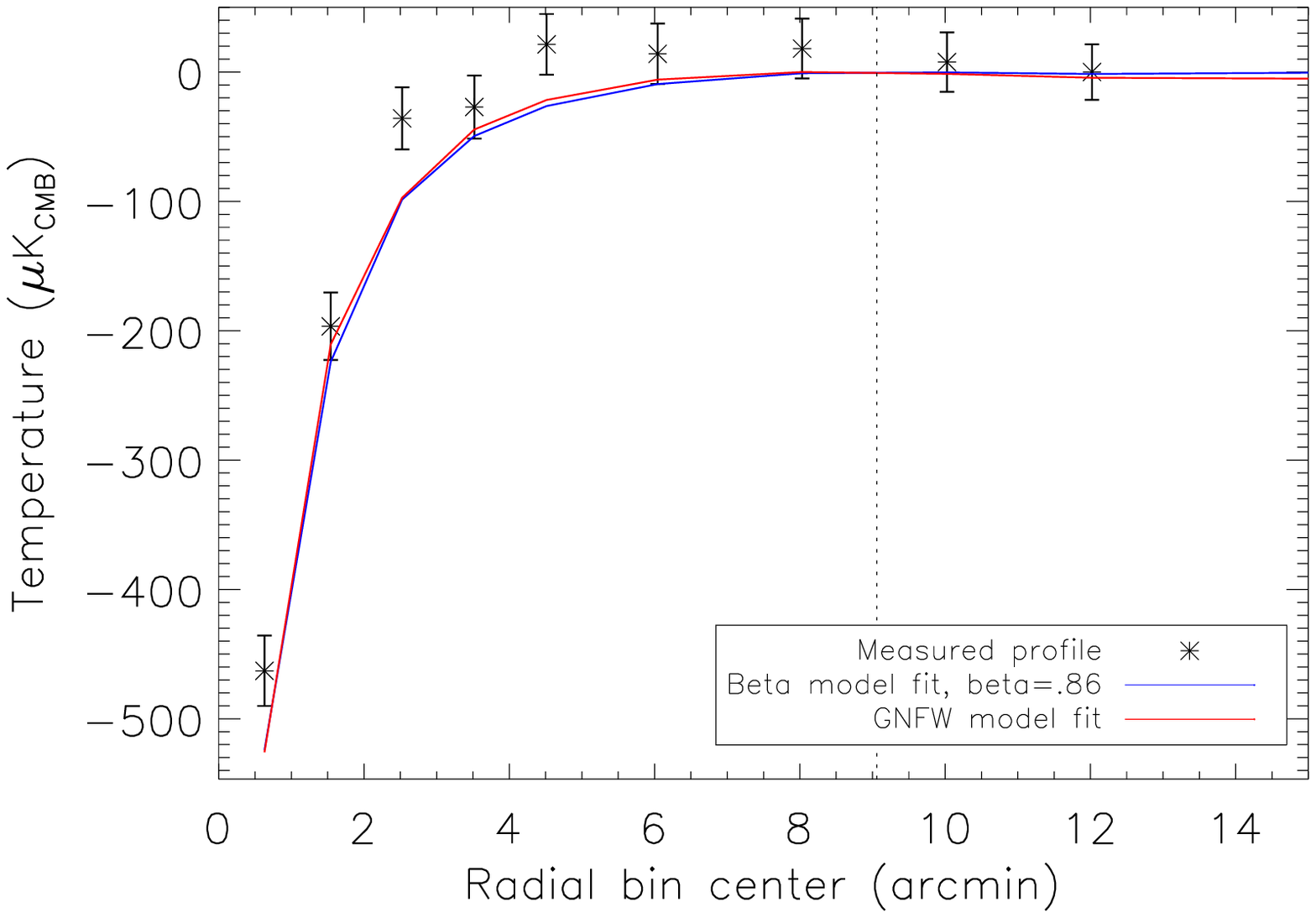} \\
\end{tabular}
\caption{MACSJ0553.4-3342 maps (left) and profile (right).  Units
are $\mu K_{\mathrm{CMB}}$.}
\end{figure*}

\begin{figure*}[ht!]
\begin{tabular}{cc}
\includegraphics[width=.4\textwidth]{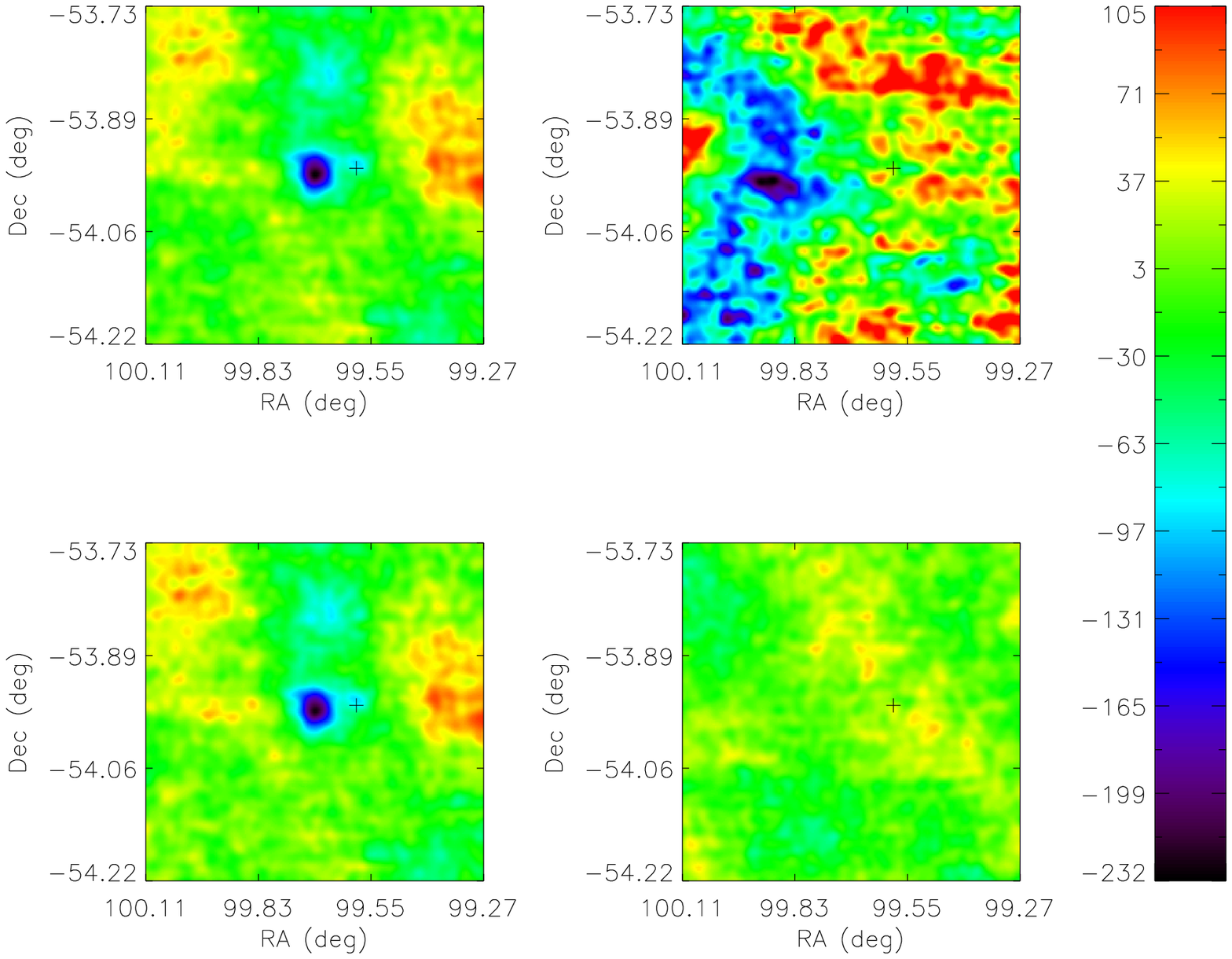} &
\includegraphics[width=.4\textwidth]{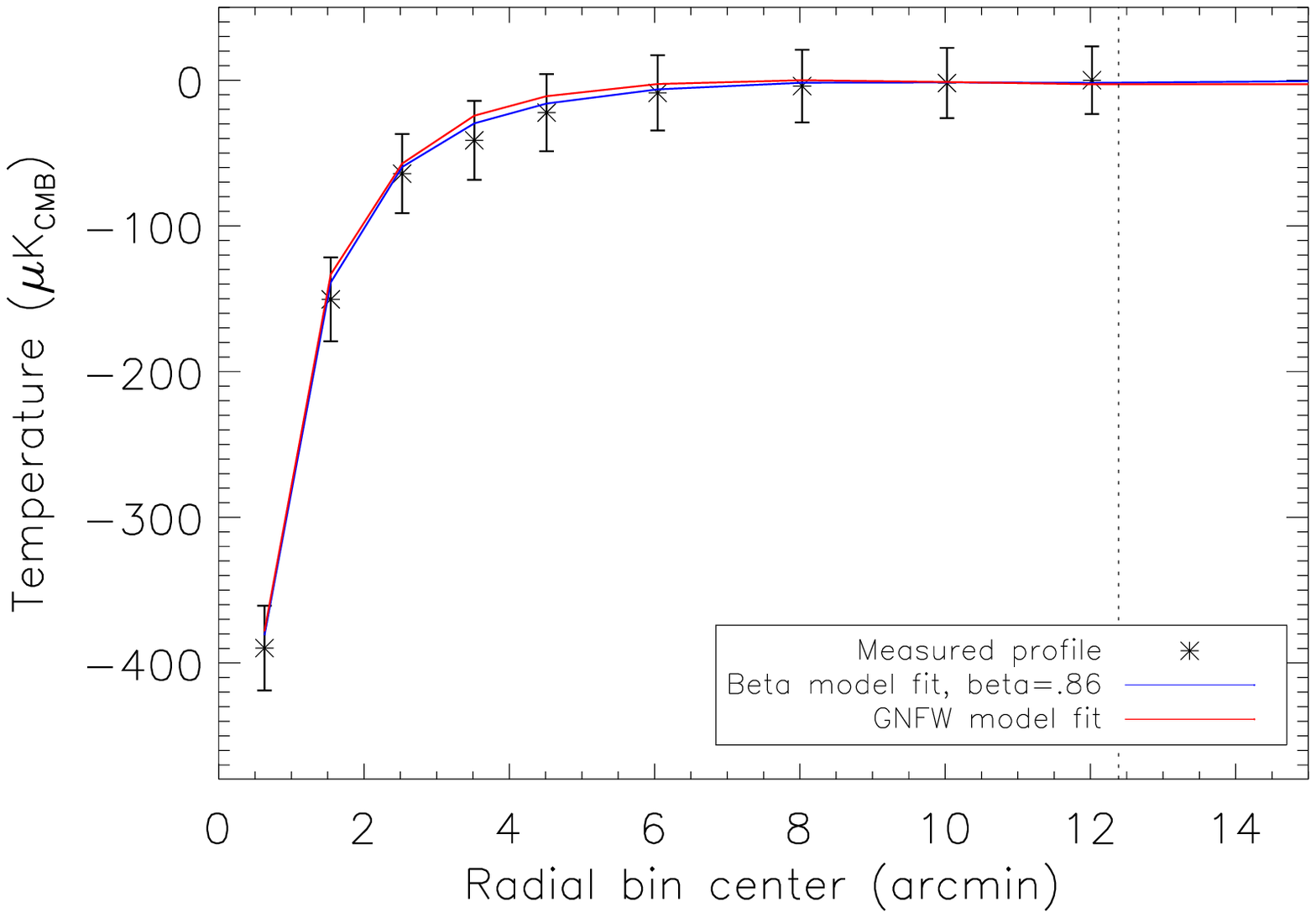} \\
\end{tabular}
\caption{AS~0592 maps (left) and profile (right).  Units
are $\mu K_{\mathrm{CMB}}$.}
\end{figure*}

\begin{figure*}[ht!]
\begin{tabular}{cc}
\includegraphics[width=.4\textwidth]{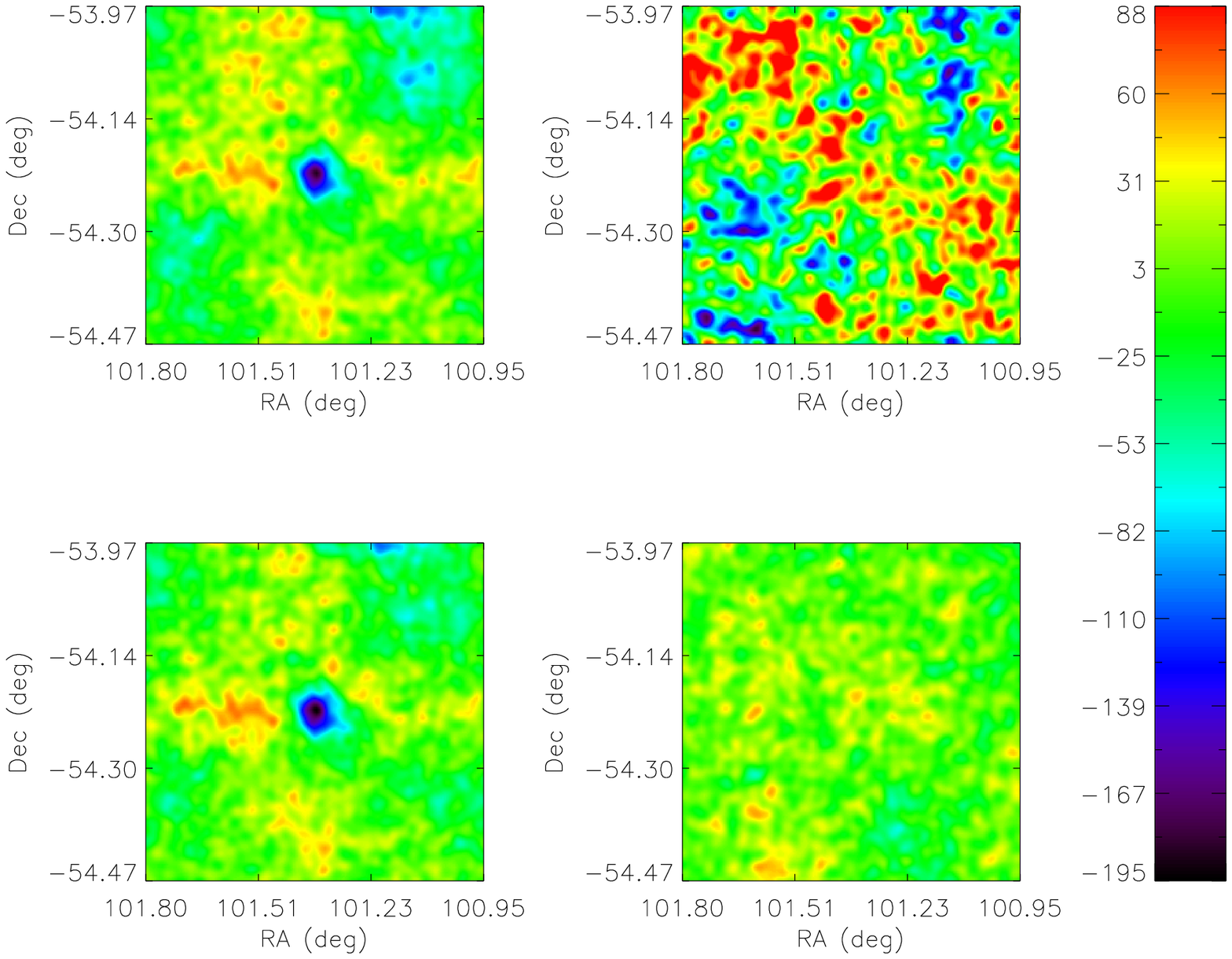} &
\includegraphics[width=.4\textwidth]{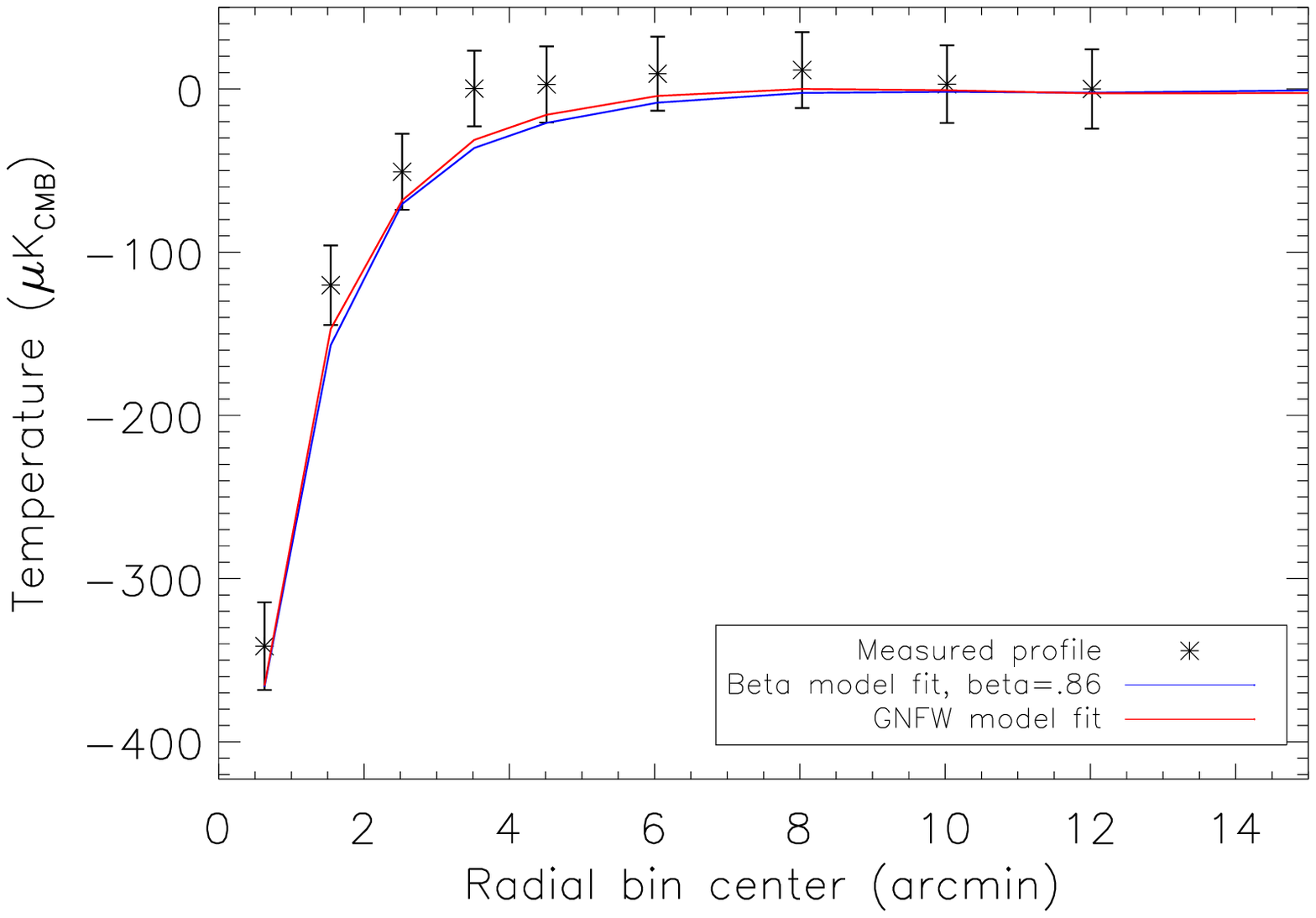} \\
\end{tabular}
\caption{A~3404 maps (left) and profile (right).  Units
are $\mu K_{\mathrm{CMB}}$.}
\end{figure*}

\begin{figure*}[ht!]
\begin{tabular}{cc}
\includegraphics[width=.4\textwidth]{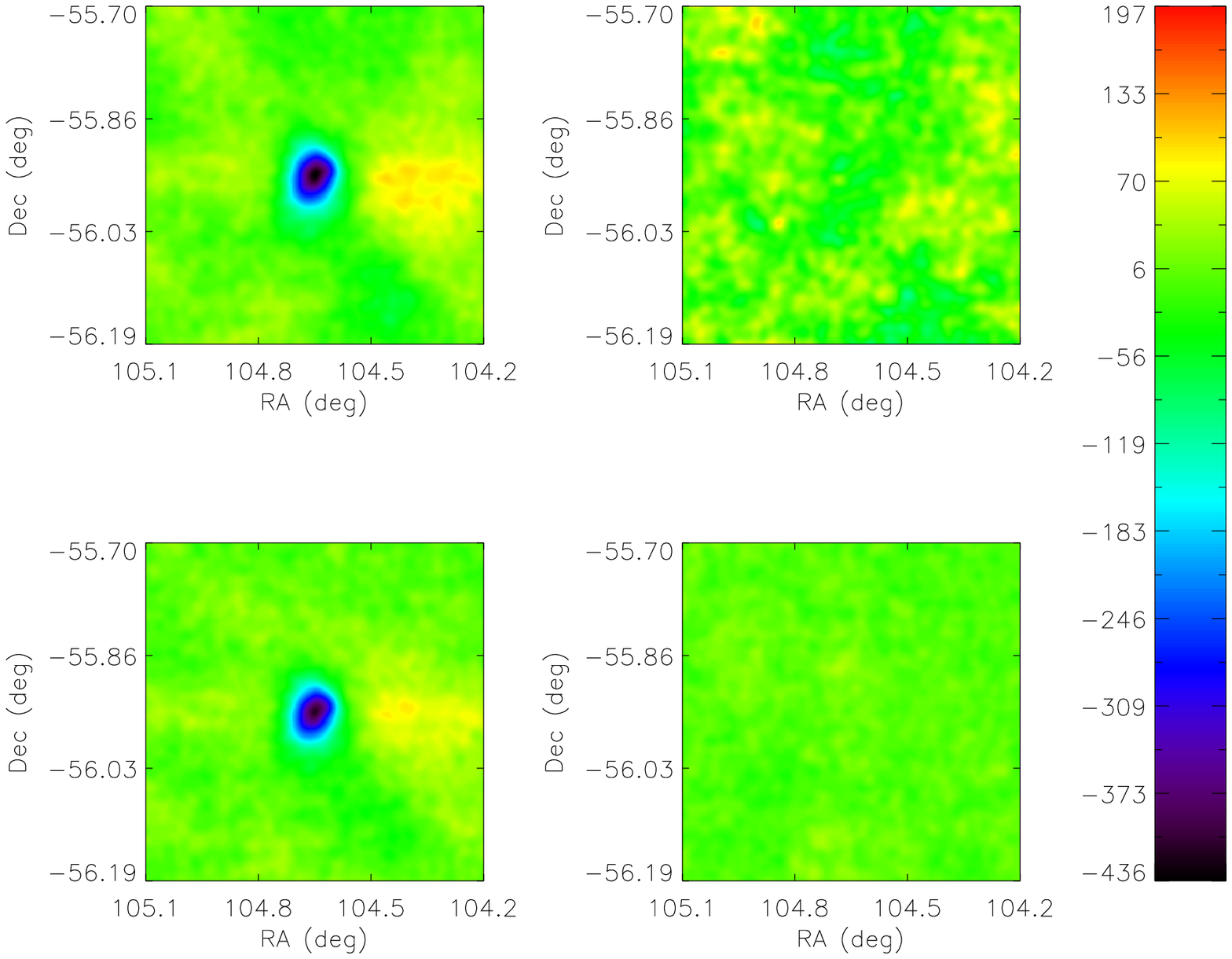} &
\includegraphics[width=.4\textwidth]{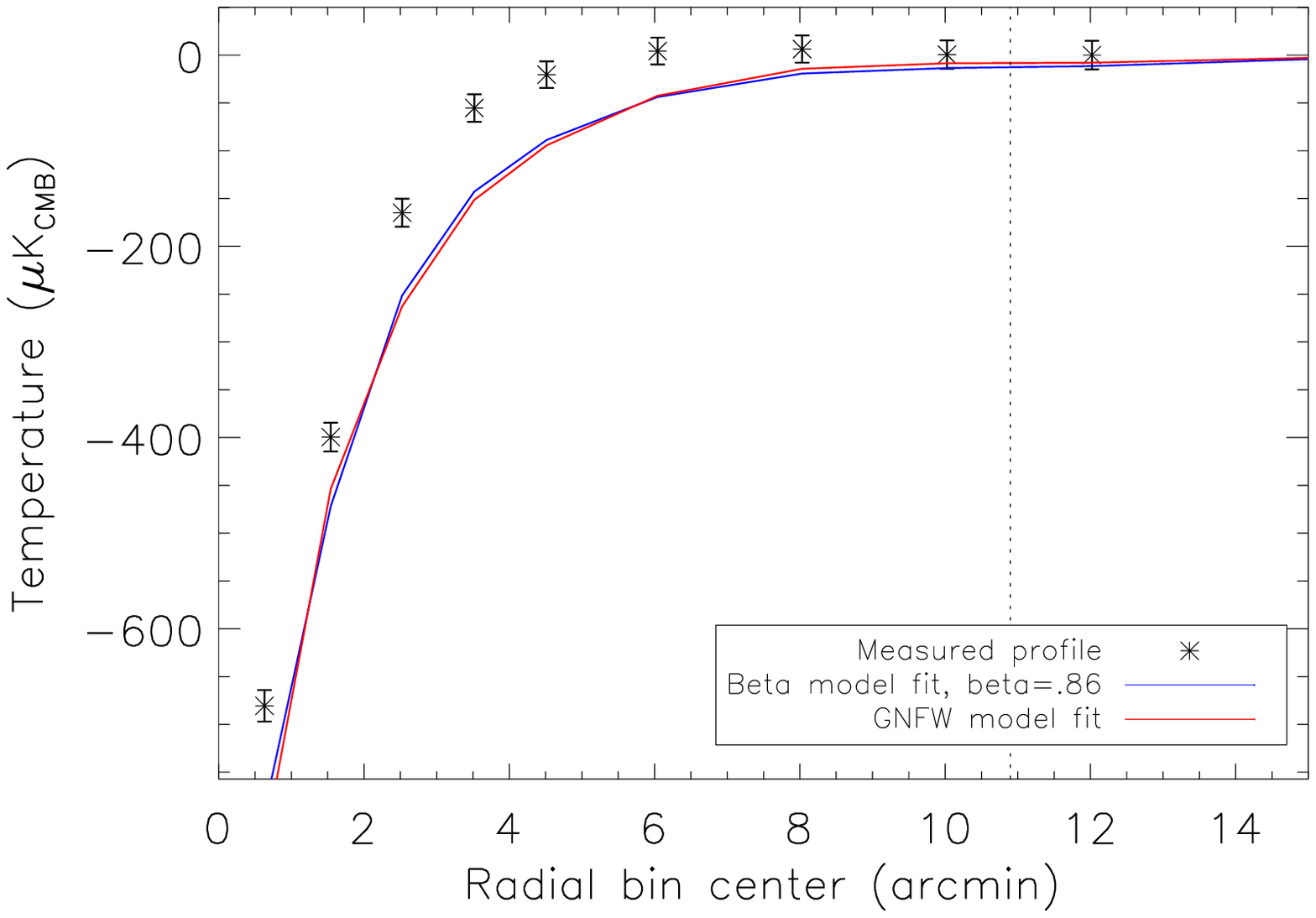} \\
\end{tabular}
\caption{1ES~0657-56 maps (left) and profile (right).  Units
are $\mu K_{\mathrm{CMB}}$.}
\end{figure*}

\begin{figure*}[ht!]
\begin{tabular}{cc}
\includegraphics[width=.4\textwidth]{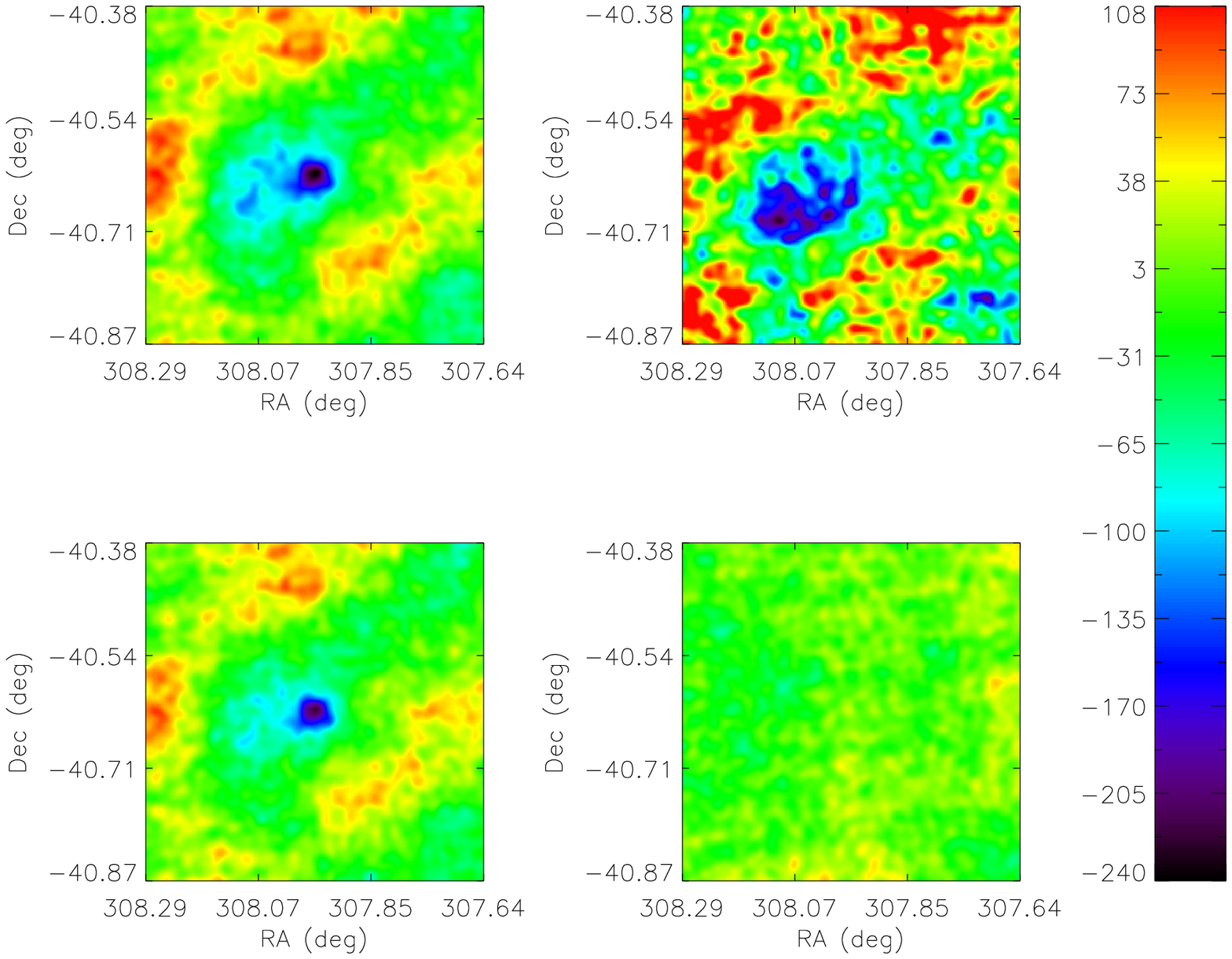} &
\includegraphics[width=.4\textwidth]{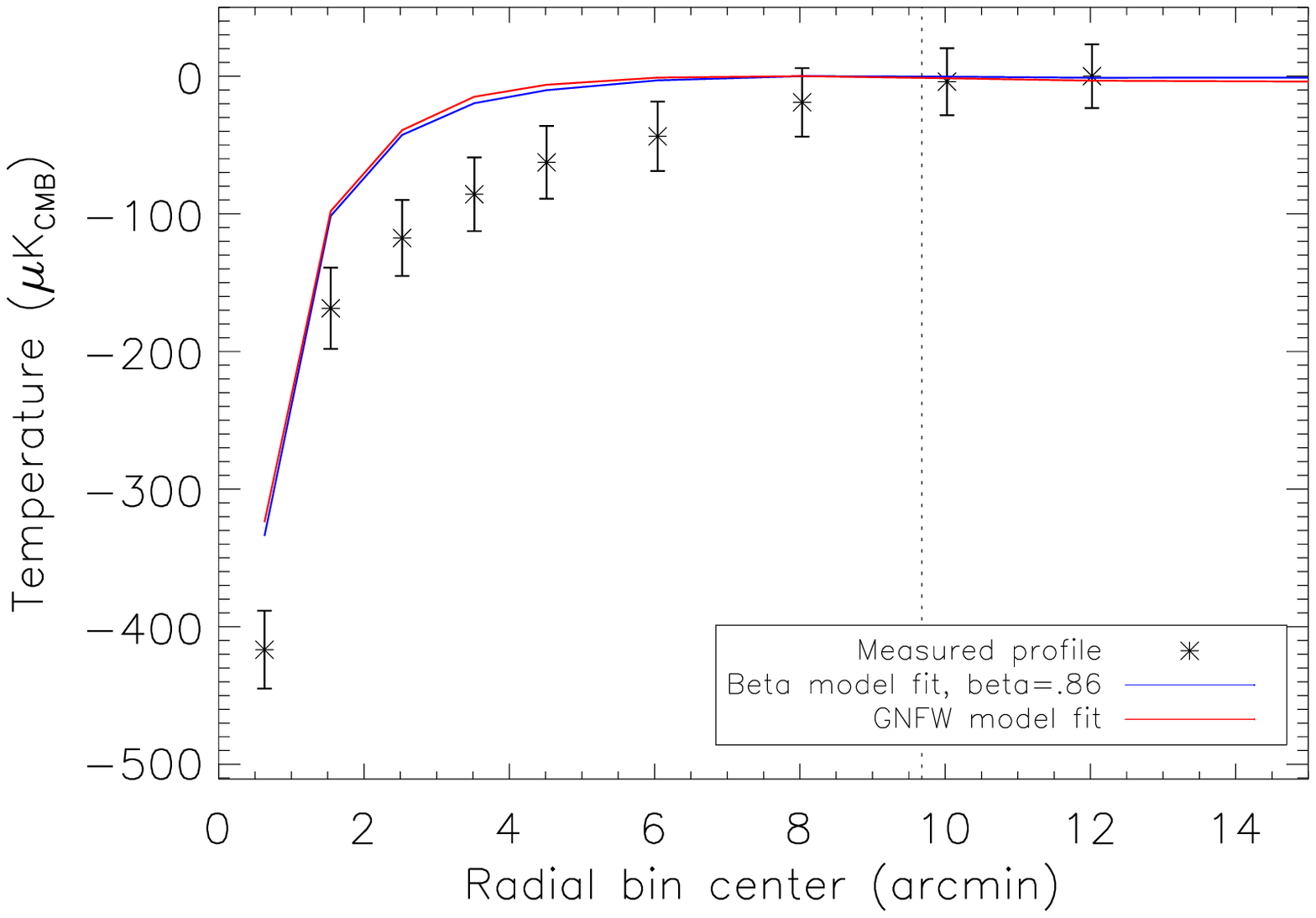} \\
\end{tabular}
\caption{RXCJ2031.8-4037 maps (left) and profile (right).  Units
are $\mu K_{\mathrm{CMB}}$.}
\end{figure*}

\begin{figure*}[ht!]
\begin{tabular}{cc}
\includegraphics[width=.4\textwidth]{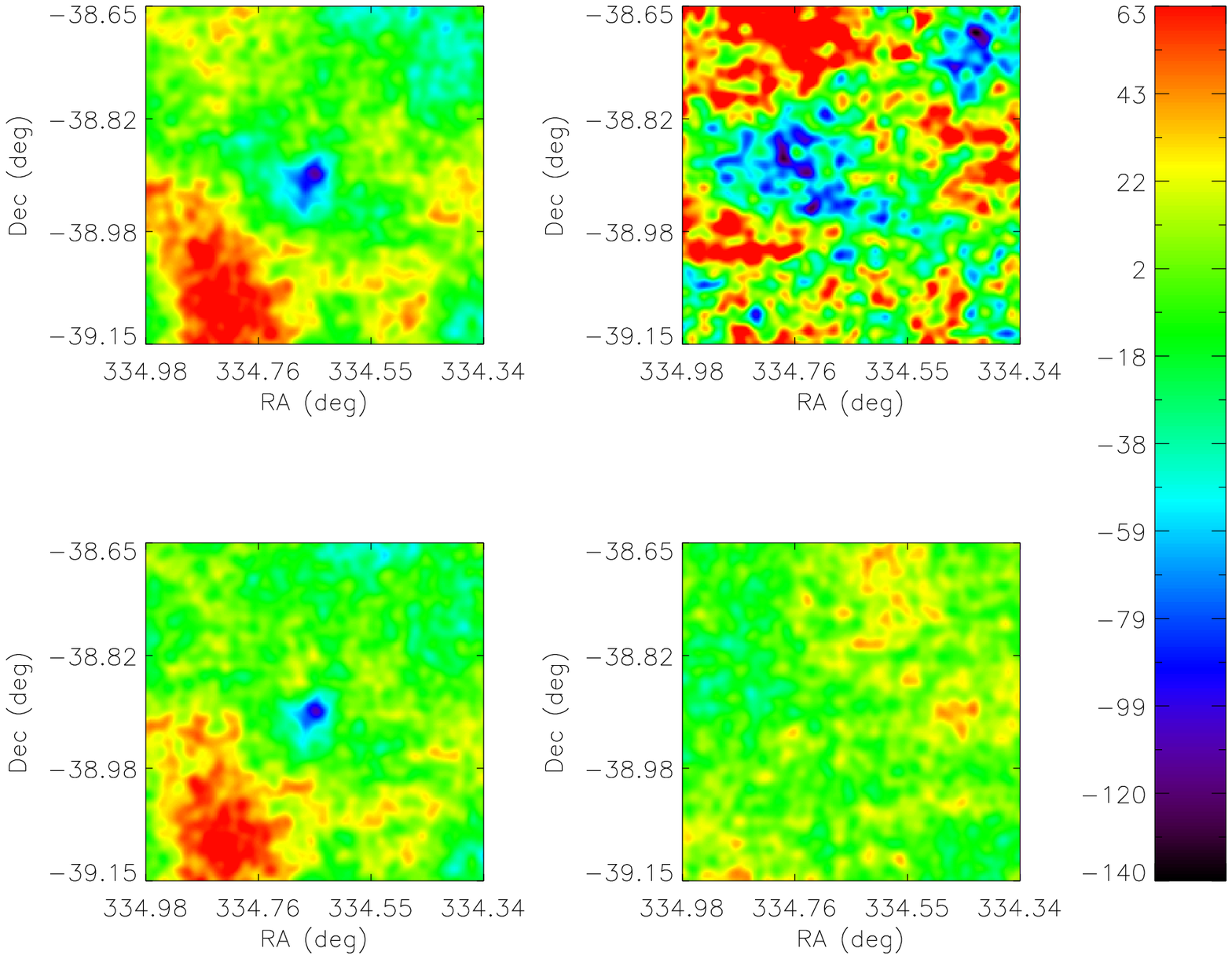} &
\includegraphics[width=.4\textwidth]{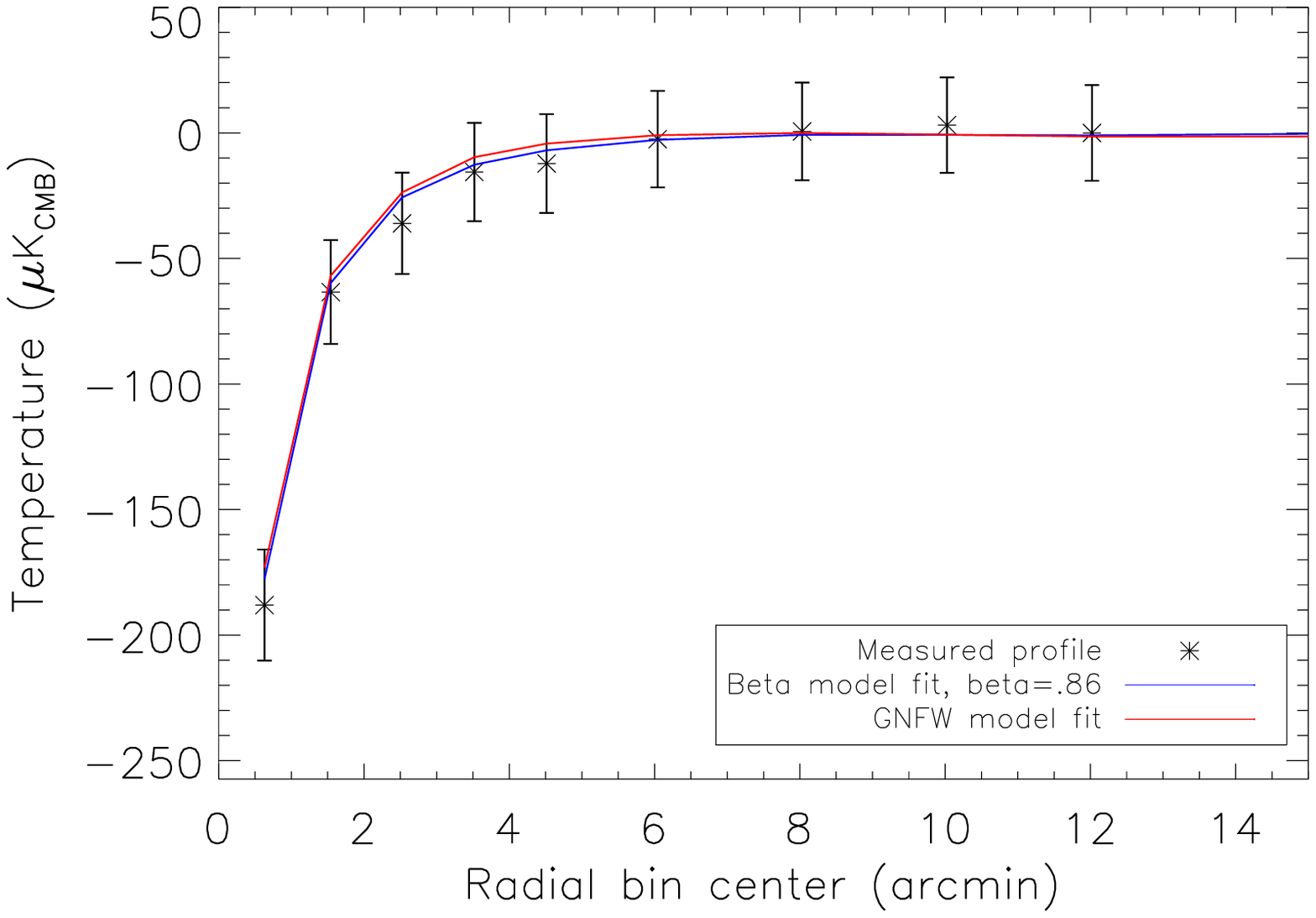} \\
\end{tabular}
\caption{A~3856 maps (left) and profile (right).  Units
are $\mu K_{\mathrm{CMB}}$.}
\end{figure*}

\begin{figure*}[ht!]
\begin{tabular}{cc}
\includegraphics[width=.4\textwidth]{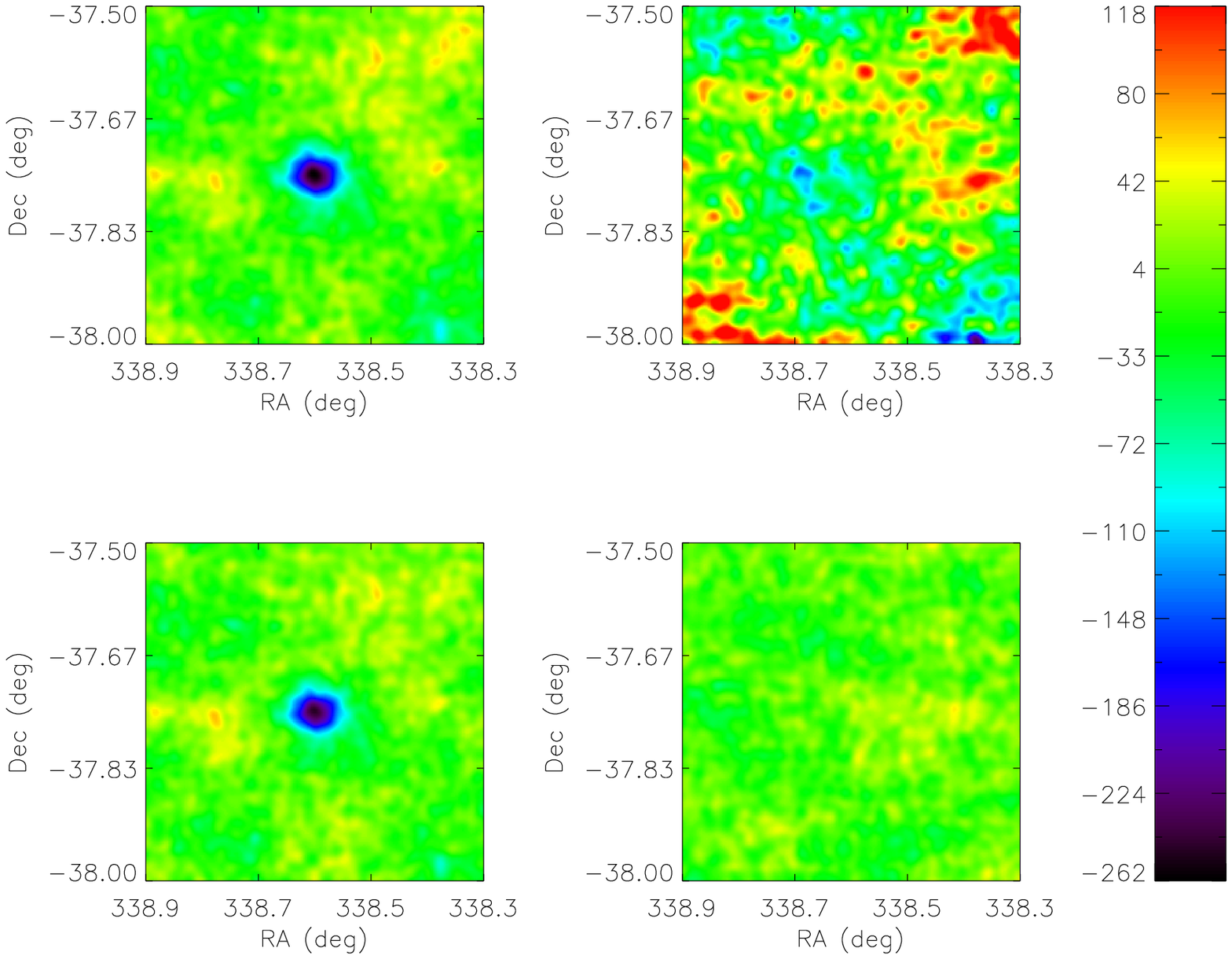} &
\includegraphics[width=.4\textwidth]{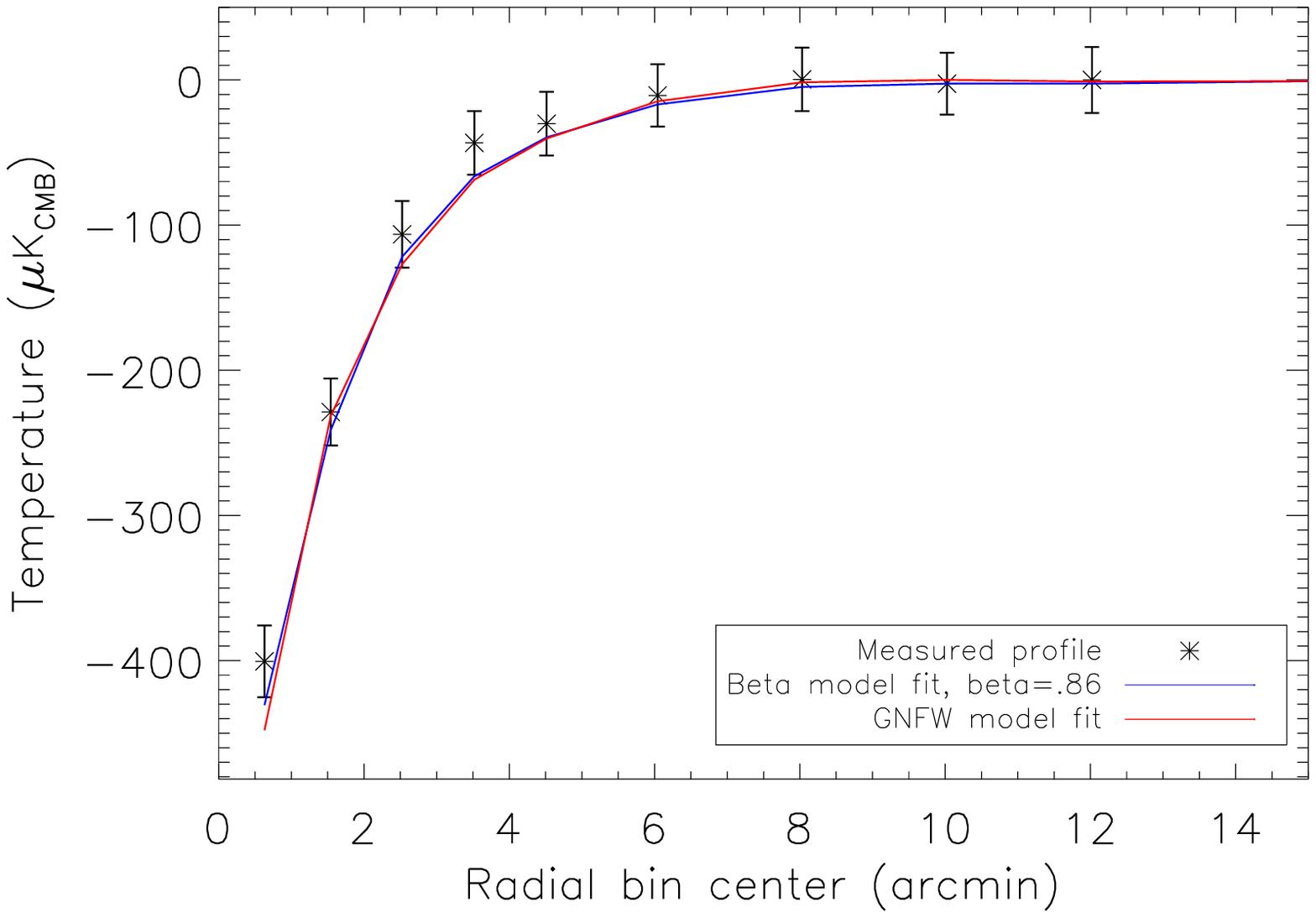} \\
\end{tabular}
\caption{A~3888 maps (left) and profile (right).  Units
are $\mu K_{\mathrm{CMB}}$.}
\end{figure*}

\begin{figure*}[ht!]
\begin{tabular}{cc}
\includegraphics[width=.4\textwidth]{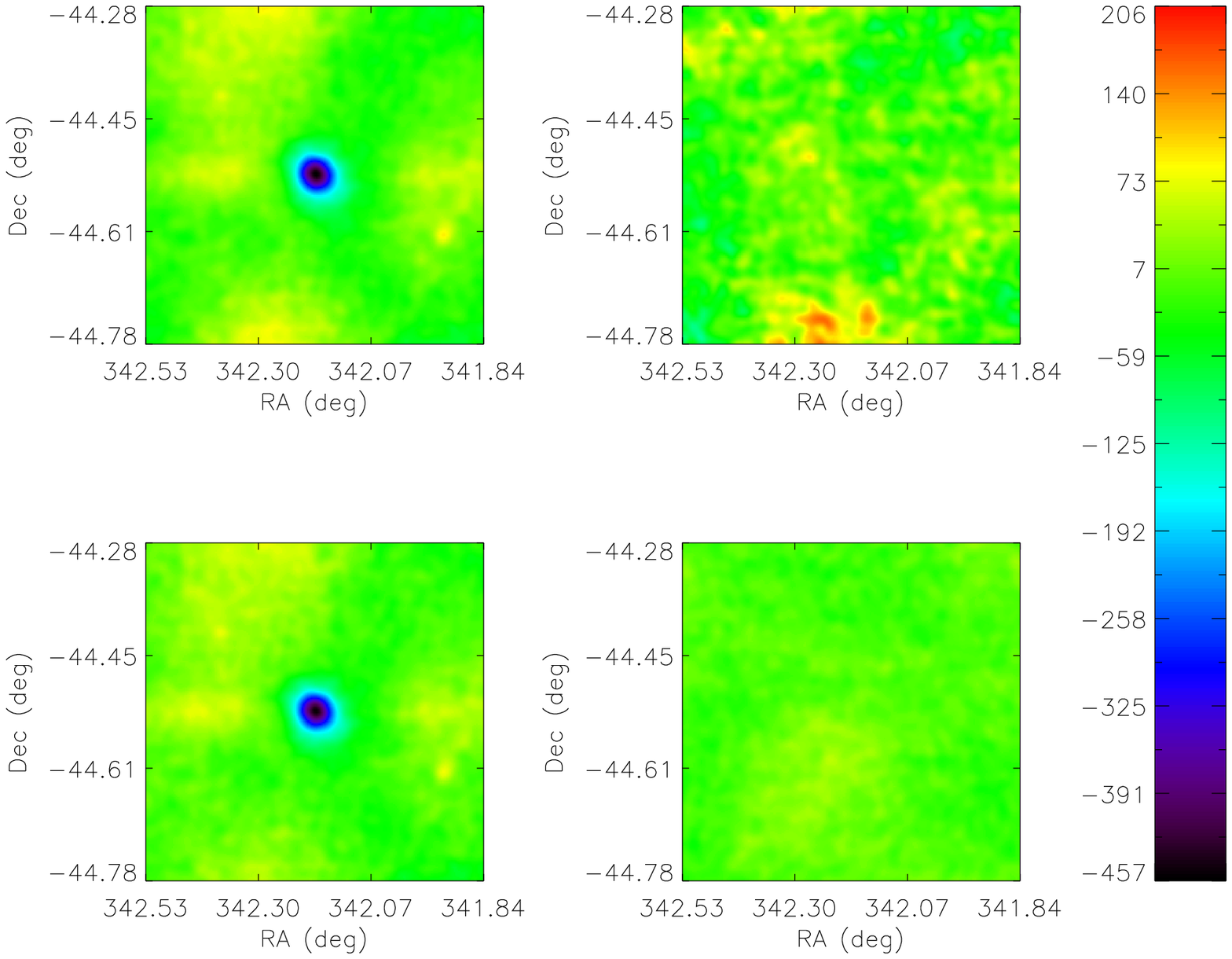} &
\includegraphics[width=.4\textwidth]{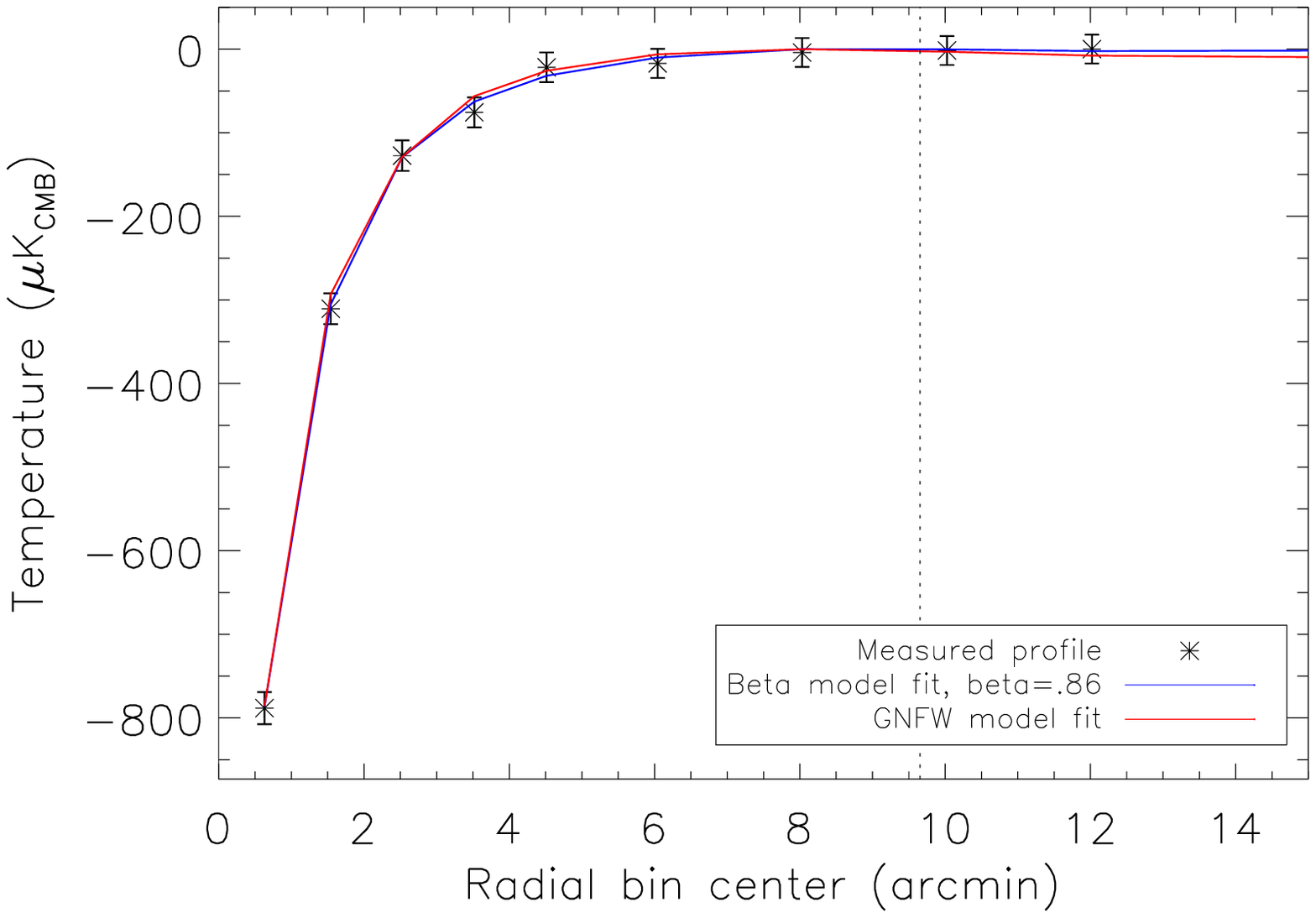} \\
\end{tabular}
\caption{AS~1063 maps (left) and profile (right).  Units
are $\mu K_{\mathrm{CMB}}$.}
\end{figure*}

\end{document}